\documentclass[iop, revtex4, twocolumn, appendixfloats, numberedappendix]{emulateapj}
\usepackage{amsfonts}
\usepackage{amsmath}
\usepackage[usenames]{color}
\usepackage{verbatim}
\usepackage{graphicx}
\usepackage{subfigure}
\usepackage{hyperref}
\usepackage{longtable}
\usepackage{lscape}
\usepackage{natbib}

\def\teff{T_{\rm eff}}
\def\logg{{\rm log}\,g}

\begin{document}

\shorttitle{Extended IRTF Library}
\shortauthors{Villaume et al.}

\title{The Extended IRTF Spectral Library: Expanded coverage in metallicity, temperature, and surface gravity}

\author{Alexa Villaume\altaffilmark{1}, 
Charlie Conroy\altaffilmark{2},
Benjamin Johnson\altaffilmark{2}, \\
John Rayner\altaffilmark{3},
Andrew W. Mann\altaffilmark{4}, and
Pieter van Dokkum\altaffilmark{5}
}

\altaffiltext{1}{Department of Astronomy and Astrophysics, University of California, Santa Cruz, CA 95064, USA, avillaum@ucsc.edu} 
\altaffiltext{2}{Harvard-Smithsonian Center for Astrophysics, Cambridge, MA 02138, USA}
\altaffiltext{3}{Institute for Astronomy, University of Hawaii, Honolulu, HI 96822, USA}
\altaffiltext{4}{The University of Texas at Austin, Austin, Texas }
\altaffiltext{5}{Yale University Astronomy Department, New Haven, CT 06511, USA}

\begin{abstract}
We present a $0.7-2.5\mu m$ spectral library of 284 stars observed with the medium-resolution infrared spectrograph, SpeX, at the 3.0 meter NASA Infrared Telescope Facility (IRTF) on Maunakea, Hawaii. This library extends the metallicity range of the IRTF Cool Star library beyond solar metallicity to $-1.7 <$ [Fe/H] $< 0.6$. All of the observed stars are also in the MILES optical stellar library, providing continuous spectral coverage for each star from $0.35-2.5\mu m$. The spectra are absolute flux calibrated using Two Micron All Sky Survey photometry and the continuum shape of the spectra is preserved during the data reduction process.  Synthesized $JHK_S$ colors agree with observed colors at the $1-2\%$ level, on average.  We also present a spectral interpolator that uses the library to create a data-driven model of spectra as a function of $\teff$, $\logg$, and [Fe/H]. We use the library and interpolator to compare empirical trends with theoretical predictions of spectral feature behavior as a function of stellar parameters. These comparisons extend to the previously difficult to access low-metallicity and cool dwarf regimes,  as well as the previously poorly sampled super-solar metallicity regime. The library and interpolator are publicly available.
\end{abstract}


\section{Introduction}

Stellar libraries, whether empirical or theoretical, are foundational to several different fields of astrophysics. In stellar population synthesis (SPS) models stellar libraries are needed to convert the stellar evolution predictions of stellar parameters: effective temperature ($T_{\rm eff}$), surface gravity (${\rm log}g$), and metallicity ([Fe/H]), into spectral energy distributions (SEDs) \citep{conroy2013}. Stellar spectra can be used to compute the line-of-site velocity distributions in galaxies \citep[][]{cappelari2004, emsellem2004}. Well-characterized stellar spectra are also a key ingredient in exoplanet studies \citep[e.g,][]{newton2014, mann2015}.

The goal for any stellar library is extensive coverage in stellar parameter space and large wavelength coverage at comparable or better resolution than typical observations to which they are compared. Large wavelength coverage is important because different wavelength regimes probe different stellar populations in the integrated light of galaxies. The optical is the most well-studied wavelength regime but there is important information that is only available in other regimes.  For example, the ultra-violet (UV) probes the populations of hot, massive stars \citep[e.g.,][and the references therein]{vazdekis2016} and the near infrared (IR) probes the populations of cool, evolved stars such as those on the asymptotic giant branch \citep[][]{athey2002, martini2013, villaume2015, simonian2016}. Furthermore, the endeavor to characterize biological signatures of exoplanets must consider the effect and characteristics of the host stars, which requires UV-IR coverage \citep{france2015}. Moreover, uniform coverage in parameter space is also necessary to ensure accurate stellar population models. Unfortunately, there is still no single stellar library that covers the entire range of wavelength and parameter space.

\begin{figure*}[t]
	\includegraphics[width=1.0\textwidth]{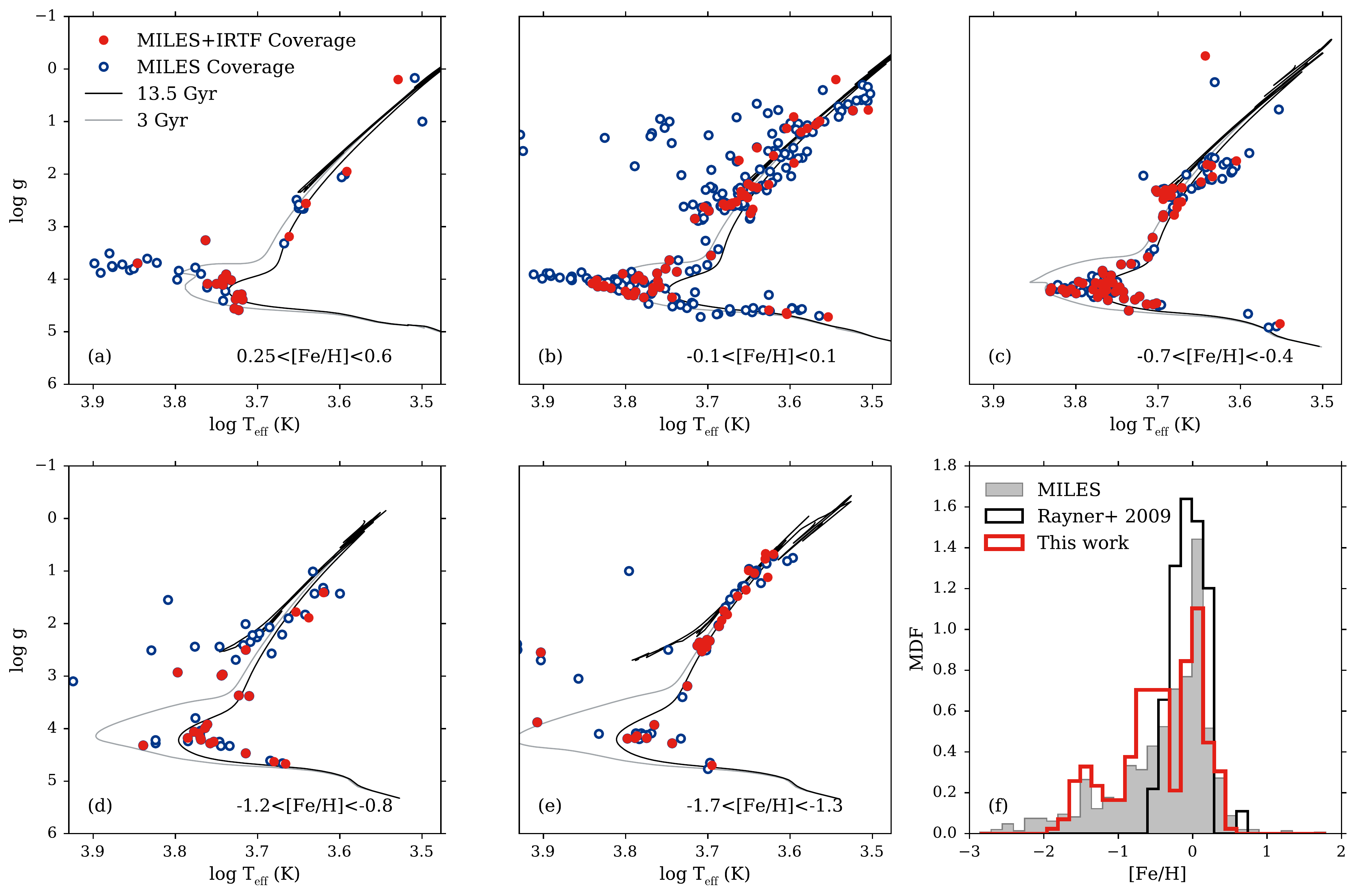}
	\caption{(Panels a-e) Comparison of the MILES spectral library (blue) with the library presented in this work (red). All the stars in this work were selected from the MILES library. Stars were chosen to sample isochrones spanning from 3 Gyr (grey line) to 13.5 Gyr (black line) with metallicities [Fe/H] = 0.25, 0.00, -0.50, -1.00, and -1.50. (f)  Comparison of the metallicity density functions (MDFs) of the MILES spectral library (solid grey), the \citet{rayner2009} (black line), and the library presented in this work (red line). The different libraries have different total number of stars so the MDFs were normalized such that the integral over the range of metallicity values is 1 for each library, indicating at any given metallicity value what the probability that a given star has that metallicity for each library.}
	\label{fig:target_selection}
\end{figure*}

Empirical stellar libraries have been available for over 30 years \citep[e.g.,][]{gunn1983, pickles1985, diaz1989, silva1992, pickles1998, worthey1994, jones1999, lancon2000, cenarro2001, prugniel2001, leborgne2003, valdes2004, sanchez2006, marm2008, rayner2009, ardila2010, sharon2010}. A key limitation of existing stellar libraries is their often limited coverage in stellar parameter space. This is in some respects a fundamental problem because not all stars of interest are close enough to enable detailed observations (e.g., hot metal poor stars).  Standard observational constraints such as atmospheric absorption and sky emission, which is especially prominent in the IR, flux calibration, wavelength coverage, and spectral resolution further challenge the development of comprehensive empirical stellar libraries.

Theoretical libraries offer the advantage of dense coverage in parameter space, arbitrarily high spectral resolution, and no need to correct for atmospheric absorption or flux calibration \citep[e.g.,][]{munari2005, martins2005, coelho2005}. However, theoretical stellar libraries are only as good as the available atomic and molecular line lists and the approximations made in computing the models, e.g., the assumption of local thermodynamic equilibrium (LTE) and 1D plane-parallel atmospheres. The effect of non-LTE is generally taken into account where it matters most \citep[e.g., hot stars and metal-poor stars][]{lanz2003, lind2012} and some 3D  theoretical models \citep[e.g,][]{magic2013} have begun to emerge but these techniques have not been widely adopted due to the fact that they are very computationally expensive. Theoretical stellar spectra are most unreliable for very cool stars and very hot stars \citep[][]{allard1997, martins2007, bertone2008, allard2013, rajpurohit2014, rajpurohit2016}. The former is a particularly acute problem for both exoplanet and galaxy studies. All the habitable zone planes found by TESS will be around M dwarfs \citep{france2015} and counting cool dwarf stars in the absorption lines of integrated spectra of galaxies has emerged as a way to constrain the initial mass function (IMF) \citep[][]{spinrad1962, wing1969, cohen1978, frogel1978, kleinmann1986, diaz1989, ivanov2004, cvd2012} in unresolved stellar populations. The limitations of the theoretical stellar spectra in the cool star regime makes it necessary to turn to empirical stellar libraries. 

\cite{sanchez2006} created a landmark empirical optical stellar library, the MILES library. The MILES library, consisting of nearly 1000 stars, covers a wide range of stellar parameter space over the wavelength range $0.35-0.75\mu m$. The MILES library enabled the creation of more precise SPS models which in turn facilitated a greater understanding of galaxies beyond the reach of resolved stellar population studies.

However, as stated previously, the optical window does not contain a complete picture of a stellar population.  A major advance occurred with the release of the \citet{rayner2009} Infrared Telescope Facility (IRTF) spectral library. The creation of this near-IR library was a great step forward for SPS models \citep[e.g.,][]{cvd2012, spiniello2012, meneses2015, rock2016}. A limitation of the IRTF stellar library is its narrow stellar parameter range, all stars being around solar-metallicity. Furthermore, there is very little overlap in the stars between the MILES and IRTF stellar libraries.  The X-Shooter Spectral Library (XSL) is observing many stars with continuous spectral coverage from $0.35-2.5\mu m$ and will be another valuable library once complete \citep[e.g.,][]{chen2014}.

\begin{figure*}[t]
	\includegraphics[width=1.0\textwidth]{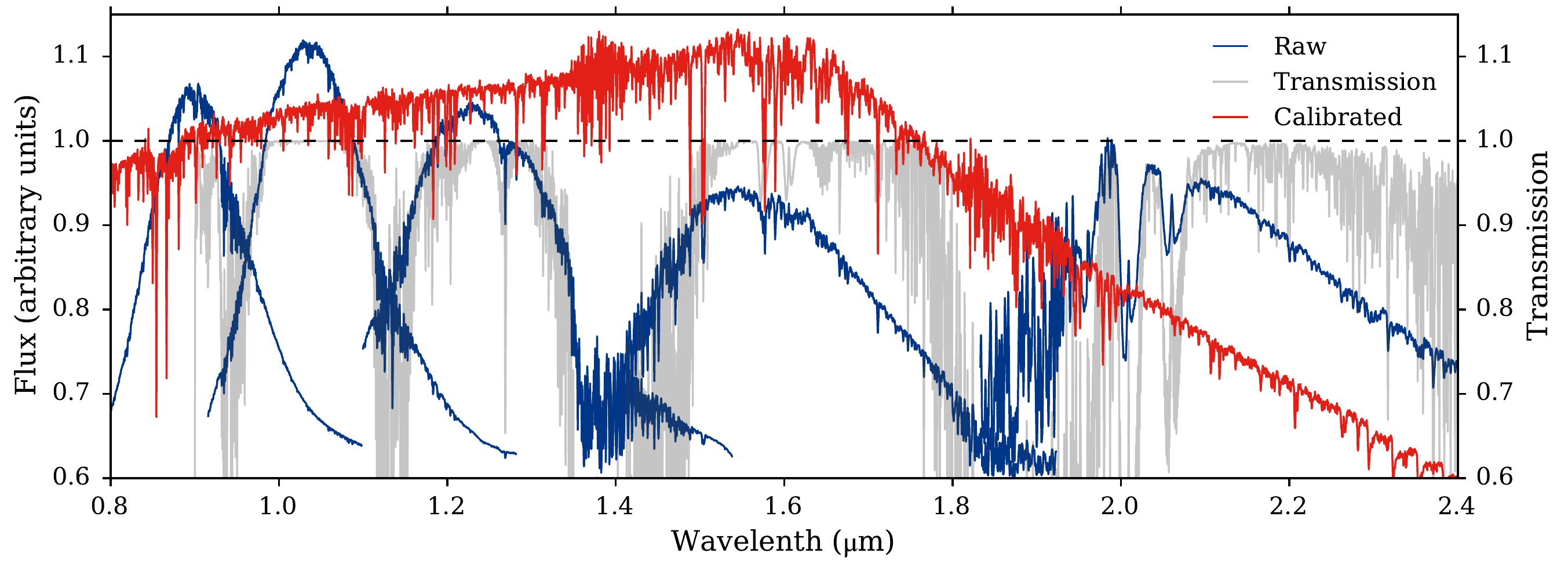}
	\caption{Demonstration of the telluric correction for the star BD+053080. We show the atmospheric transmission spectrum (grey), the spectrum for BD+053080 before the telluric correction (blue), and the spectrum after telluric correction and flux calibration (red).}
	\label{fig:tc_quality}
\end{figure*}

In this work we present the Extended IRTF Library. The extension is two-fold: we expanded the metallicity coverage to $-1.7 < $[Fe/H] $< 0.6$ a large expansion from the just solar-metallicity coverage of the original IRTF library, and we selected all of our objects from the MILES stellar library to provide continuous coverage from the optical through near-IR. In addition to the new stellar library we also present an interpolator that uses a data-driven model created from the new stellar library. The interpolator generates a stellar spectrum as a function of $T_{\rm eff}$, log$g$, and [Fe/H]. This interpolator provides smooth variation in stellar spectra across parameter space, and is an important component in the creation of stellar population models.

The rest of this paper is organized in the following manner: Section 2 details our sample selection, observational strategy, data reduction, and characteristics of the objects and data. In Section 3 we describe our interpolator and assess its quality. In Section 4 we explore the behavior of the library and the interpolator and compare both to theoretical predictions. Finally, in Section 5 we summarize the main points of this work.

\section{The Extended IRTF Library}

\subsection{Sample Selection}

We selected our target stars from the MILES stellar library \citep{sanchez2006}, avoiding known spectroscopic binary stars, to provide continuous coverage from the optical to the near-IR. We selected our targets to span a stellar parameters that would enable the creation of stellar population models for intermediate and old population ages as determined with stellar isochrones from the MIST project \citep[][]{choi2016} .

\begin{figure*}[t]
	\includegraphics[width=1.0\textwidth]{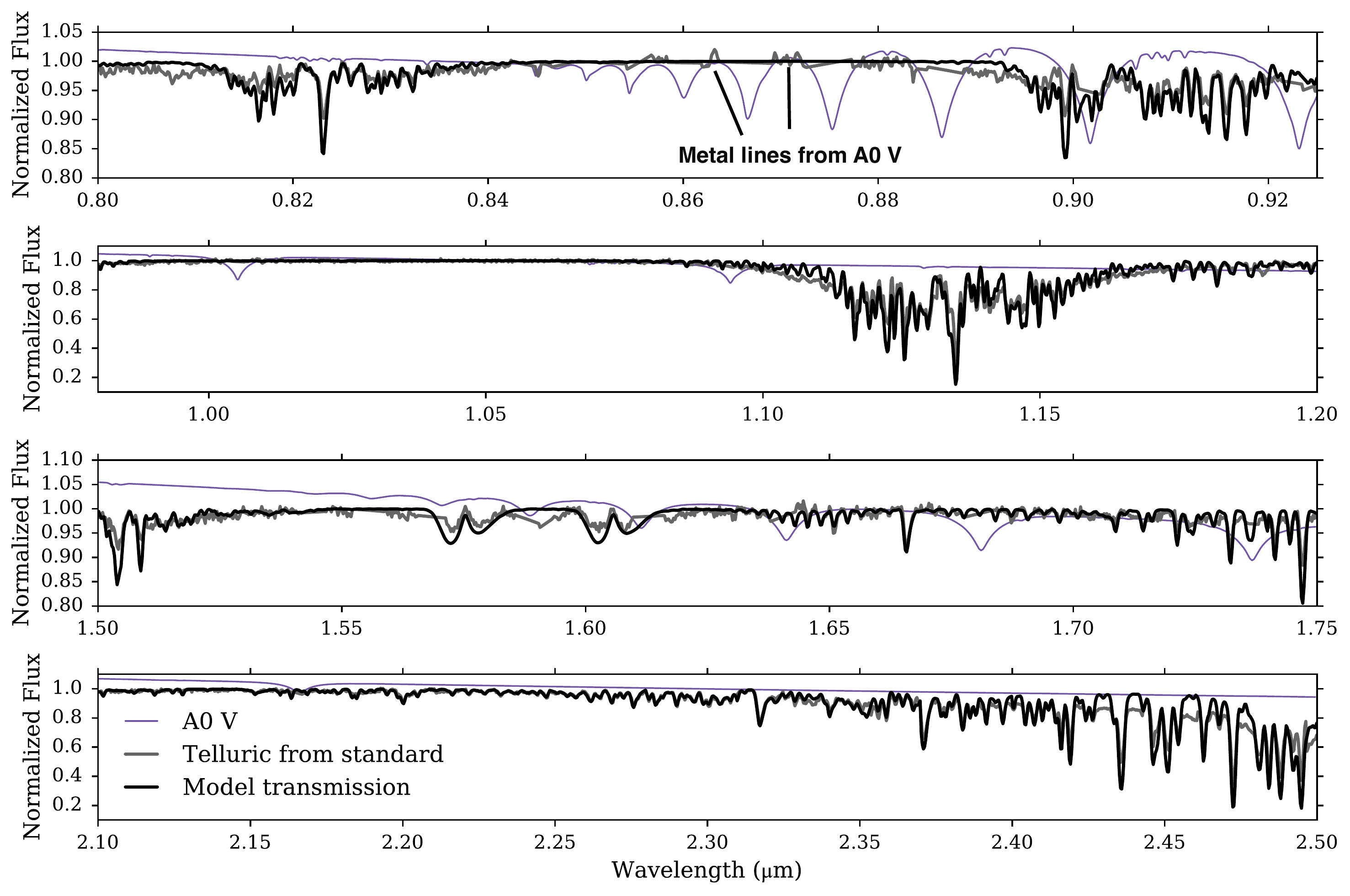}
	\caption{Comparison of a theoretical atmospheric transmission spectrum (black) to an example of a telluric spectrum (grey) used to telluric correct the library spectra and a theoretical spectrum of an A0 V star (purple). In regions of the A0 V spectrum containing strong hydrogen and metallic lines (top panel), the telluric spectrum includes unphysical features. We indicate a few of these features introduced by the metal lines in the A0 V stars in the top panel. Note that no attempt was made to fit the model transmission spectrum to the observations; the model is meant only to guide the eye. }
	\label{fig:tc_check}
\end{figure*}

In Figure~\ref{fig:target_selection} we compare the MILES coverage in $T_{\rm eff}$-${\rm log}g$ space and the stars observed with IRTF for this work. Each panel shows the stars in a given metallicity bin where the stars from the MILES library are represented as blue circles and the stars observed for this library are represented as red circles. Each panel also displays 13.5 Gyr (black line) and 3 Gyr (grey line) MIST isochrones. Selecting targets by eye, we achieved relatively uniform coverage along the intermediate age isochrones for all but the lowest metallicity bins and the old age isochrones across all the metallicity bins. In particular, we have well sampled turn-offs for all except the intermediate-age, low-metallicity stellar tracks. The metallicity distribution of the library is summarized in panel f of Figure~\ref{fig:target_selection}.

\begin{figure*}
	\includegraphics[width=1\textwidth]{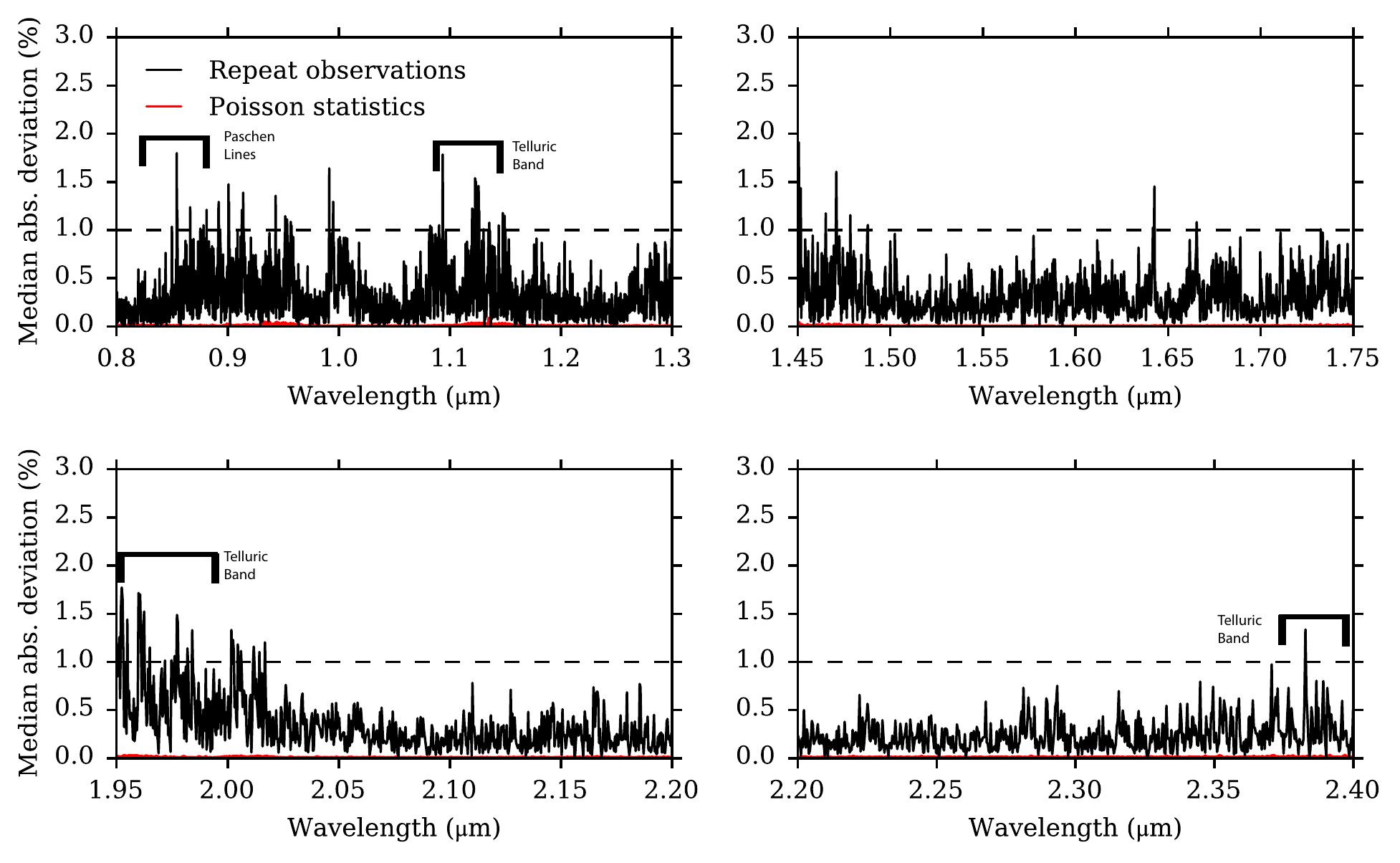}
	\caption{Comparison of the median absolute deviation of the Poisson and empirical uncertainty values of five library stars with repeat observations with the same standard star. We indicate 1\% uncertainty with the dashed line as a guide. The Poisson uncertainty is $\sim$.2\% throughout the wavelength range and underpredicts the empirical uncertainty (difference between the two observations) but nonetheless the median empirical uncertainty is generally $<1\%$. The regions where the uncertainty exceeds $1\%$ are regions contaminated by telluric absorption or where there exist prominent hydrogen absorption features in the A0 V standard stars.}
	\label{noise}
\end{figure*}

\begin{figure*}
	\includegraphics[width=1.0\textwidth]{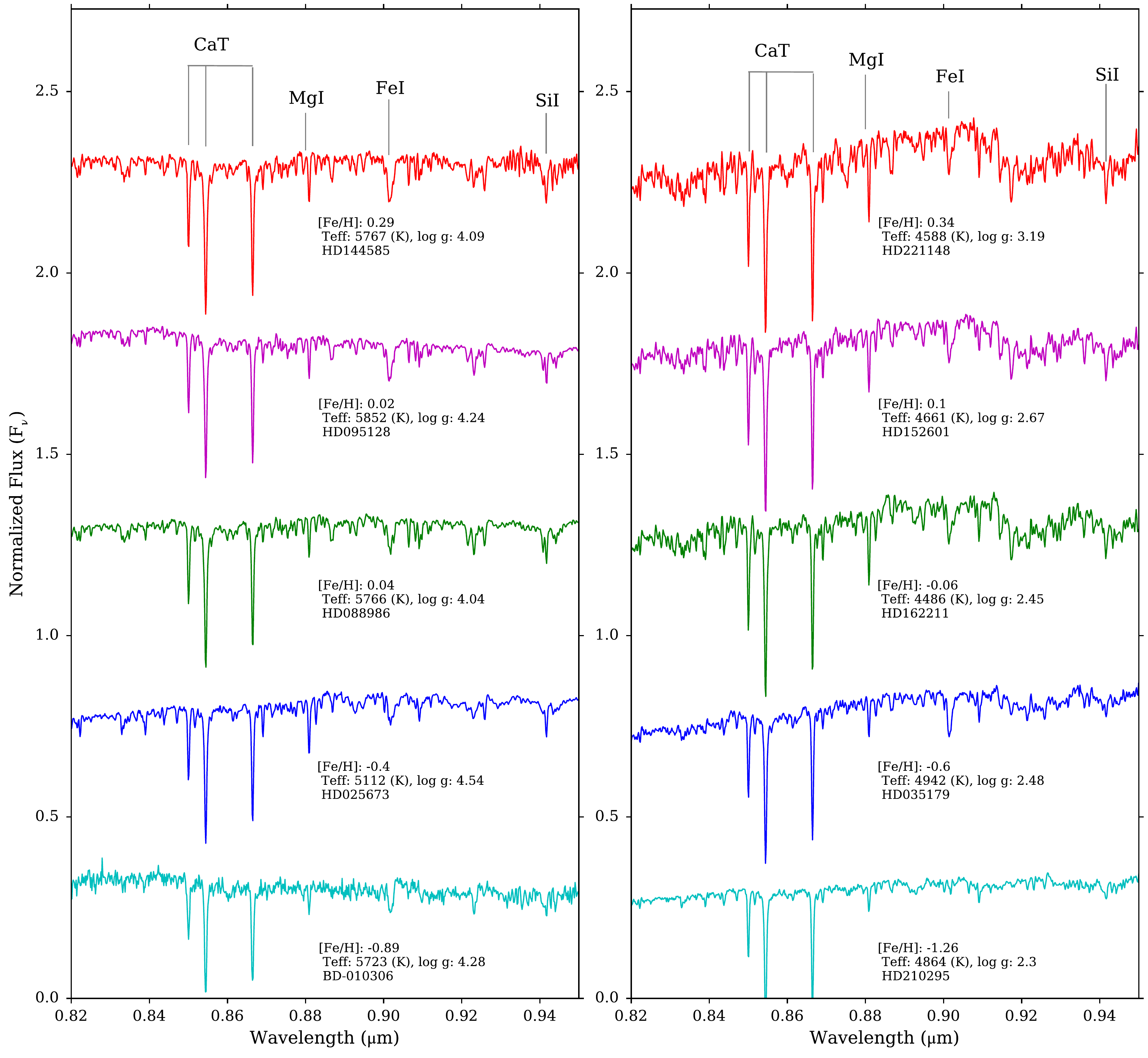}
	\caption{Sequence of stars on the main sequence turn-off (left) and subgiant branch (right) plotted as a function of metallicity metallicity over the $I$ band (0.82-0.95$\mu$m).The spectra have been divided by the median flux value over the wavelength region shown and offset by constants.}
	\label{zoom1}
\end{figure*}

\begin{figure*}
	\includegraphics[width=1.0\textwidth]{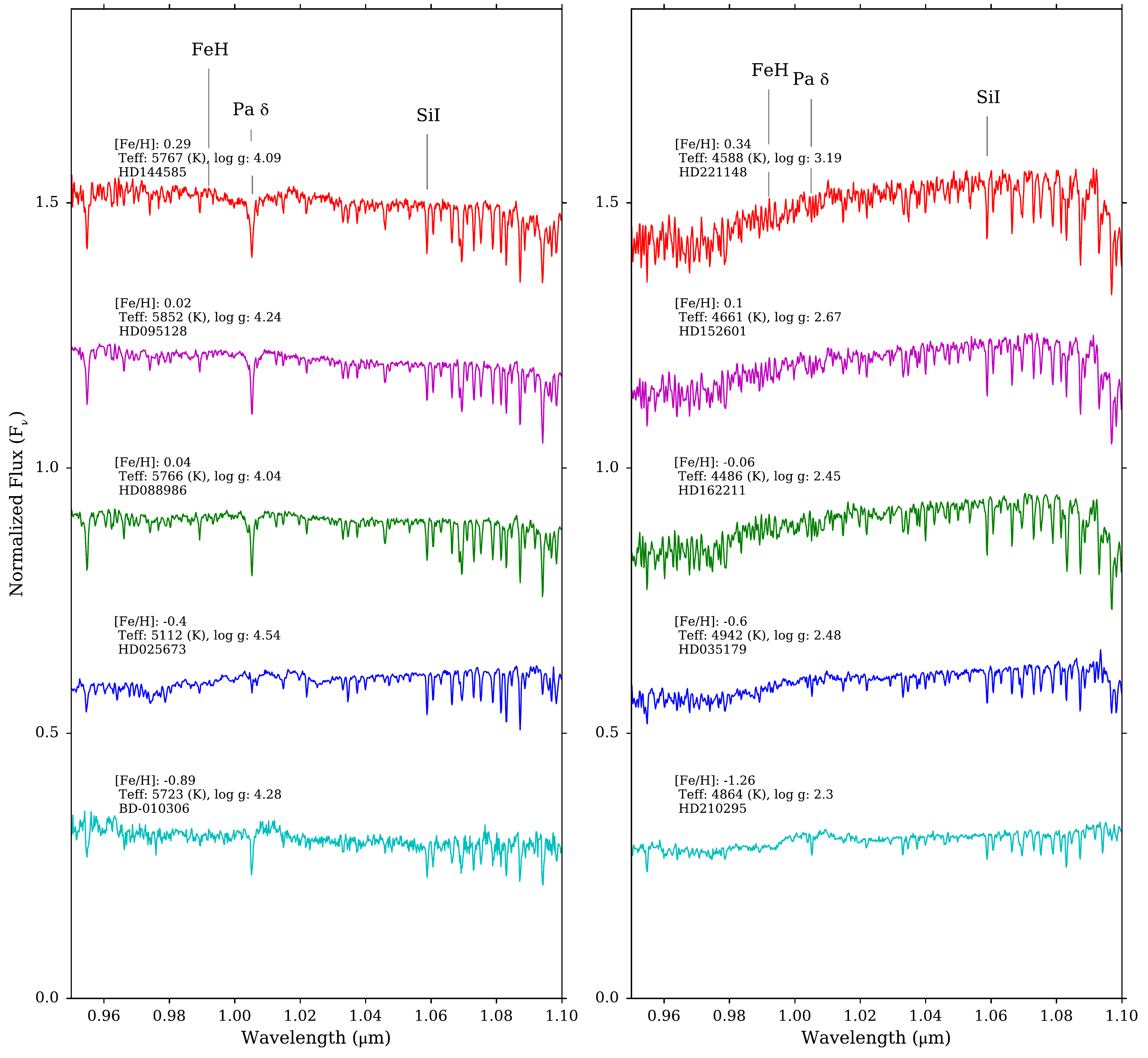}
	\caption{Same as Figure~\ref{zoom1} except over the $Y$ band (0.95-1.10$\mu$m).}
	\label{zoom2}
\end{figure*}

\begin{figure*}
	\includegraphics[width=1.0\textwidth]{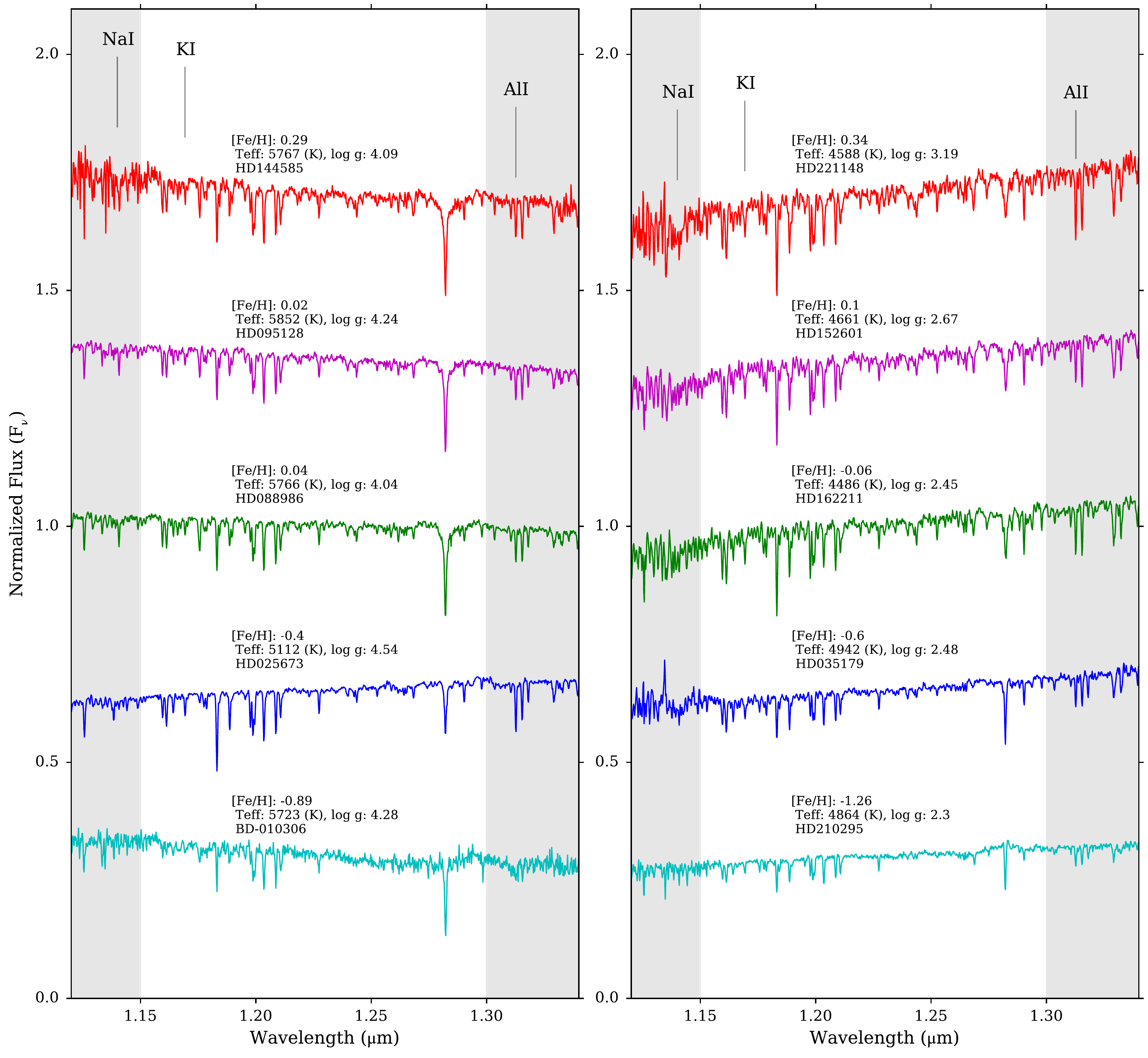}
	\caption{Same as Figure~\ref{zoom1} except over the $J$ band (1.12-1.34$\mu$m). The vertical grey bands mark regions of poor transmissivity due to telluric absorption. }
	\label{zoom3}
\end{figure*}

\begin{figure*}
	\includegraphics[width=1.0\textwidth]{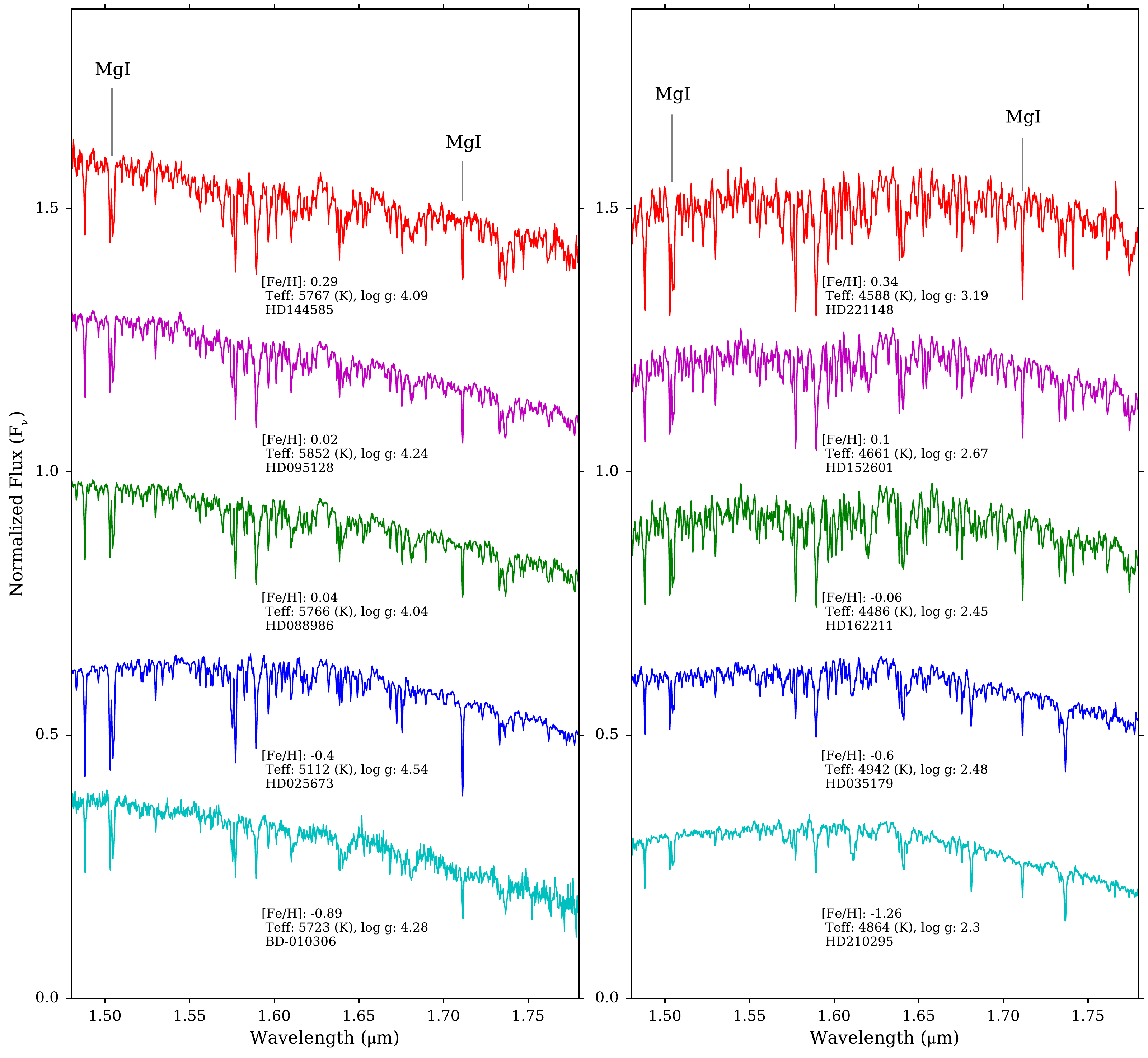}
	\caption{Same as Figure~\ref{zoom1} except over the $H$ band (1.48-1.78$\mu$m).}
	\label{zoom4}
\end{figure*}

\begin{figure*}
	\includegraphics[width=1.0\textwidth]{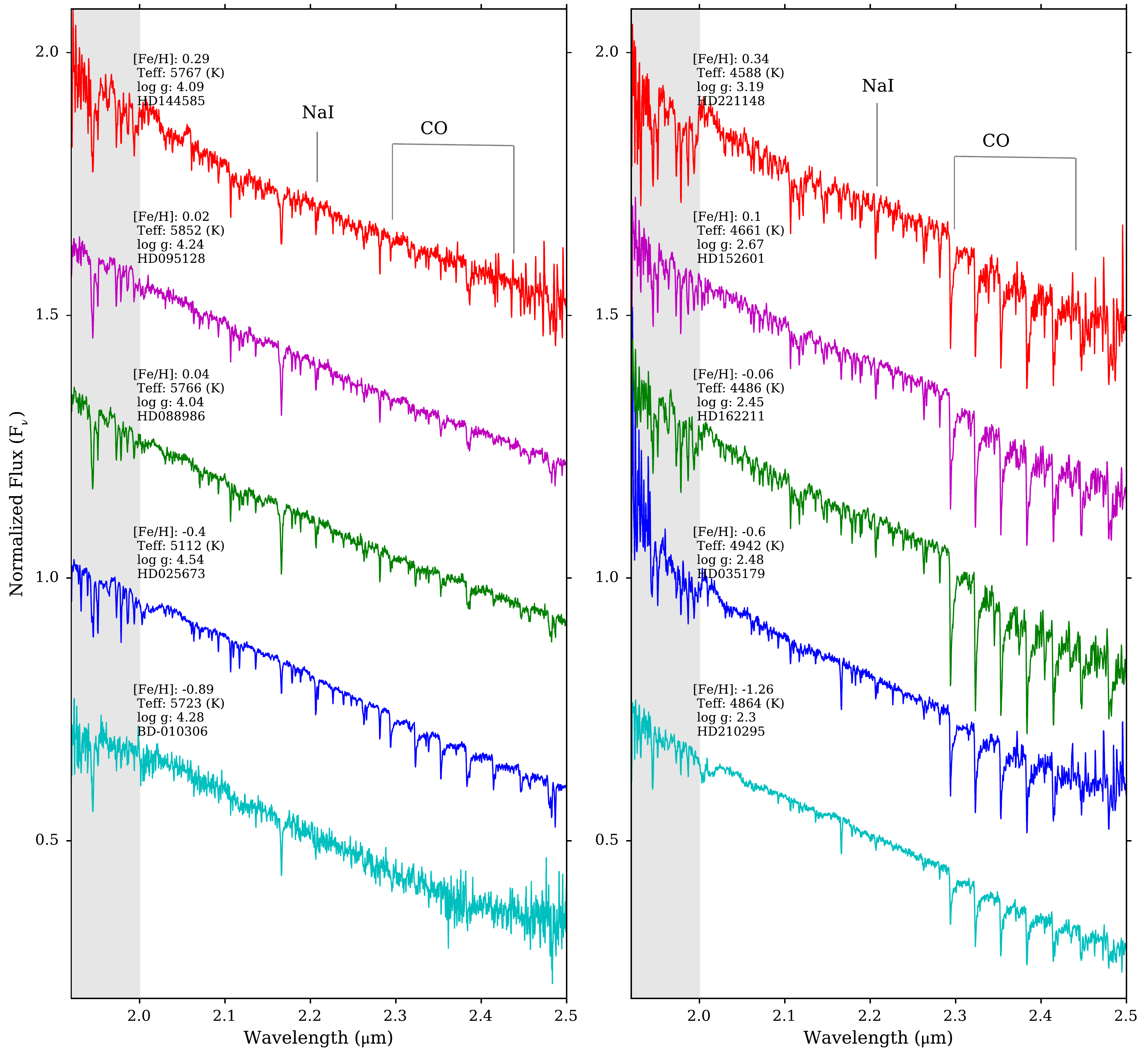}
	\caption{Same as Figure~\ref{zoom1} except over the $K$ band (1.92-2.50$\mu$m). The vertical grey band marks a region of poor transmissivity due to telluric absorption.}
	\label{zoom5}
\end{figure*}

Throughout this work we use the stellar parameters determined by \cite{prugniel2011} and \cite{sharma2016} instead of those reported by \citet{cenarro2007}. The stellar parameters determined by \cite{cenarro2007} were determined heterogeneously, depending on which additional literature data was available for a given star.  As a result, the stellar parameters from \citet{cenarro2007} vary in quality and trustworthiness. \cite{prugniel2011} sought to revise the stellar parameters of the stars in the MILES library in a more robust and homogeneous way. They used ELODIE spectra to perform full-spectrum $\chi^2$ minimization fits \citep[][]{koleva2009} between the MILES spectra and templates built from the ELODIE 3.2 library \citep[][]{wu2011}. \cite{sharma2016} revised the stellar parameter values of MILES library stars with $T_{\rm eff} < 4800$ K with an improved interpolator.

It is important to note that the accuracy of SPS models is predicated upon the accuracy of the stellar parameters. \cite{percival2009} found that a change of 100 K in the effective temperature of stars in a stellar library can propagate into a 20\% error in absolute ages estimated from stellar population models. Furthermore, derived exoplanet properties are a direct function of the stellar parameters of the host star \citep[e.g.,][]{mann2015}. These examples underly the need for accurate stellar parameters in empirical stellar libraries.

We do not make use of the \citet{rayner2009} library (except for comparison with re-observed stars) for two reasons. First, as we will discuss later, the detectors on SpeX have been upgraded since the IRTF library of \citet{rayner2009}.  A significant component of the upgrade included a blueward extension to $0.70\mu m$.  The original IRTF library had a blue cutoff of $0.8\mu m$, which meant that there was no overlap with the red cutoff of the MILES library.  Second, there were few stars that were in both the MILES stellar library and the \citet{rayner2009} library.  Our requirement for continuous coverage from the optical to the near-IR for a common set of stars led us to not make use of the original IRTF library.

\subsection{Observations}

\begin{figure*}[t]
	\includegraphics[width=1.0\textwidth]{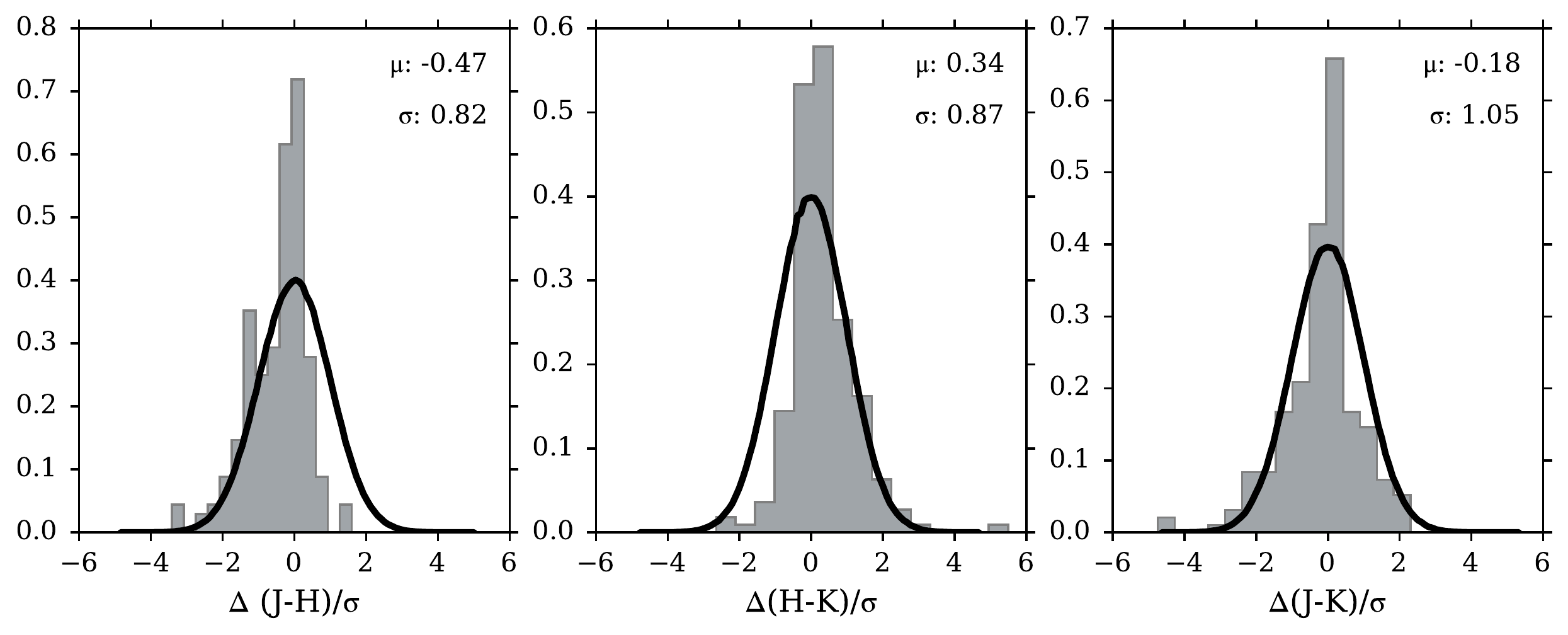}
	\caption{Normalized histogram of the residuals between the observed and synthesized 2MASS  $J-H$, $H-K_S$, and $J-K_S$ colors for a subset of the library stars without quality issues, shape issues, or stellar parameter issues (see later discussion and Table~\ref{table:obs_log}) divided by the uncertainty. We have indicated the mean offset, $\mu$, and standard deviation, $\sigma$. A Gaussian distribution with $\sigma=1$ is also shown.}
	\label{fig:shape_errors}
\end{figure*}

\begin{figure*}[t]
	\includegraphics[width=1.0\textwidth]{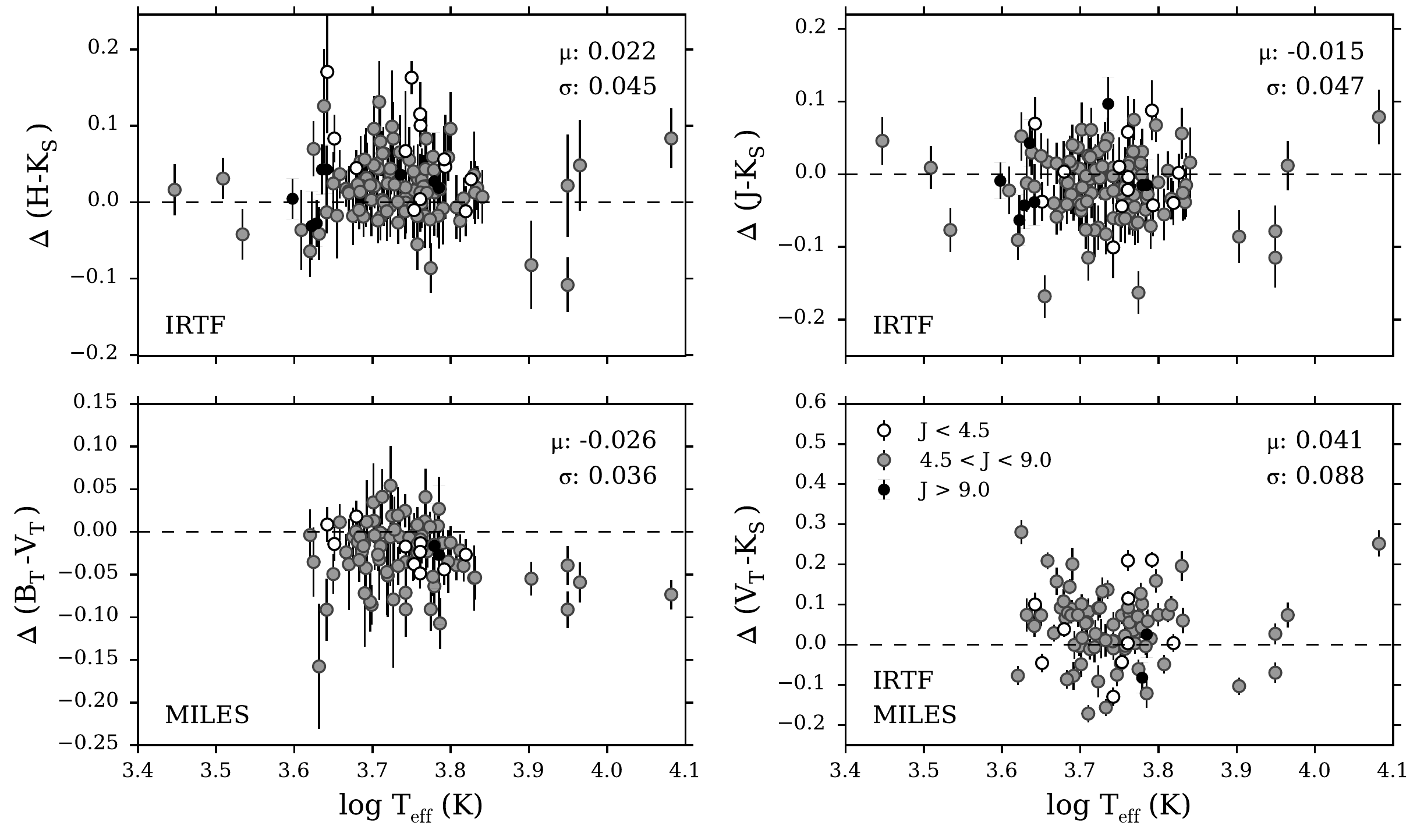}
	\caption{Computed residuals between the observed and synthesized 2MASS $H-K_S$, $J-K_S$, optical $B_{\rm T}-V_{\rm T}$,  and $V_{\rm T} - K_S$ colors of 124 library stars with relatively good (uncertainty $ \leq 3\%$) 2MASS photometry. The plotted error bars show the 2MASS and Tycho errors in magnitudes. The $V_{\rm T} - K_S$  colors are a test of how well the MILES and IRTF stars were stitched together.}
	\label{fig:shape_check}
\end{figure*}

We carried out 24 nights of observations from August 2014 to June 2016 using the upgraded SpeX instrument on IRTF. For details on SpeX we refer the reader to \cite{rayner2003}. In short, SpeX is a medium-resolution, cross-dispersed spectrograph equipped with a 2048x2048 Hawaii-2RG array.  All stars were observed in the short-wavelength cross-dispersed mode (SXD) using the $0\arcsec.3$ (2 pixel) slit, providing a nominal resolving power of R$\sim$2000. We observed every star at the parallactic angle. The upgraded detectors extended the wavelength coverage further into the blue, giving a wavelength range from $0.7 - 2.55 \mu m$. The upgraded detectors also remove the 0.06 $\mu$m gap that used to exist between the $H$ and $K$ bands \citep[][]{rayner2009}.

An A0 V star (or a star of similar spectral type) was observed either before or after each science object to correct for absorption due to the Earth's atmosphere and to flux calibrate the science object spectra. For most stars the airmass difference between the target star and the standard star was less than 0.1. Standard stars were also chosen to be located within 10 degrees of the target object whenever possible, to minimize the effects of any differential flexure in the instrument between the observations of the target and standard.   Due to a shortage of standard stars near some targets, in a few cases airmass differences between the two reached 0.15.

We took a set of internal flat field and argon arc lamp exposures with each target-standard pair. This procedure helps minimize the possible effects of flexure of the detector on the quality of the flat-fielding and the final wavelength solution.

\begin{figure*}[t]
	\includegraphics[width=1.0\textwidth]{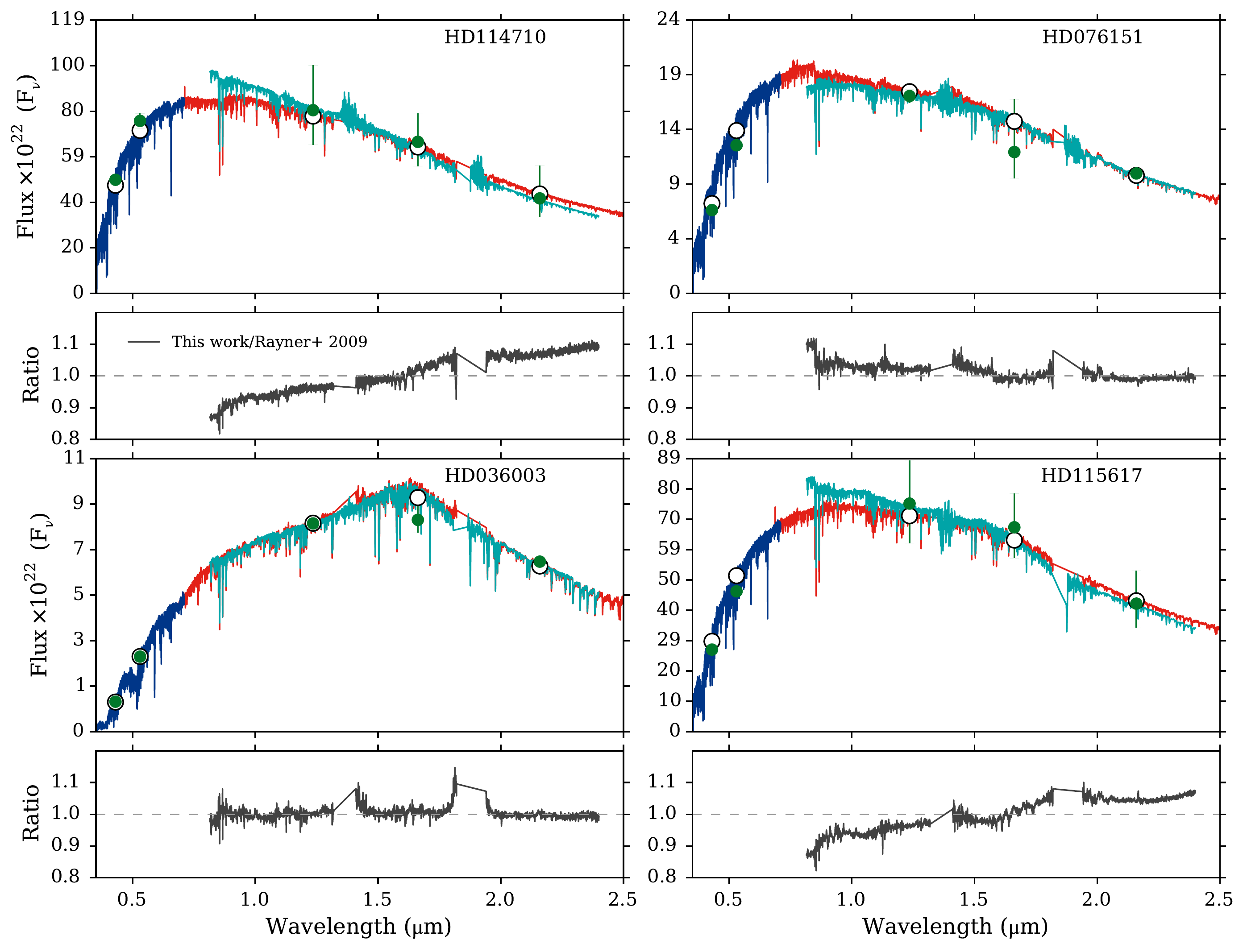}
	\caption{Comparison of new spectra presented in this work (red), spectra from the \citet{rayner2009} library (turquoise), and the MILES spectra (blue) for four stars. Also plotted is the observed photometry (green circles) and synthetic photometry (open circles) derived from the spectra. The shape of the spectra presented in this work match well with the shape of the MILES spectra in the overlap region, and also with the observed  $JHK_S$ photometry. Below each spectral comparison we plot the ratio of the IRTF spectra presented in this work and the spectra from \citet{rayner2009}.}
	\label{fig:oldlib_compare}
\end{figure*}

A table summarizing the content of the library including the object names, $K_S$-band magnitudes, associated telluric standard stars, and various other characteristics for all the 284 library stars is presented in Table~\ref{table:obs_log}.

\subsection{Data Reduction}

\begin{figure}[t]
	\includegraphics[width=0.5\textwidth]{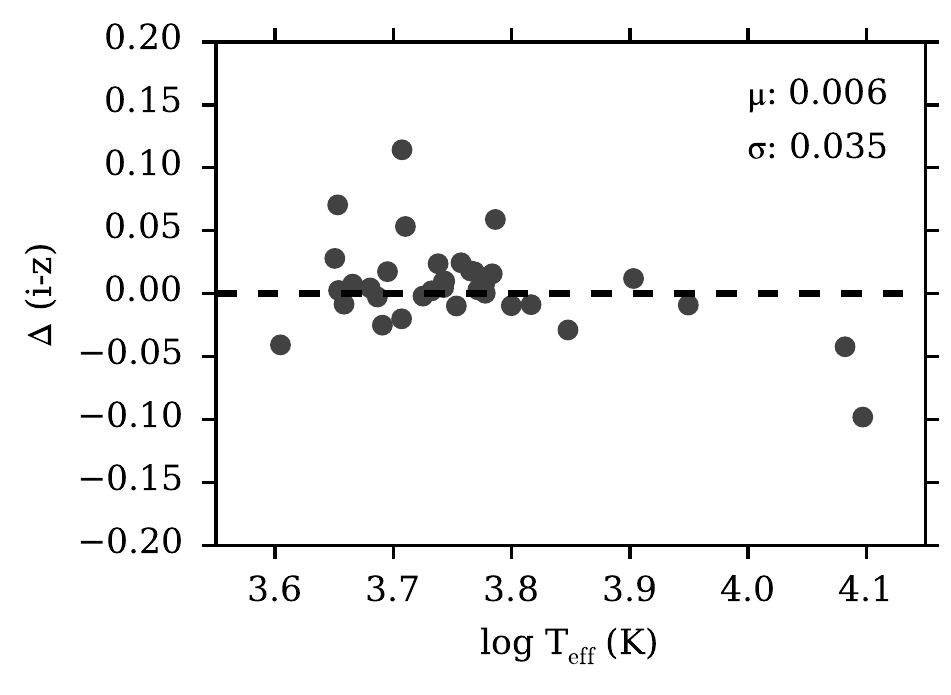}
	\caption{ Computed residuals between the NGSL and IRTF synthetic SDSS $i-z$ colors for 35 stars that are in both libraries.}
	\label{fig:blue_shape_check}
\end{figure}

Data were reduced following the same procedure described in \citet{rayner2009} using the facility IDL-based reduction package for SpeX, Spextool v4.0.4 \citep{cushing2004}. \citet{cushing2004} provides a comprehensive explanation of the data reduction process and so here we summarize and highlight the main steps.

The science images were processed by correcting for nonlinearity, subtracting the pairs of images taken at two different slit positions, and dividing the pair-subtracted images by a normalized flat field. Any residual sky signal left over from the pair subtraction is removed from the image and the spectra in individual orders were then optimally extracted. Argon lines were used to wavelength calibrate the data. For the SXD observing mode, several arc exposures are combined to increase the signal-to-noise ratio (SNR).

We used the A0 V standard star observed with each science object to correct for the telluric absorption and flux calibrate the science object using the method described by \citet{vacca2003}. In brief, the method uses a theoretical spectrum of Vega to determine the intrinsic spectrum of any A0 V star by scaling and reddening the Vega model to match the near-IR magnitudes of the observed A0 V stars and modifying the model to account for differences in line strengths, radial and rotational velocities, and spectral resolution. This method also corrects for instrument throughput and flux calibrates the spectra of the target stars.

In Figure~\ref{fig:tc_quality} we show the SpeX spectrum for the library star BD+0503080 before (blue) and after (red) the telluric correction. The before and after spectra have been vertically scaled to clarify the comparison. Also shown is atmospheric transmission at Maunakea for an airmass of 1.5 and precipitable water vapor of 3mm \citep{lord1992}. The before spectrum has areas of very low signal and features in wavelength regions that align with the deepest parts of absorption in the atmospheric transmission spectrum. The red line demonstrates that much of the noise in this region is decreased after the telluric correction. We subsequently mask the regions in the IRTF spectra where the atmospheric absorption is most prominent, 1.32 $\mu$m-1.41$\mu$m and 1.82$\mu$m-1.94$\mu$m.

This telluric correction procedure is quite effective in removing the effects of the atmosphere.  However, A0 V stars have prominent hydrogen absorption lines and weak metal features, which can complicate the telluric correction. In Figure~\ref{fig:tc_check} we show a theoretical spectrum of an A0 V star (purple) comparef to an atmospheric transmission (black, the model for the first three panels is from \citet{lord1992} and the model for the fourth panel is from \citet{clough2005}), and an example of a telluric spectrum used to correct the library stars (grey). Both the A0 V and transmission spectra have been degraded to a resolution of $R\equiv 2000$. The top panel shows the region around the Paschen break, which can be particularly difficult to fit with models at the percent level. We also see that there are features in the telluric spectrum that are likely spurious and are associated with the metal lines in A0 V stars. We have labeled several of the metal features in Figure~\ref{fig:tc_check}. Our telluric correction method also relies on modeling and removing the A0 V spectrum by modifying the model spectrum of Vega to match the A0 V star. This method is better than interpolating over the hydrogen lines, as there are telluric features at those wavelengths, but there are imperfections to this process. We address the uncertainties associated with these difficulties shortly. The telluric correction appears to be well-behaved in regions away from hydrogen lines and the weak metal lines.

After the telluric correction the spectra extracted from different orders were merged into a single, continuous spectrum. In \citet{rayner2009} an additional scale factor was needed to match the flux levels of the different orders but the updates to the SpeX detector made this step unnecessary. All wavelengths are in vacuum. Spextools provides a measure of uncertainty for each reduced spectrum and the typical formal SNR of the library is very high with a mean value $\sim 500$. However,  the quoted SNR only includes statistical (photon counting) uncertainty, not correlated noise or any source of systemic uncertainty. Most importantly, we are interested in the uncertainty our telluric correction method introduces (see above).

To empirically measure the uncertainty of the library stars we compare five stars with repeat observations: HD007106, HD021197, HD138776, HD201891, and HD204613. For all of these observations we followed the same observing and reduction strategies as described above. The repeat observations were done with the same standard star. We computed the difference between the normalized spectra of the same star and divided by a factor of $\sqrt{2}$ (black line in Figure~\ref{noise}). The Poisson uncertainty is the inverse of the wavelength-dependent average of the SNR for each pair of target observations (red line in Figure~\ref{noise}).

While Figure~\ref{noise} demonstrates that the Poisson uncertainty underpredicts the empirical uncertainty for most of the wavelength range, the empirical uncertainty is almost always $ \sim0.2 - 0.3\%$. The exceptions to this are in regions heavily affected by telluric absorption (upper-right and lower left panels) or where there are many broad hydrogen absorption and metal lines in the A0 V standard stars (upper-left panel). Still, even in these regions the uncertainty is 1-2\%.

The empirical uncertainty presented here encompasses differences in weather and/or observing conditions, potential human error in the reduction process, and correlated noise. It does not assess the uncertainty associated with using different standard stars or using different techniques for telluric correction \citep[e.g.][]{vd2012, kausch2015}. It is not known what effect using a different telluric correction technique would have on the quality of SpeX spectra. Assessing the impact of these additional sources of uncertainty is beyond the scope of this work but will be pursued in the future. 

To demonstrate the quality of the individual features, noise in the continuum after the full reduction, and qualitative metallicity trends we show a sequence of stars ordered by metallicity over the $IYJHK$ photometric bands in Figures~\ref{zoom1}-\ref{zoom5}. The left-panel shows stars roughly on the main-sequence turn-off and the right-panel shows stars roughly on the subgiant branch. We highlight a selection of prominent lines in each bandpass. We qualitatively see general trends with metallicity. However, we are not completely controlling for temperature and surface gravity effects and so it is difficult to say anything definitively about potential trends. In Section 4 we look at equivalent width trends as a function of temperature and surface gravity.

\begin{figure*}[t]
	\includegraphics[width=1.0\textwidth]{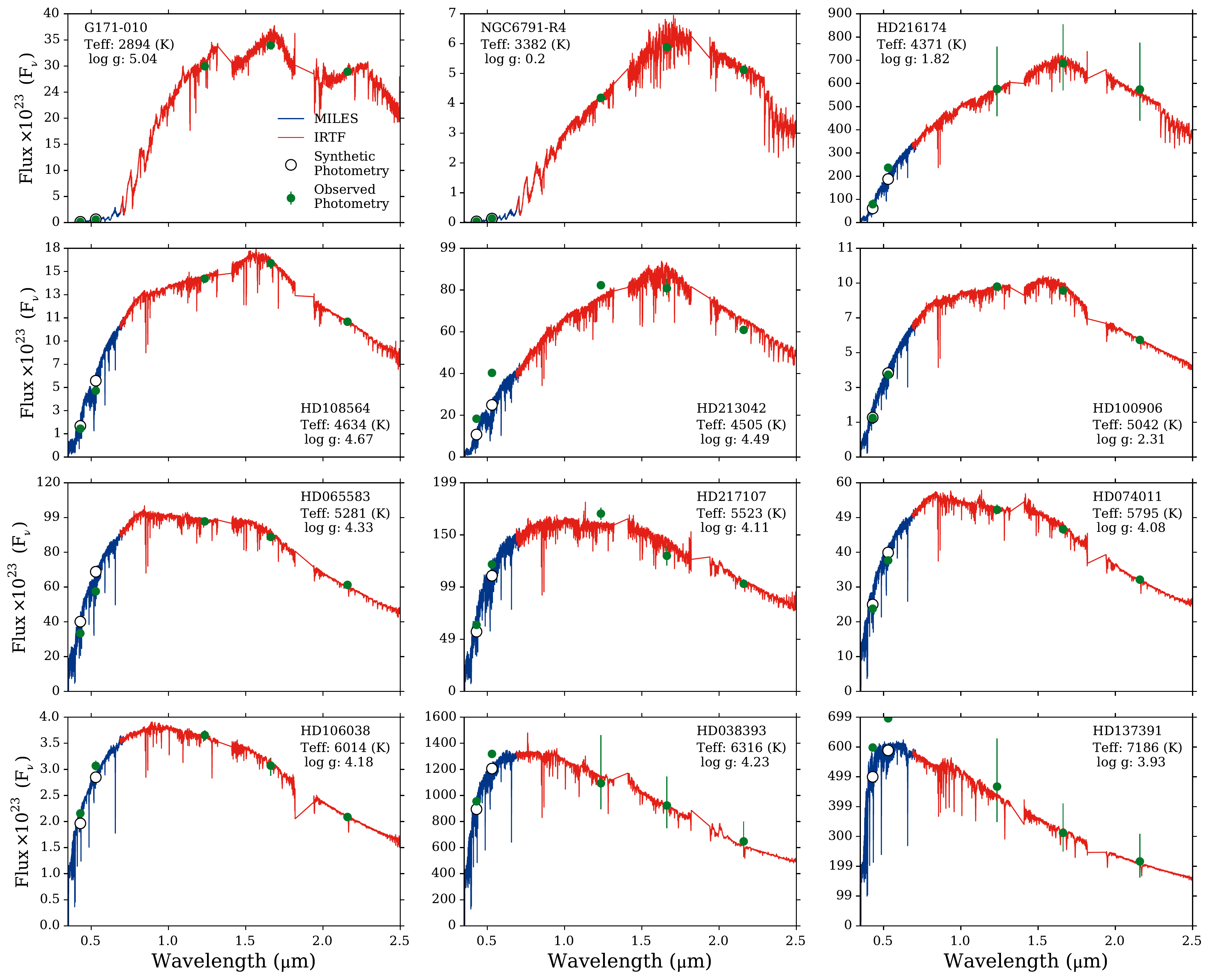}
	\caption{IRTF spectra (red) for a small subset of the library stars plotted with the MILES spectra (blue), observed photometry (green circles), and synthetic photometry (open circles).}
	\label{fig:gallery}
\end{figure*}

\subsubsection{Correcting to Restframe Wavelength}

We used the code \texttt{Prospector} (Johnson et al. in prep)\footnote{https://github.com/bd-j/prospector} to determine line-of-sight velocities for each star, which were used to correct the final reduced spectra to the restframe. Prospector is a code for inference of physical parameters from spectroscopic data via MCMC sampling of the posterior probability distributions.  To obtain estimates of the posterior velocity distribution using prospector, a spectral model is constructed for each star by linearly interpolating the C3K theoretical spectral models\footnote{This is the theoretical library described in \citet{cvd2012} with some minor updates to the line lists.} to the \citet{prugniel2011} and \citet{sharma2016} parameters, and then broadened to the resolution of IRTF. At each MCMC step this spectrum is shifted in velocity and the likelihood of the data given the redshifted model is calculated, after masking regions of strong telluric lines.  

\begin{figure}[t!]
	\includegraphics[width=0.5\textwidth]{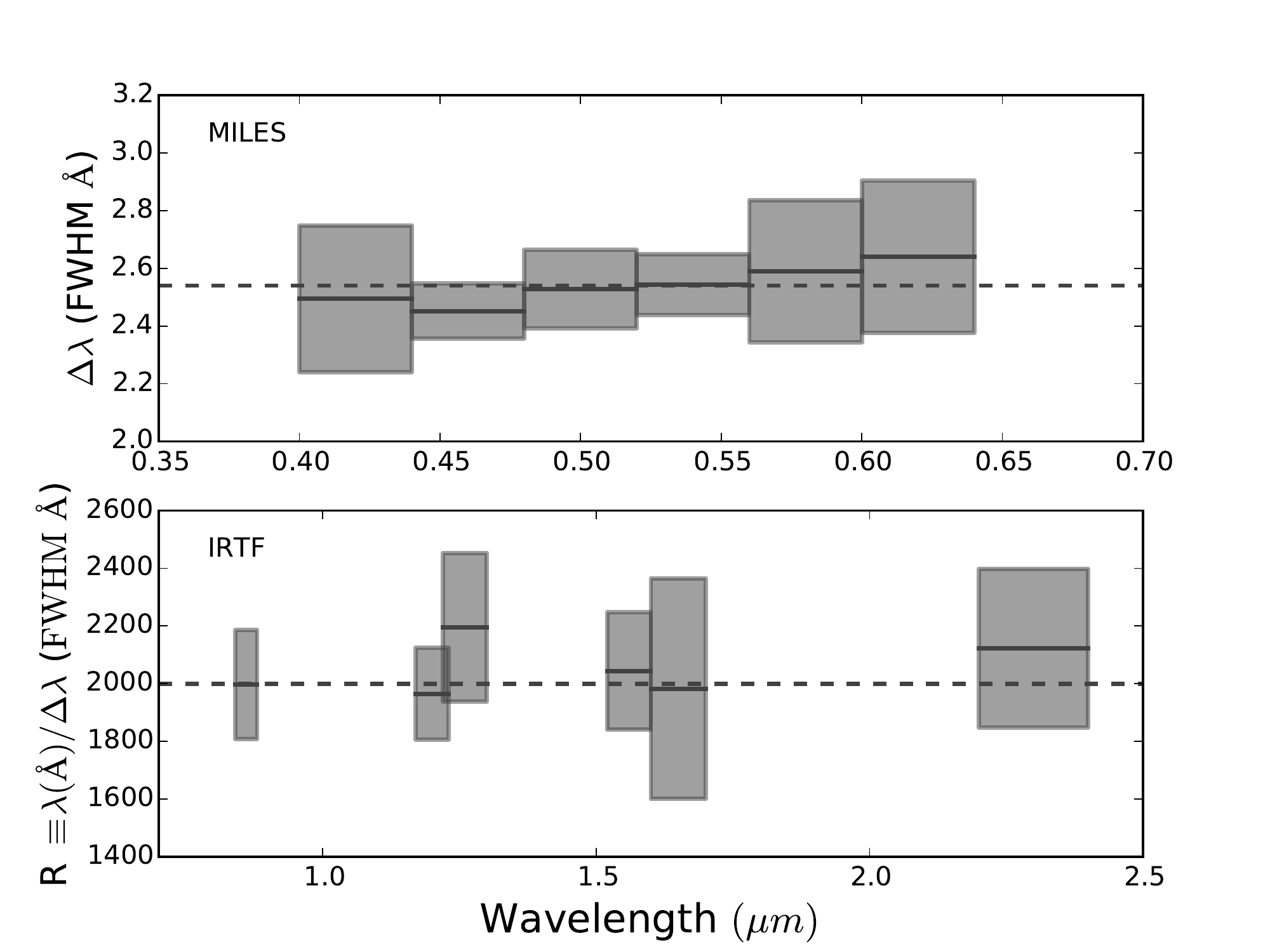}
	\caption{Median resolution (solid line) for 135 stars and the scatter (shaded region) over twelve wavelength segments. The top panel is the wavelength range of the MILES spectra with the resolution quoted as $\Delta \lambda$ (FWHM ${\rm \AA}$), the commonly quoted unit for MILES spectral resolution. We plot a constant line at $\Delta \lambda = 2.54 {\rm \AA}$, the revised resolution found in \citet{beifiori2011}. The bottom panel is the wavelength range of the IRTF spectra and the resolution is quoted in $R\equiv \lambda/ \Delta \lambda$. We plot a constant line at $R = 2000$, the quoted resolution for SpeX. For the MILES spectra we measure a median $\Delta \lambda = 2.54 {\rm \AA} \pm 0.19$ and for the IRTF spectra we measure a median $R = 2020 \pm 230$.}
	\label{fig:resolution}
\end{figure}

\subsection{Flux Calibration}

To de-redden the flux calibrated spectra we used the $E(B-V)$ values given in the MILES stellar library (P. S{\'a}nchez-Bl{\'a}zquez, private communication), 
\noindent
\begin{equation}
f_{\lambda}^{\rm cor}(\lambda) = f_{\lambda}(\lambda) \times 10^{(0.4\times A_\lambda)},
\end{equation}
\noindent
where $ f_{\lambda}(\lambda)$ is the observed spectrum and $A_\lambda$ is the extinction law as a function of wavelength. We adopted the near-IR law given by \cite{fitzpatrick2007} with $R_V = 3.1$,
\noindent
\begin{equation}
A_\lambda = 1.056 \times E(B-V)\lambda^{-1.84}. 
\end{equation}
\noindent
The MILES spectra were de-reddened using the same $E(B-V)$ values.

We absolutely flux calibrated the IRTF spectra using a correction factor based on 2MASS  $JHK_S$ photometry. We computed the correction factor for each star by using the reported 2MASS magnitude, $\mu$, and error, $\sigma$, to create a random normal distribution for each bandpass, $\mathcal{N}\left(\mu, \sigma^2 \right)_{X_{\rm obs}}$, where $X$ denotes a given bandpass. We created a distribution of calibration factors using the following expression, 
\noindent
\begin{equation}
f(C_X) = 10^{(X_{\rm syn} - \mathcal{N}_{X_{\rm obs}})/2.5},
\end{equation}
\noindent
 $X_{\rm syn}$ is the AB synthetic magnitude computed from the spectrum. The observed spectrum for each star is multiplied by a single correction factor,

\begin{equation}
\left<C\right> = \frac{\sum_{X} w_X{\rm Peak}[f(C_X)]}{\sum_X w_X}, 
\end{equation}

\noindent where the weight, $w_X$, is the inverse variance given by the $1\sigma$ deviations from $f(C_X)$, and ${\rm Peak}[f(C_X)]$ is the peak of the distribution of calibration factors.  As in \cite{rayner2009}, this scaling has the effect of shifting the entire spectrum up or down so that the overall absolute flux level is correct, while simultaneously preserving the relative flux calibration of each spectral order derived from the observations and telluric correction procedures. 

The publicly available MILES spectra are normalized to unity so in addition to absolutely flux calibrating the IRTF spectra we needed to absolutely flux calibrate the MILES spectra. We corrected the MILES spectra in a similar manner to the IRTF spectra using the average of the available $BV$ photometry. We preferentially used Tycho \citep[][]{hog2000a, hog2000b} $B_TV_T$ photometry and for stars where that is not available we used Johnson $BV$ photometry taken from Simbad. However, uncertainties in the observed photometry and the different epochs at which the optical and near-IR photometry were taken means that for most stars this correction of the IRTF and MILES spectra does result in perfect agreement in the overlap region. Since we are eventually going to stitch the two spectra together we need them to be at the same flux level, and any incongruity would affect the derived bolometric luminosities and spectral shapes around $0.7\mu m$. We force the MILES and IRTF spectra by computing the percent difference of the flux in the overlap region assuming that the IRTF spectra is at the correct level. Then, to ensure that we do not have flux going into the negative, if the difference is negative we shift the IRTF spectrum by that factor. If the difference is positive we shift the MILES spectrum by that factor.

In the case where 2MASS photometry is not available (e.g. HD 134439) but there is available near-IR photometry from Simbad we used the same basic method as described above but only considered the $K_S$-band photometry, using the raw difference between the observed and synthetic magnitudes. For variable stars in the sample, the absolute flux calibration is only approximate since the 2MASS photometry was obtained at an earlier epoch than the SpeX spectroscopy. However, consistency between the MILES spectra and IRTF spectra is a natural check of the calibration. 

\subsubsection{Quality of Flux Calibration}

We assess the quality of the spectral shape of the IRTF spectra in Figure~\ref{fig:shape_errors}. Here we show a histogram of the error-normalized color differences between the observed and synthesized  2MASS \citep{skrutskie2006} $J-H$ (left), $H-K_S$ (middle), and $J-K_S$ (right) colors. We also show in each panel of Figure~\ref{fig:shape_errors} a Gaussian distribution of $\sigma=1$. Overall the differences between observed and synthesized colors are consistent with the errors in the observed colors. There is some tension in the $J-H$ and $H-K_S$ colors but not in $J-K_S$, suggesting that there might be some modest issue with the $H-$band normalization.

We further examine the quality of the flux calibration by computing the residuals between observed and synthetic photometric colors.  For this exercise we choose stars with 2MASS photometry better than 3\% for the $JHK_S$ bands and with good quality spectra and well-determined stellar parameters (see Table~\ref{table:obs_log} and later for how we determine this); 124 stars in total. In Figure~\ref{fig:shape_check} we show these residuals as a function of effective temperature for the colors $H-K_S$ (upper-right), $J-K_S$ (lower-left), $B_{\rm T}-V_{\rm T}$ (we only display stars with Tycho photometry in this plot) and $V_{\rm T} - K_S$ (lower-right). The observed photometry have been redden corrected using the same $E(B-V)$ values used to correct the spectra. Points are color-coded by $J$ band magnitudes because at  $J\sim9$  and $J\sim4.5$  the 2MASS observing strategy changed to avoid non-linearity and saturation for bright stars. This will affect the photometry and corresponding error bars.

 For each color we find mean residuals of $0.022$, $-0.015$, $-0.026$, and $0.041$, respectively. The residuals for the near-IR colors are consistent with those of \cite{rayner2009} and the $B_{\rm T}-V_{\rm T}$ residuals are consistent with the result from \citet{sanchez2006}. The $V_{\rm T} - K_S$ residuals are a reflection of the overall uncertainty of the absolute flux calibration and stitching of the MILES and IRTF spectra.  Given the large wavelength baseline and the complications associated with stitching together different spectral datasets, it is not surprising that this color shows the largest offsets between synthesized and observed data. 

We explore the reliability of the IRTF spectra further in Figure~\ref{fig:oldlib_compare}, where we compare our new IRTF spectra to the spectra in the \citet{rayner2009} IRTF library for four stars in common between the two libraries.  In this figure we also compare to $BVJHK_S$ photometry and the MILES spectra.  The bottom panel for each star shows the ratio between the old and new IRTF spectra. The agreement in the shape is excellent for HD076151 and HD036003, except at the blue end for the former where we see a small, unphysical bump in the flux at $\sim 0.8$ $\mu m$. This is caused by difficulty modeling the Paschen break in the A0 V standard and occurs in a small subset (4\%) of our sample. We found that the stars showing this bump were flux calibrated with standard stars that show nebulosity (labelled A0 Vn in Simbad). This was not an issue in the \citet{rayner2009} library since the detectors did not extend to as blue wavelengths. This issue is addressed further in Section 3.

We explore the reliability of the flux calibration of the blue end of the SpeX spectra by comparing the Extended IRTF library stars with the Next Generation Spectral Library \citep[NGSL][]{gregg2001} which spans $1680$-$10000 {\rm \AA}$. The NGSL spectra were observed on the STIS instrument on the Hubble Space Telescope (HST) so we can use these spectra as flux standards. There are 35 Extended IRTF Library stars that are also in NGSL. To measure how consistent the flux calibration of the Extended IRTF Library is with NGSL we synthesized SDSS $i$ and $z$ photometry (the two passbands that span the blue end of the Extended IRTF Library spectra) for both the NGSL and Extended IRTF Library spectra for the 35 overlap stars. In Figure~\ref{fig:blue_shape_check} we show the residuals between the NGSL $i$-$z$ colors and the IRTF $i$-$z$ colors. The mean residual is sub-1\% with a scatter on the order of a few percent, indicating consistency between the flux calibration of the NGSL spectra and the IRTF spectra.

The shape of the spectra is important for computing bolometric luminosities and in creating stellar population models. Consistency at the level of absorption lines between the old and new IRTF spectra is also important as that will change equivalent widths and affect the accuracy of fitting stellar features using these spectra. From Figure~\ref{fig:oldlib_compare} we can see that the differences between the \citet{rayner2009} stellar features and the stellar features in the spectra from this work are 1-2\%, much smaller than the over all shape and flux level differences.

\begin{figure*}[t]
	\includegraphics[width=1\textwidth]{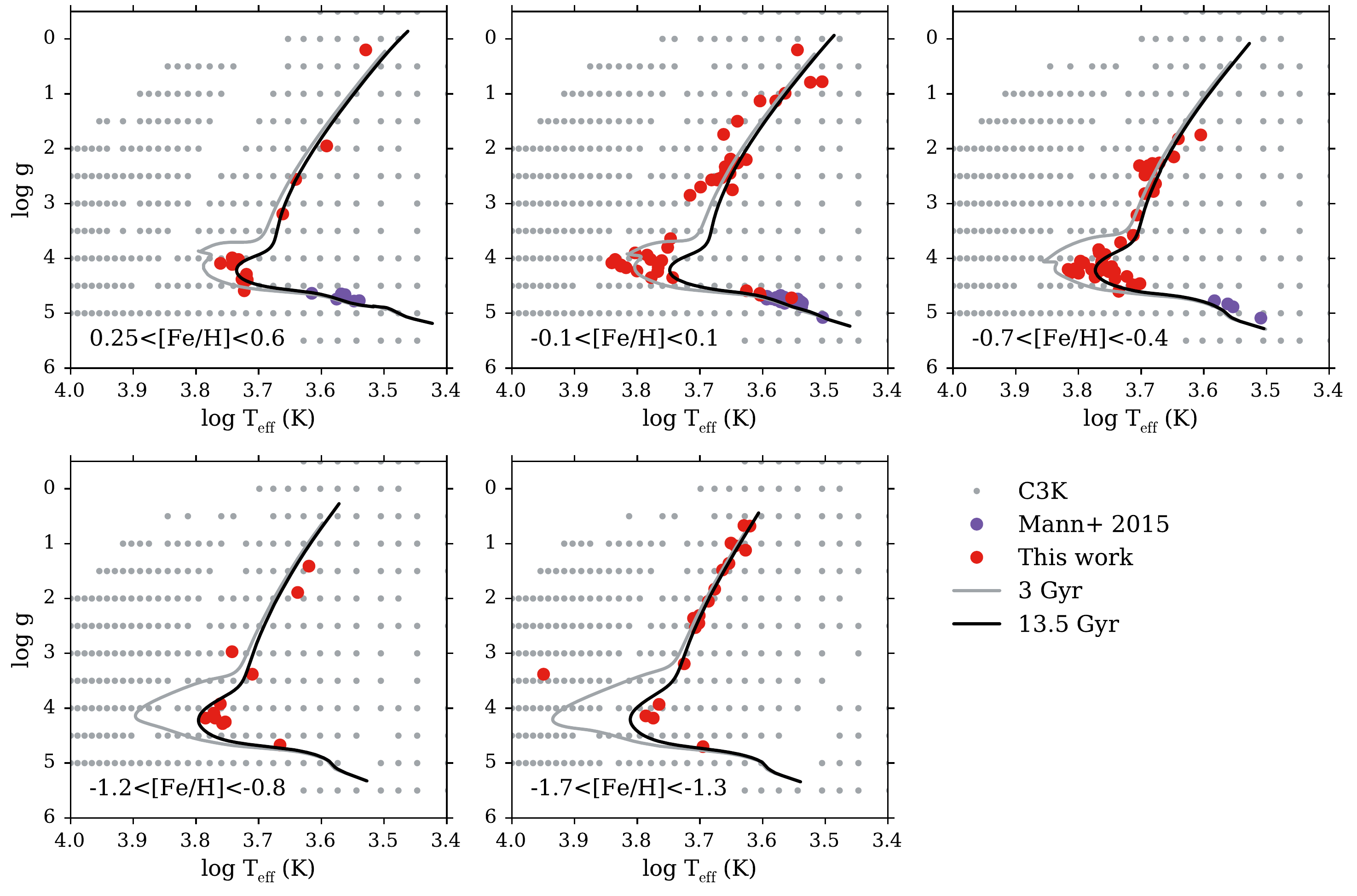}
	\caption{Location of all the stars included in the SPI training set in $T_{\rm eff}$ vs log$g$ space including the stars from the Extended IRTF library (red circles), the M dwarfs from \citet{mann2015} (purple circles), and the C3K theoretical spectra (grey dots). We only use the C3K theoretical spectra outside the convex hull determined by the empirical spectra. Isochrones are the same as in Figure~\ref{fig:target_selection}.}
	\label{fig:psi_train}
\end{figure*}

In Figure~\ref{fig:gallery} we show a small sample of stars in the library. They are ordered from cool (upper left corner) to hot (lower right corner). The reduced, redshift corrected, and absolutely flux-calibrated spectra for all the stars observed as part of the Extended IRTF library are available at the IRTF website \url{http://irtfweb.ifa.hawaii.edu/~spex/IRTF_Spectral_Library/}.

\subsection{Combining the Extended IRTF Library with MILES}

The wavelength solution for the SpeX data is not linear and so there is no constant $\Delta \lambda$ between pixels. In practice this means that the spectra for the different stars in the library are on different wavelength grids. The first step to combining the SpeX spectra with the MILES spectra is to put all the Extended IRTF Library stars on the same wavelength grid. We did this by finding the wavelength range shared by all the stars in the library, $\lambda_{\rm min} = 0.713$ $\mu m$ and $\lambda_{\rm max} = 2.559$ $\mu m$ with 7408 wavelength points (with spacing between the pixels as $\Delta\lambda=2.495 {\rm \AA}$), and interpolated every spectrum onto that grid. 

With the Extended IRTF Library spectra all on a uniform wavelength grid we combine them with the MILES spectra (where we have converted the MILES wavelengths from being in air to being in vacuum) as follows: for each star, we took the weighted average between the MILES and the Extended IRTF spectra between 0.713$\mu m$ and 0.743$\mu m$ using a linear ramp function. The Extended IRTF Library spectra between 0.713$\mu m$-0.743$\mu m$ were placed on the MILES wavelength grid. At 0.728$\mu m$ the blend is half MILES and half IRTF. The final, combined spectrum is a concatenation of the MILES spectrum at $\lambda \leq 0.713 \mu m$, the blended section, and the IRTF spectrum at $\lambda \geq 0.743 \mu m$.

\subsection{Bolometric Luminosity} \label{luminosity}

To compute the bolometric luminosity for each star we extended the combined MILES and IRTF spectrum to $0.03 \mu{\rm m}$ and $ 20 \mu{\rm m}$ using the C3K stellar grid, and integrated the extended spectrum to obtain the bolometric flux. For most of the stars we used parallax based distances to convert from bolometric flux to bolometric luminosity. For the cluster stars we used cluster distances as a proxy for distance to the star \citep[for NGC 6791,][2010 edition, for the rest]{grundahl2008, harris1996}.

14 stars or 5\% have neither parallaxes nor known distances and so the bolometric luminosity cannot be computed for them. 

\subsection{Resolution}

The nominal resolution of the SpeX instrument is $R\sim2000$.   In this section we characterize the wavelength-dependent SpeX resolution, which is essential for using these spectra to model observational data. We also re-measure the MILES spectral resolution for comparison.

The spectral resolution of the MILES/IRTF data was measured by fitting theoretical stellar spectra to these data with the Prospector code.  Briefly, for each wavelength regime of each star we calculate a posterior probability distribution for the line spread function (LSF) width  and residual velocity, marginalized over the stellar parameters. The log-likelihood for the data given these parameters is calculated by simple $\chi^2$ between the data and the smoothed theoretical model.  This model is constructed by tri-linear interpolation of a high resolution version of the C3K theoretical stellar library in the stellar parameters log T$_{\rm eff}$, log$g$, and [Fe/H], which is then smoothed to a fixed resolution in $\Delta \lambda$(FWHM $\rm \AA$) using fast Fourier transforms, and interpolated to the wavelength scale of the data after shifting in velocity.  This smoothing assumes a gaussian LSF with rms width $\Delta \lambda/2.355$. To minimize template mismatch, the stellar parameters log$g$, log T$_{\rm eff}$, and [Fe/H] are allowed to vary from the \citet{prugniel2011} parameters within some tolerance.  The velocity and instrumental resolution $\Delta \lambda$ (${\rm \AA}$) have uniform priors within some reasonable range for each segment. The residual velocities are consistent with zero. This posterior probability is then sampled using MCMC techniques, specifically the ensemble sampling algorithm \citep[][]{goodman2010, fm2013}. The resulting maximum a-posteriori values are reported, corrected for the resolution of the theoretical library.

These fits are done separately for several wavelength regions of the spectrum of each star; the optical spectra are split into 6 regions each 400${\rm \AA}$ wide, while 6 regions in the IRTF spectra are defined around prominent spectral features and avoiding regions of low atmospheric transmission.

In Figure~\ref{fig:resolution} we show the median resolution (solid line) for 135 warm ($3980 K < T_{\rm eff} < 6300$K) stars and the scatter (shaded region) over twelve wavelength segments spanning the full Extended IRTF spectra. In the top panel we show the wavelength segments for the part of wavelength space covered by the MILES spectra and in the bottom panel we show the wavelength segments for the wavelength space covered by the IRTF spectra. In the top panel we show a constant line at $\Delta \lambda = 2.54 {\rm (\AA)}$, the revised resolution found by \citet{beifiori2011}. The median resolution we find over the different wavelength regions of the MILES spectra is $\Delta \lambda = 2.54 {\rm (\AA)} \pm 0.19$ (R $\sim 1970$ at 0.5$\mu m$), consistent with the \citet{beifiori2011} value. In the bottom panel we show a constant line for $R\equiv \lambda/\Delta \lambda =2000$. The median resolution we find over the different wavelength regions for the IRTF spectra is $R = 2020 \pm 230$,  consistent with the quoted value for the SpeX instrument.

\section{Spectral Polynomial Interpolator} \label{spi}

Interpolators have long been used in conjunction with stellar libraries. What began as ``fitting functions'' for specific indices \citep[e.g.,][]{gorgas1993, worthey1994} has now evolved to full spectral interpolators \citep[e.g.,][]{vazdekis2003, prugniel2011, ness2015, sharma2016, dries2016}. Broadly, interpolators are either ``global'', where a polynomial is fit to the input sample points, or ``local'', which essentially averages the nearby data e.g., the linear or spline interpolation is a weighted average of only the data closest in parameter space. Global interpolation is appropriate when the surface is smooth and can be approximated by a (relatively) simple function.  The interpolators of \citet{vazdekis2003} and \citet{dries2016} are local interpolators while the interpolators described in \citet{prugniel2011}, \citet{ness2015}, and \citet{sharma2016} are global.

In this same spirit, as a companion to the Extended IRTF library we created the Spectral Polynomial Interpolator (SPI \footnote{http://github.com/bd-j/spi}). With SPI we create a data-driven, global interpolator which we can use to retrieve a spectrum for a set of arbitrary stellar parameters. In this section we will describe the input to SPI, how we construct the model, the quality of interpolation, and example uses for SPI.

\subsection{The Training Set}

SPI works by fitting polynomial functions of $T_{\rm eff}$, log$g$, and [Fe/H] to the $L_{\rm bol}$-normalized flux of all the stars included in the training set. These fits are carried out independently of the fluxes at each wavelength. Then, for any set of stellar parameters, these polynomial functions can be used to predict or approximate the flux at a given wavelength. 

The primary input into SPI is the IRTF Extended Library, limiting the sample to stars with computed bolometric luminosities (see section 2.4.1), high-quality spectra, and accurate stellar parameters (more on the latter two later; see also Table~\ref{table:obs_log}). The library stars that made the final cut to be included in the training set are shown as red points in Figure~\ref{fig:psi_train}. We have augmented the SPI training set in a couple of ways, which we describe below. In total 194 Extended IRTF Library stars were included in the final SPI training set.

We include the wavelength dependent uncertainties for every star in the training set and weight the fluxes in the training sample by the corresponding uncertainty. 

\subsubsection{Additions to the Cool Dwarf Regime}

We can see from Figure~\ref{fig:target_selection} that there is a paucity of cool dwarf stars in the MILES library and subsequently the library presented in this work. Having few cool dwarf stars would have hindered our ability to build an accurate polynomial model in this regime. We therefore included 76 M dwarf stars presented in \citet{mann2015} to the SPI training set (purple points in Figure~\ref{fig:psi_train}), excluding stars with low SNR. We also found that 4 stars had strong chromospheric Balmer emission lines which we removed from the training set.

The spectra from \citet{mann2015} are a combination of SNIFS ($0.3-0.95\mu m$) and pre-detector upgrade SpeX ($0.8-2.4\mu m$). The resolution of the SNIFS data is coarser (R$\sim 1000$) than the resolution of the MILES data. To put the SNIFS data on the same spectral resolution scale as the MILES data we used a high resolution (R$\sim$10,000) version of C3K to deconvolve the SNIFS portion of the \citet{mann2015} data to the resolution of the MILES data. We did this by producing a spectrum using the stellar parameters of a given star in the \citet{mann2015} sample and made two models, one each for the MILES and SNIFS resolutions. We took the ratio of these two models, interpolated it onto the MILES wavelength grid and multiplied the \citet{mann2015} spectrum by this correction factor.  Visual comparison of a SNIFS and MILES star of the same stellar parameters led us to conclude that this procedure was effective.

We use the stellar parameters as presented in \citet{mann2015} with surface gravity given by,
\begin{equation}
{\rm log} g = {\rm log}_{10}\left(\frac{6.6743\times 10^{-8}\times M\times1.989\times10^{33}}{(R\times6.955\times10^{10})^2}\right),
\end{equation}

\noindent where $M$ and $R$ represent the stellar mass and radius in solar units.

We note that while we use surface gravity to make the parameterization of the M dwarf stars consistent with the other regimes, surface gravity is not a commonly used metric in M dwarf research. M dwarfs do not age on a Hubble time and so surface gravity can be uniquely determined by effective temperature and metallicity.

\subsubsection{Support for extrapolation}  
\label{c3k}

To preserve the quality of the interpolator at the edges of the empirical parameter space, we supplement the training set with spectra from the theoretical C3K library (grey points in Figure~\ref{fig:psi_train}) degraded to the SpeX resolution. We only added C3K spectra for stellar parameters outside the convex hull (a hull being the set of planes that encloses all the training points) of the MILES stars. The C3K spectra are normalized in the same way as the observed spectra.  

Our goal is for the C3K library to keep the polynomial terms ``well-behaved'' at the boundaries of parameter space for the empirical spectra, but we do not want the fits to be driven by the large numbers of C3K stars.  We therefore weight the C3K spectra in the fits by a factor of $10^{-2}$ times the median SNR of the empirical spectra.  This factor was chosen after considering a range of values and inspecting the resulting polynomial behavior.

\begin{figure*}[t]
	\includegraphics[width=1.0\textwidth]{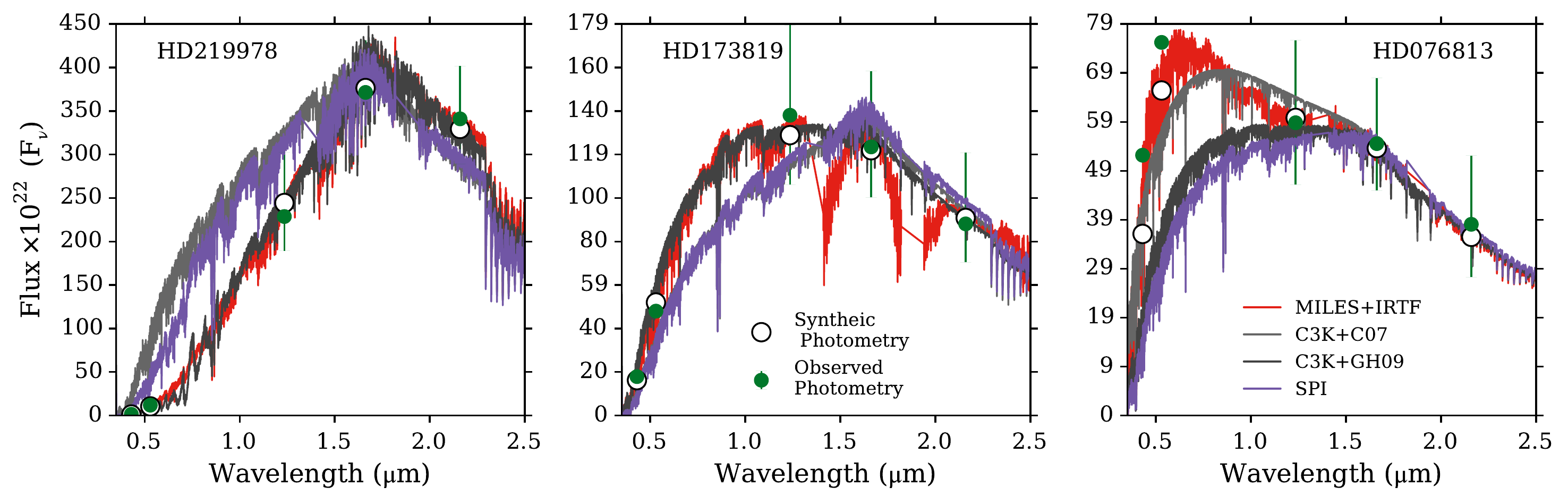}
	\caption{ Comparison of data (spectroscopy in red and photometry as green circles) with the SPI model (purple) and theoretical spectra using two choices for effective temperature (grey lines). Synthetic photometry is shown for clarity (open symbols). For the dark grey line the temperature used was derived from the metallicity-color relations of \citet{gonz2009}. The temperature used for the light grey line is the value from \citet{cenarro2007}. This shows the spread in T$_{\rm eff}$ values, and thus the spread in expected spectral shape, from different methods for each star. For the stars shown the spectral shape expected from the  \citet{prugniel2011} and \citet{sharma2016} values is not consistent with the observed spectral shape.}
	\label{fig:bad_sp}
\end{figure*}

\subsection{Training the Model}

The library contains a wide range of spectral types, with stellar temperature being the primary driver of the shape of the spectra. Modeling all the stars together is not feasible since it is difficult to specify a polynomial model that can encompass such a diverse set of stars. We therefore partition the library into five subsets based on temperature and surface gravity, essentially making five global interpolators that we use in conjunction with each other to span the entirety of parameter space.  For each subset we define the polynomial terms we use to create the model ($t\equiv{\rm log}\,T_{\rm eff}$, $z\equiv$[Fe/H], $g\equiv{\rm log}\,g$):

\begin{itemize}
\item Cool Dwarfs
\begin{align}
\begin{split}\label{}
{\rm log} F_\nu = {} & a_0 + a_1t + a_2z + a_3g+ a_4z^2 + a_5t^2 + a_6g^2 + \\
			      & a_7 (t \times z) + a_8(t \times g) + a_9(z\times g) + a_{10}z^3 + \\
			      & a_{11}t^3 + a_{12} g^3 +  a_{13}(t^2\times z) + a_{14}(z^2 t) + \\
			      & a_{15}(g \times t^2) + a_{16}t^4 + a_{17}z^4 +  a_{18}(t^2\times z^2) +\\
			      & a_{19}(t^3\times z) + a_{20}t^5.
\end{split}
\end{align}
\item Cool Giants
\begin{align}
\begin{split}\label{}
{\rm log} F_\nu = {} & a_0 + a_1t + a_2z + a_3g + a_4t^2  + \\
	     & a_5g^2 + a_6z^2 + a_7(z \times g) + a_8(t \times g) + \\
              & a_8(t \times z) + a_9t^3 +  a_{10}g^3 +  a_{11}z^3  + \\
              & a_{12}(t \times g \times z) +  a_{13}(t \times t \times z) + \\
              &  a_{14}(t \times t \times g) +  a_{15}(z^2\times t) +  a_{16}(z^2\times g) + \\
              & a_{17}(t \times g^2) + a_{18}(z \times g^2) + a_{18}t^4.
\end{split}
\end{align}
\item Warm Dwarfs 
\begin{align}
\begin{split}\label{}
{\rm log} F_\nu = {} & a_0 + a_1t + a_2z + a_3g + a_4t^2 + a_5g^2 + a_6z^2 + \\
			      & a_7(t \times z) + a_8(t \times g) + a_9t^3 + a_{10}(t\times g^2 ) + a_{11}z^3 + \\
			      & a_{12}(t^2\times g) + a_{13}(t^2\times z) + a_{13}( t\times z^2) + \\ 
			      & a_{14}(t \times g \times z) + a_{15}( g \times z^2) + a_{16}t^4 + a_{17}g^4 + \\
			      & a_{18}(t^3 \times g) + a_{20}(z \times t^3) + a_{21}(z^2 \times t^2) +  \\
			      & a_{21}(z^3 \times t) + a_{22}(t^2\times g)^2) +  a_{23}(z \times t^2 \times g ) + \\
			      & a_{24}t^5.
\end{split}
\end{align}
\item Warm Giants
\begin{align}
\begin{split}\label{}
{\rm log} F_\nu = {} & a_0 + a_1t + a_2z + a_3g +  a_4t^2 + a_5g^2 + a_6z^2 + \\
			      & a_7(t \times z) + a_8(t \times g) +  a_9(g \times z) + a_9t^3 + a_{10}g^3 + \\
			      & a_{11}z^3 + a_{12}(t^2\times z) + a_{13}(t \times z^2) + a_{14}(g \times t^2) + \\
			      & a_{15}(g^2\times t) + a_{16}t^4 + a_{17}z^4 + a_{18}(t^2 \times z^2) + \\
			      & a_{19}(t^2 \times g^2) + a_{20}(z^2 \times g^2) + a_{21}t^5.
\end{split}
\end{align}
\item  Hot Stars 
\begin{align}
\begin{split}\label{}
{\rm log} F_\nu = {} & a_0 + a_1t + a_2z + a_3g + a_4t^2  + a_5z^2 + a_6g^2 + \\
			      & a_7(t \times g) + a_8(t \times z) +  a_9(g \times z) + a_{10}t^3 + a_{11}g^3 + \\
			      & a_{12}z^3 + a_{13}( t \times g \times z) +  a_{14}(t^2 \times z) +  a_{15}(t^2 \times g) + \\
			      & a_{16}(z^2 \times g) + a_{17}(t \times g^2) +  a_{18}(z\times g^2) + a_{19}t^4.
\end{split}
\end{align}
\end{itemize}

The maximum likelihood coefficients for each regime are determined by weighted linear least squares. The glut of polynomial terms is a classic problem in polynomial regression modeling. As we are concerned with the ability to predict spectra (which we assess later in Section \ref{spi_quality}) rather than the values of the coefficients themselves we do not make an effort to simplify the polynomial functions using, e.g., L1 regularization. Furthermore, the oscillatory behavior of extrapolations that can result from unregularized high order polynomial regression is mitigated by our use of theoretical spectra near the boundaries of the valid parameter space.

In Table~\ref{table:spi_limits} we show the ``training'' ranges for $T_{\rm eff}$, log$g$, and [Fe/H], i.e., the stellar parameter limits for the training set stars in each regime and the ``Interpolating'' ranges for $T_{\rm eff}$, log$g$, and [Fe/H], i.e., the stellar parameter limits for safe interpolation. We have the same range in metallicity for the training bounds of all the stellar regimes. The ranges were determined by minimizing the rms difference between the observed and interpolated spectra in the training set. The overlap in the $T_{\rm eff}$  training ranges is meant to mitigate the effect of five separate, disjoint hulls having a smaller volume than the hull for all the library points. For most regimes the overlap in T$_{\rm eff}$ is 500 K but for the cool dwarf regime we extend the training sample to 5500 K. We do this to compensate for the paucity of low-metallicity cool dwarfs in the empirical library. If we did not extend the training sample to 5500 K the metallicity dependence for the cool dwarf regime would end at [Fe/H] $\sim$ -0.7, which would bias the interpolation to lower metallicities. With the extended temperature range of the training sample SPI is able to pull information from the warmer low-metallicity stars which mitigates the effects of the lack of observed low-metallicity cool dwarfs.

\begin{figure*}[t]
	\includegraphics[width=1\textwidth]{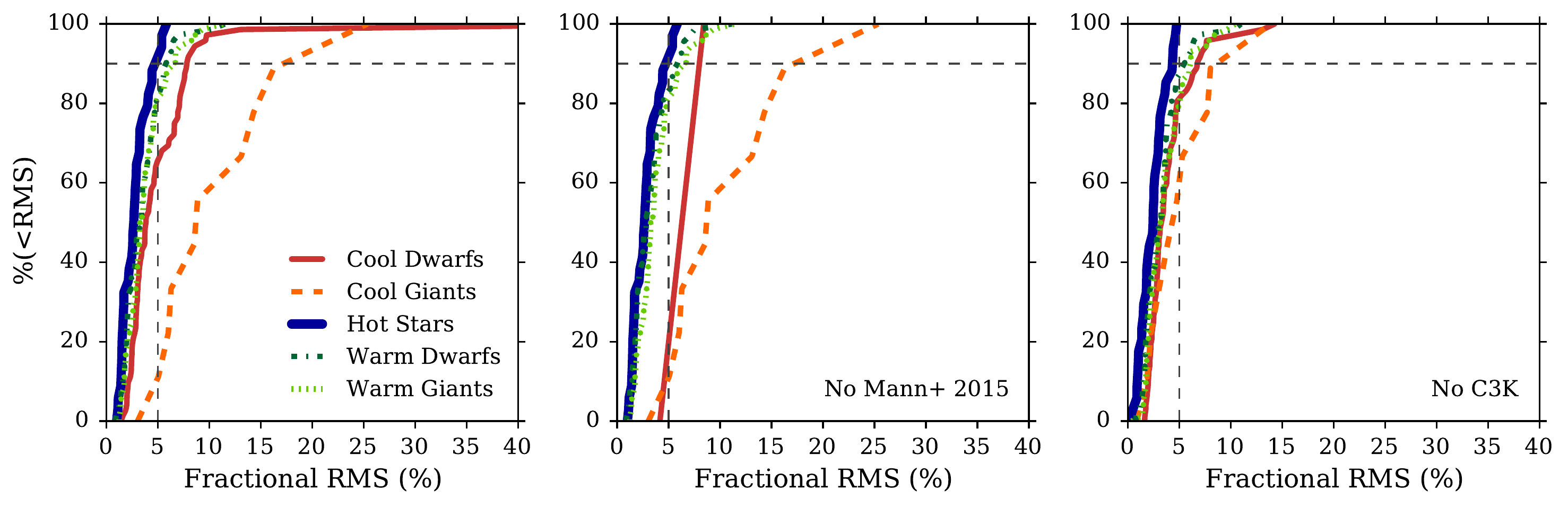}
	\caption{Cumulative distribution functions of the fractional rms differences between the observed spectra and interpolated spectra in the cool dwarf (red), cool giant (orange), hot (blue), warm dwarf (green), and warm giant (lime) regimes. To aid interpretation, a fractional rms difference of 5\%  is marked with a vertical dashed line and 90\% of the sample is marked with a horizontal dashed line. }
	\label{fig:recovery}
\end{figure*}

\begin{figure*}[t]
	\includegraphics[width=1\textwidth]{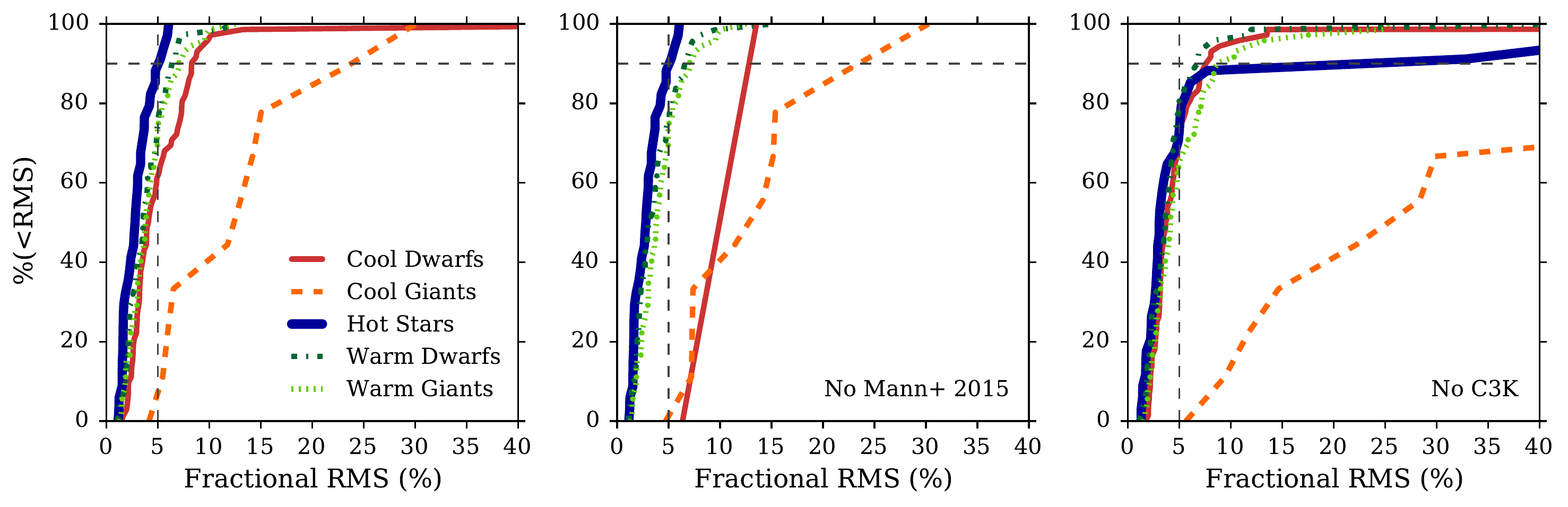}
	\caption{Same as Figure~\ref{fig:recovery} but now the interpolated spectra are the result of the leave-one-out (``jack knife'') test where each star in the training sample was removed from the model in turn before the interpolation. Including the C3K spectra reduces the dependence of the model on the presence of any one star, especially the cool giant stars and hot stars. The inclusion of the \citet{mann2015} M dwarfs also helps mitigate issues in the cool dwarf regime.}
	\label{fig:leave1}
\end{figure*}

\begin{deluxetable*}{llllllll}
\tabletypesize{\footnotesize}
\tablecolumns{5}
\tablewidth{0pt}
\tablecaption{Valid ranges for SPI interpolation \label{table:spi_limits}}
\tablehead{
\colhead{} & \colhead{Training} & \colhead{} & \colhead{} & \colhead{Interpolating} & \colhead{} \\
\colhead{Regime} & \colhead{$T_{\rm eff}$} & \colhead{log$g$} & \colhead{[Fe/H]} & \colhead{$T_{\rm eff}$} & \colhead{log$g$} & \colhead{[Fe/H]} 
} 
\startdata 
\\

Cool Dwarfs & 1100-5500 & 3.5-6.5 & -2.5-0.6  & 2500-4000  & -0.5-3.5 & -2.0-0.6 \\
Cool Giants & 1100-4500 & -1.0-2.75 & -2.5-0.6  & 2500-4000  & $>$3.5 &  -2.0-0.6 \\
Warm Dwarfs & 3000-6500 & 3.0-5.75 &  -2.5-0.6  &  4000-6000 & -0.5-3.5 & -2.0-0.6 \\
Warm Giants & 3500-6500 & -0.75-4.0 & -2.5-0.6  & 4000-6000 & $>$3.5 &  -2.0-0.6 \\
Hot Stars & 5500-12,500 & 2.5-5.5 & -2.5-0.6 & 6000-12,000  & $<$5.0 &  -2.0-0.6 \\ 

\end{deluxetable*}

\subsubsection{Culling the Extended IRTF Library}\label{section:culling}

Only the highest quality spectra should be included in the SPI training set. Any star used in the training set needs to have accurate stellar parameters and a reliably flux-calibrated observed spectrum.  Based on visual inspection we removed stars with spectra that were either of poor quality or appeared to have flux calibration issues that severely affected the shapes. We removed 51 stars following this criteria (see the Quality Flag and Shape Flag columns in Table~\ref{table:obs_log}).

Furthermore, since the interpolation relies on having like-spectra grouped by like-stellar parameters, we need to ensure that our stellar parameters are accurate. SPI provides an opportunity to check the accuracy of the stellar parameters associated with each of the stars in the library. If there is significant discrepancy between the observed spectrum of the star and the interpolated spectrum given by SPI it might be indicative of an issue with the stellar parameters associated with that star. A similar technique was used by \citet{vazdekis2010} to remove 60 stars from the MILES library in creating stellar population models.

In Figure~\ref{fig:bad_sp} we show an example of three stars where the SPI predicted spectra (purple) using the stellar parameters from \citet{prugniel2011} and \citet{sharma2016} are not consistent with the observed data (both spectroscopy, red line, and photometry, green circles). Consistency between the observed photometry and observed spectra for these stars suggests that the problem is not exclusively with the flux calibration but with the stellar parameters. To further emphasize this point we also show two C3K spectra corresponding to different effective temperatures in each panel. For the dark grey spectra we used the effective temperatures computed using the \citet{gonz2009} $J-K_S$-metallicity relations. For the light grey spectra we used the effective temperatures from \citet{cenarro2007}. For HD219978 (left) and HD173819 (middle) the overall shape of the spectra using the $J-K_S$ derived temperatures are more consistent with the observed data than SPI and the spectra corresponding to the \citet{cenarro2007} spectra. For HD076813 (right) the spectrum corresponding to the temperature from \citet{cenarro2007} is most consistent. 

This test both demonstrates the possible range in derived stellar parameter values and that there are available stellar parameters that would better match the observed spectra and photometry and thus the problem is with the \citet{prugniel2011} and \citet{sharma2016} stellar parameters. These stars were flagged for having an exceptionally high rms difference between the observed and SPI predicted spectra.  We flagged 9\% of our observed sample as having incorrect stellar parameters and do not include them in the SPI training set. The differences in the temperatures determined using different methods for these stars are large compared to the bulk of the library stars. This indicates that in general the stellar parameters from \citet{prugniel2011} and \citet{sharma2016} are consistent with the observations.

This test does not mean that any issue with the data is exclusively an issue with the stellar parameters. Several of the stars flagged as potentially having incorrect stellar parameters also have flux calibration issues or other quality issues as indicated by their Quality and Shape Flags (see Table A1). For example, we still see the effects of prominent telluric absorption for HD173819 (the middle) panel which is indicated in Table A1. However, the SPI predicted spectrum is not consistent with the shape of the observed spectrum but the predictions based on the  \citet{cenarro2007} temperature and temperature from the \citet{gonz2009} relations are consistent with the observed shape.

Thus, we can use SPI as a way to flag stars that need more accurate stellar parameters but otherwise have good data (in the future we will use SPI to re-derive stellar parameters in an iterative approach). We removed 27 stars following this criteria and re-derived the SPI parameters with these stars removed (see `Parameter Flag' column in Table~\ref{table:obs_log}).

\subsection{Quality of Interpolation} \label{spi_quality}

We can assess the quality of the interpolation by examining how well SPI can recover the spectra of the stars in the training sample (including both the data presented in this work and the \citet{mann2015} data).  For each star we compute the fractional rms between an observed spectrum in the empirical training set and the corresponding interpolated spectra from SPI. In Figure~\ref{fig:recovery} we show the cumulative distribution of this fractional rms. In each panel we show results separately for the cool dwarf stars (red), cool giant stars (orange), hot stars (blue), warm dwarf stars (dark green), and warm giant stars (light green). Also plotted in each panel is a horizontal line showing where 90\% of the stars are placed on the cumulative distribution and a vertical line marking 5\% fractional rms.  In the left panel we show the distributions when using the complete training set in the model, in the middle panel we show the distributions when we exclude the \citet{mann2015} spectra from the training set, and in the right panel we show the distributions that result when we exclude the C3K spectra from the training set. Note that the rms is insensitive to overall offsets between the two spectra. Note that SPI is not an interpolator in the strictest sense of exactly reproducing the input spectrum at the input points, and so there is no guarantee that the rms should be small.

The most important take away from Figure~\ref{fig:recovery} is that for all the regimes the recovery of the training set spectra is very good. For cool dwarfs, warm dwarfs, warm giants, and the hot stars the recovery is better than 10\% for 90\% of the stars. When we exclude the \citet{mann2015} M dwarfs from the training set the recovery for the cool dwarfs is worse. The recovery of the cool giants is worse than the other regimes and counterintuitively improves when we exclude the C3K spectra from the training set. This could be an effect of the problems theoretical spectra have in the cool giant regime \citep[see][]{bertone2008}. 

Since SPI relies on stars with similar stellar parameters having similar stellar spectra we can look at stars where SPI fails to recreate its spectrum as stars with possible issues with the stellar parameters. Since we see in Figure~\ref{fig:recovery} that for most of the library ($\sim 90\%$) the rms difference between the observed and interpolated spectra is $< 5\%$ we can feel confident that the stellar parameters from \citet{prugniel2011} and \citet{sharma2016} are internally self-consistent. We looked at Figure~\ref{fig:recovery}  separately for the MILES and IRTF spectra and found that the recovery is about the same for the majority of the stars.

Figure~\ref{fig:leave1} is the same as Figure~\ref{fig:recovery} but now the interpolated spectra are the result of leave-one-out validation. The leave-one-out test consists of going through all the stars in training sample, taking one out of the training sample at a time, retraining the function such that the information from that star is no longer included in the model, then comparing the SPI prediction for that spectrum to the actual spectrum.

\begin{figure}[t]
	\includegraphics[width=0.5\textwidth]{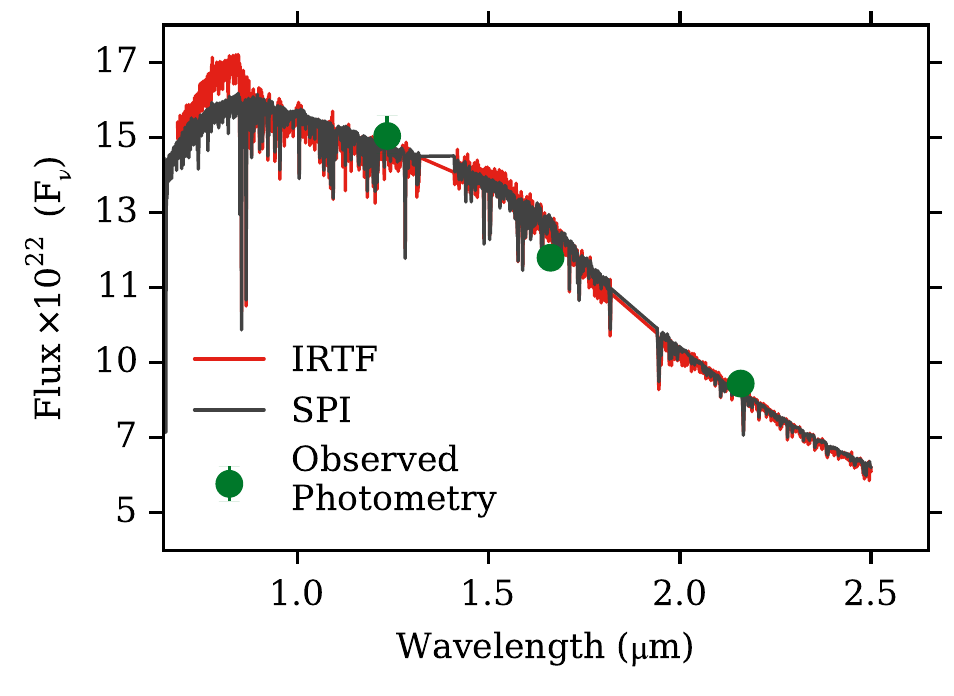}
	\caption{Demonstration of the ``self-calibration'' possibilities with SPI. The observed spectrum (red) for the star, HD004307, has an unphysical artifact at $\sim 0.8 \mu m$ due to nebulosity near the standard star. Since this issue only affected a small subset, 4\%, of the stellar library we can use SPI to obtain a spectrum for HD004307 and others like it without the unphysical feature. This is shown in the interpolated spectrum (black) that is largely the same as the observed spectrum but without the bump at $\sim 0.8 \mu m$. }
	\label{fig:self_cal}
\end{figure}

The leave-one-out test is an assessment of how sensitive the model is to the presence of any one star and demonstrates the utility of including the C3K spectra in the training set. This figure shows that by including the C3K spectra in the training set we are mitigating the bias a single star can introduce in the model, especially for the cool giant stars and hot stars, where the number of empirical stars is low. We see that the \citet{mann2015} data has a similar effect in the cool dwarf star regime. This figure demonstrates that the inclusion of the \citet{mann2015} and C3K spectra is important for the interpolation of stars outside the confines of the training set.

\subsection{Applications of SPI}

As discussed previously, SPI can be used to flag stars with potentially inaccurate stellar parameters. More generally, stars that are outliers with respect to their SPI prediction could be used to uncover other unusual behavior such as variability or peculiar abundance patterns.

SPI also allows for ``self-calibration'' of the observations. By this we mean that because SPI uses the information of all the stars in the library it is possible to remove artifacts that affect a small fraction of the library stars. As we mentioned earlier, some of the stars observed for this library have an unphysical bump from standard stars with unusual spectra. Since most stars were not affected in this way we can remove this artifact by using the model to interpolate for a spectrum for the stars that are affected. In Figure~\ref{fig:self_cal} we show the observed spectrum (red) for the star HD004307 and the interpolated spectrum (black). The interpolated spectrum is consistent with the observed spectrum except for the bump seen at 0.8 $\mu m$ in the observed spectrum, where the interpolated spectrum produces more sensible behavior. In future work we will use the self-calibration capabilities of SPI to re-derive the stellar parameters for the stars presented in this work.

We emphasize that any library can be used as input for SPI. The new metallicity coverage of the Extended IRTF Library allows us to interpolate to a larger metallicity range than is possible with the original IRTF library \citep{rayner2009}. Furthermore, the library presented in this work can be augmented with, for example, NGSL. The inclusion of UV data into SPI could potentially help with modeling planet atmospheric properties.

\section{Behavior of the Stellar Libraries}

The main feature of the Extended IRTF Library is the expansion of near-IR coverage into the sub- and super-metallicity regimes. We would like to highlight various spectral features in the data and explore how these features depend on not just surface gravity and effective temperature but also on metallicity. This is also an opportunity to examine the behavior of SPI beyond its ability to simply reproduce the training set spectra. Also of interest is how the empirical trends compare with the theoretical models.  To explore the behavior of SPI and C3K we computed spectra for stellar parameters along a 3 Gyr  (for metallicities $ > -0.7$) and a 13.5 Gyr (for metallicities $ < -0.7$) MIST isochrone. 

Sharp boundaries where the five different polynomial models that make up SPI join together is a concern. To ensure smoothness we took the weighted average of the different predicted fluxes for the evolutionary points (EPs) that have temperatures that are in the overlap between the cool and warm dwarf and the cool and warm giant training bounds. This means that for EPs with log$g$ $> 4.0$ and temperatures between 3500-4500 K we generated a flux using both the warm giant and dwarf giant models and averaged the fluxes together weighted depending on the temperature. We did the same for the EPs with log$g$ $< 4.0$ and temperatures between 3000-5500 K.

In this section we analyze the metallicity-dependence of key stellar features by using equivalent widths. The limitations of equivalent widths are well known -- they are sensitive to the definition of the psuedocontinuum and each index is a blend of features from more than one element. Here we use equivalent widths as a way to compress the information to explore broad trends. All the equivalent widths quoted in this work are in units of ${\rm \AA}$ and all wavelengths are in vacuum. We present a combination of a selection of the Lick indices defined in Table 1 \citet{worthey1994} and Table 1 of \citet{cvd2012}.

\begin{figure*}[!th]
	\centering
	\includegraphics[width=\textwidth]{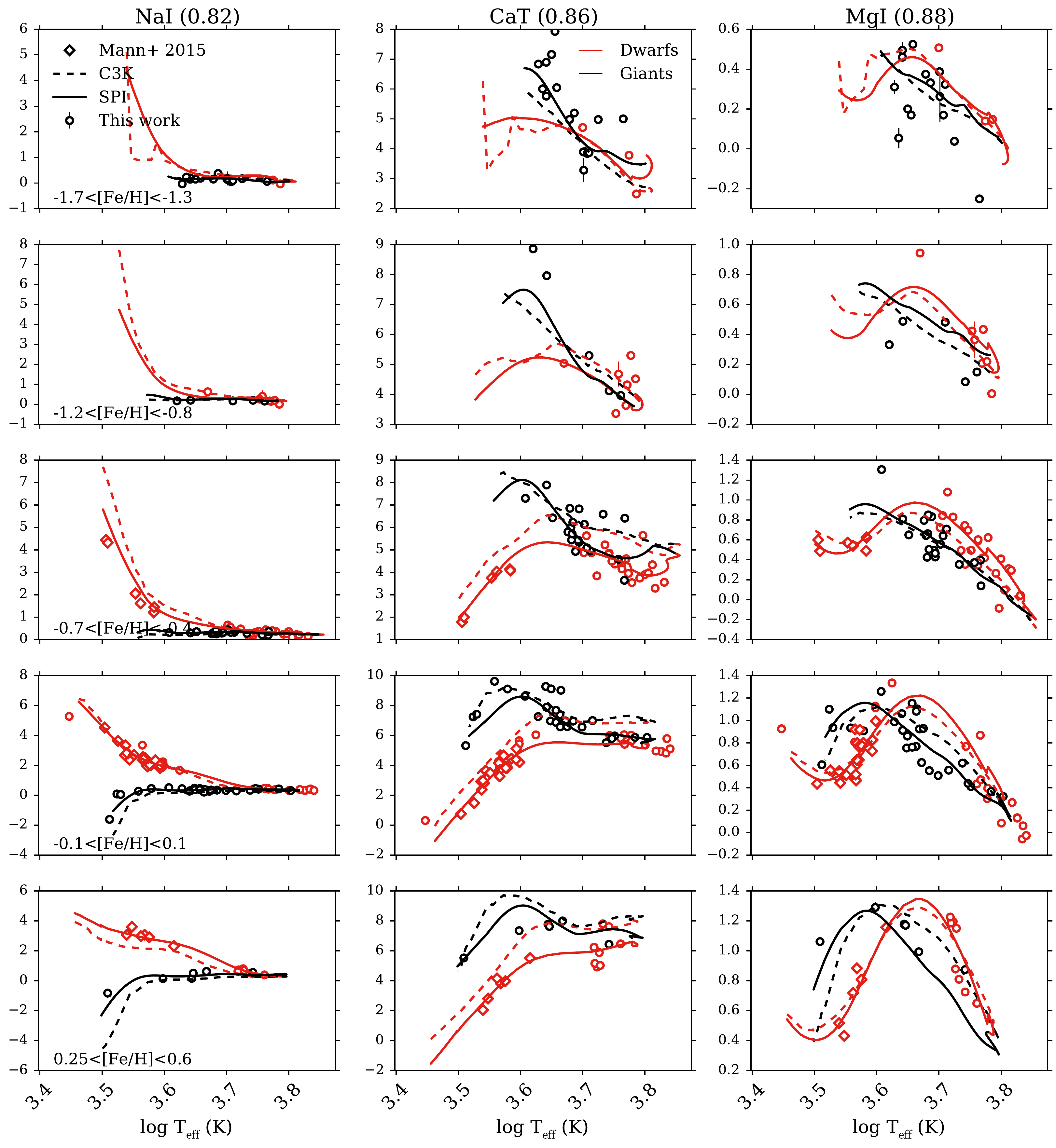}
	\caption{Dependence of selected spectral indices on effective temperature for dwarfs (log$g$ $> 4.0$, red) and giants (log$g$ $\leq 4.0$, black). Plotted are index strengths using the IRTF data from this work (open circles), data from \citet{mann2015} (open diamonds), empirical prediction from SPI (solid lines), and theoretical predictions from C3K (dashed lines).}
	\label{fig:ew1}
\end{figure*}

\begin{figure*}[!th]
	\centering
	\includegraphics[width=1\textwidth]{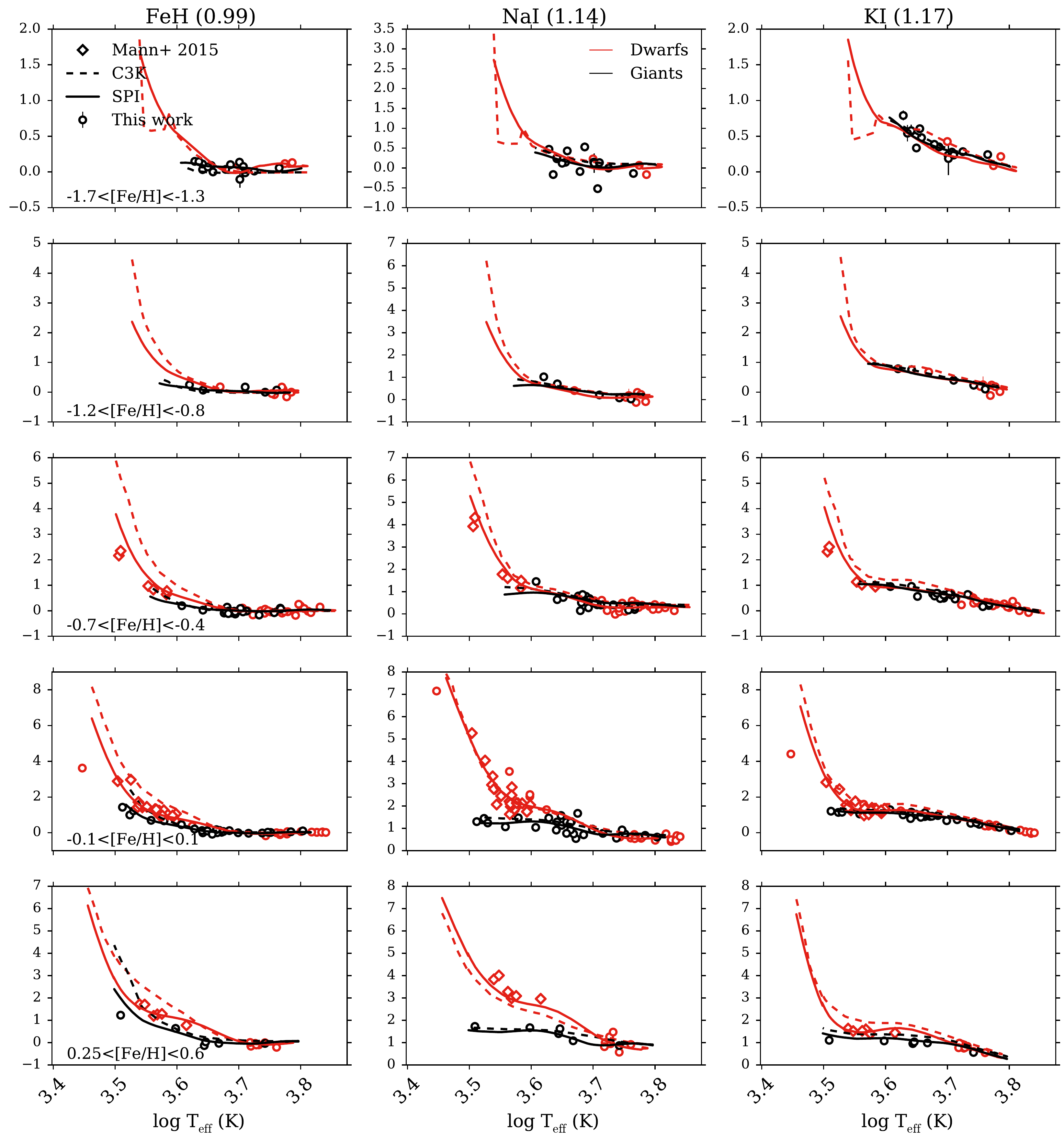}
	\caption{Continuation of Figure~\ref{fig:ew4}.}
	\label{fig:ew2}
\end{figure*}

\begin{figure*}[!th]
	\centering
	\includegraphics[width=1\textwidth]{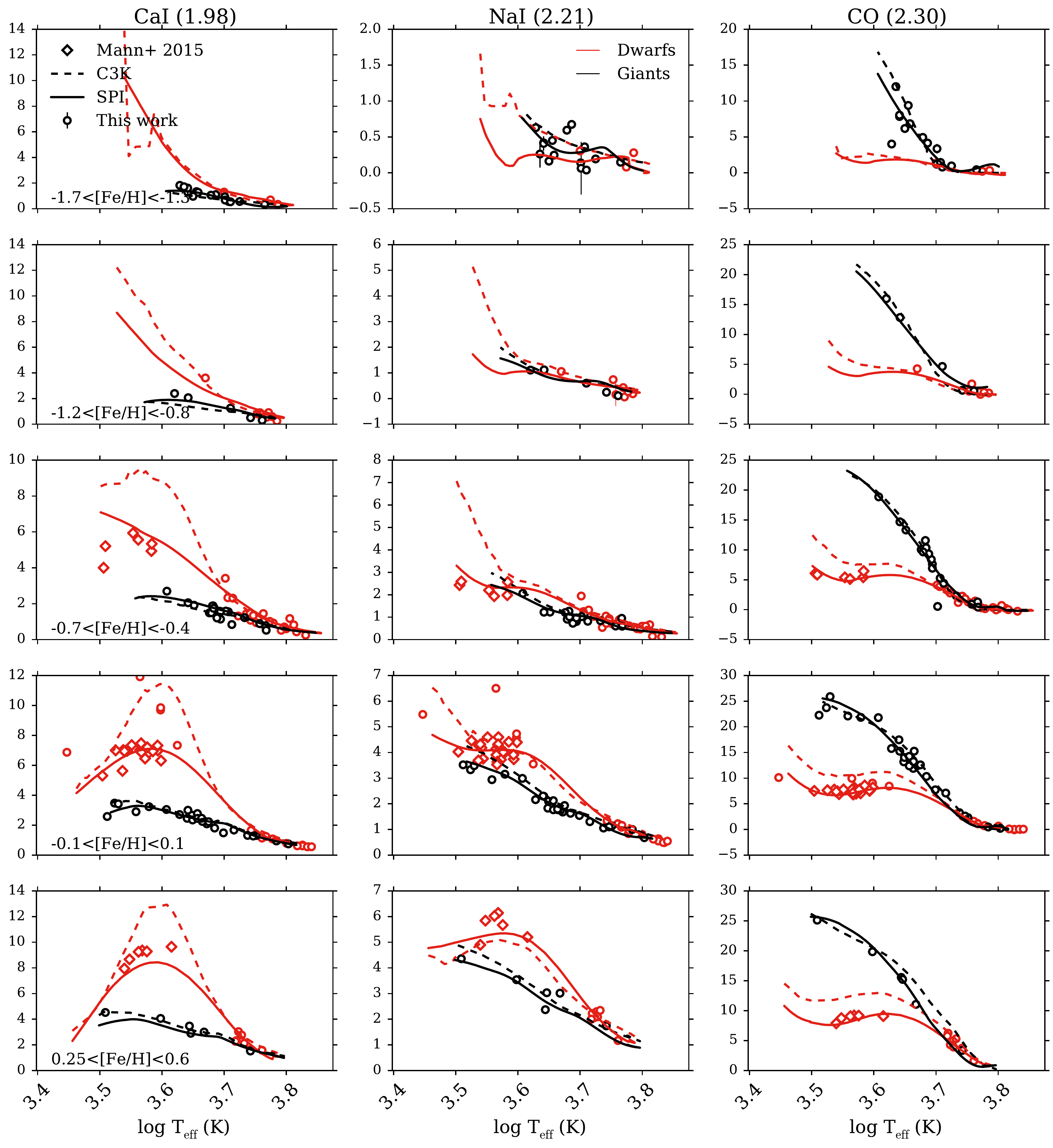}
	\caption{Continuation of Figure~\ref{fig:ew1}.}
	\label{fig:ew3}
\end{figure*}

\begin{figure*}[!th]
	\centering
	\includegraphics[width=1\textwidth]{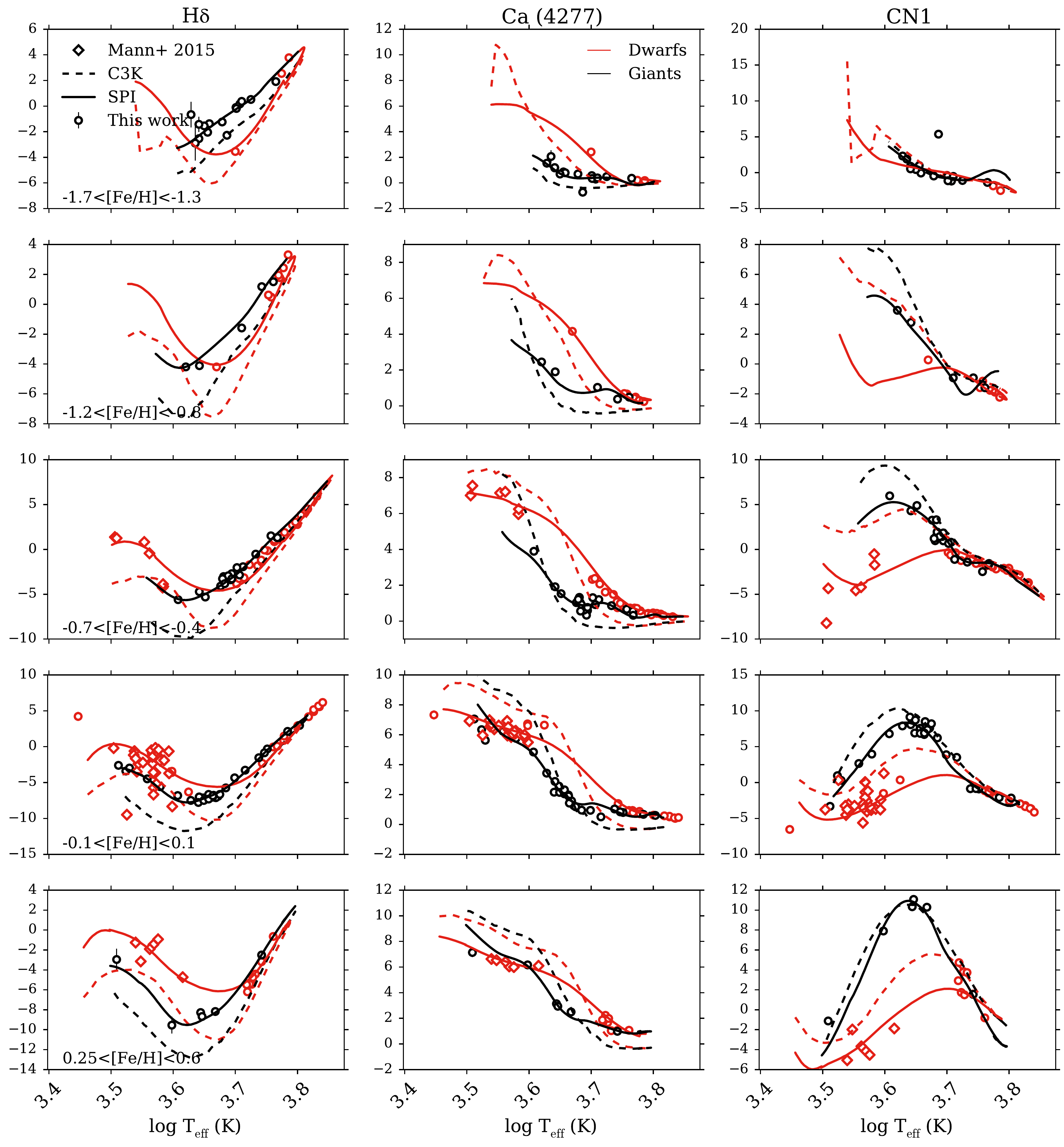}
	\caption{Continuation of Figure~\ref{fig:ew1} except now using the corresponding MILES spectra for the stars in the Extended IRTF Library..}
	\label{fig:ew4}
\end{figure*}

\begin{figure*}[!th]
	\centering
	\includegraphics[width=1\textwidth]{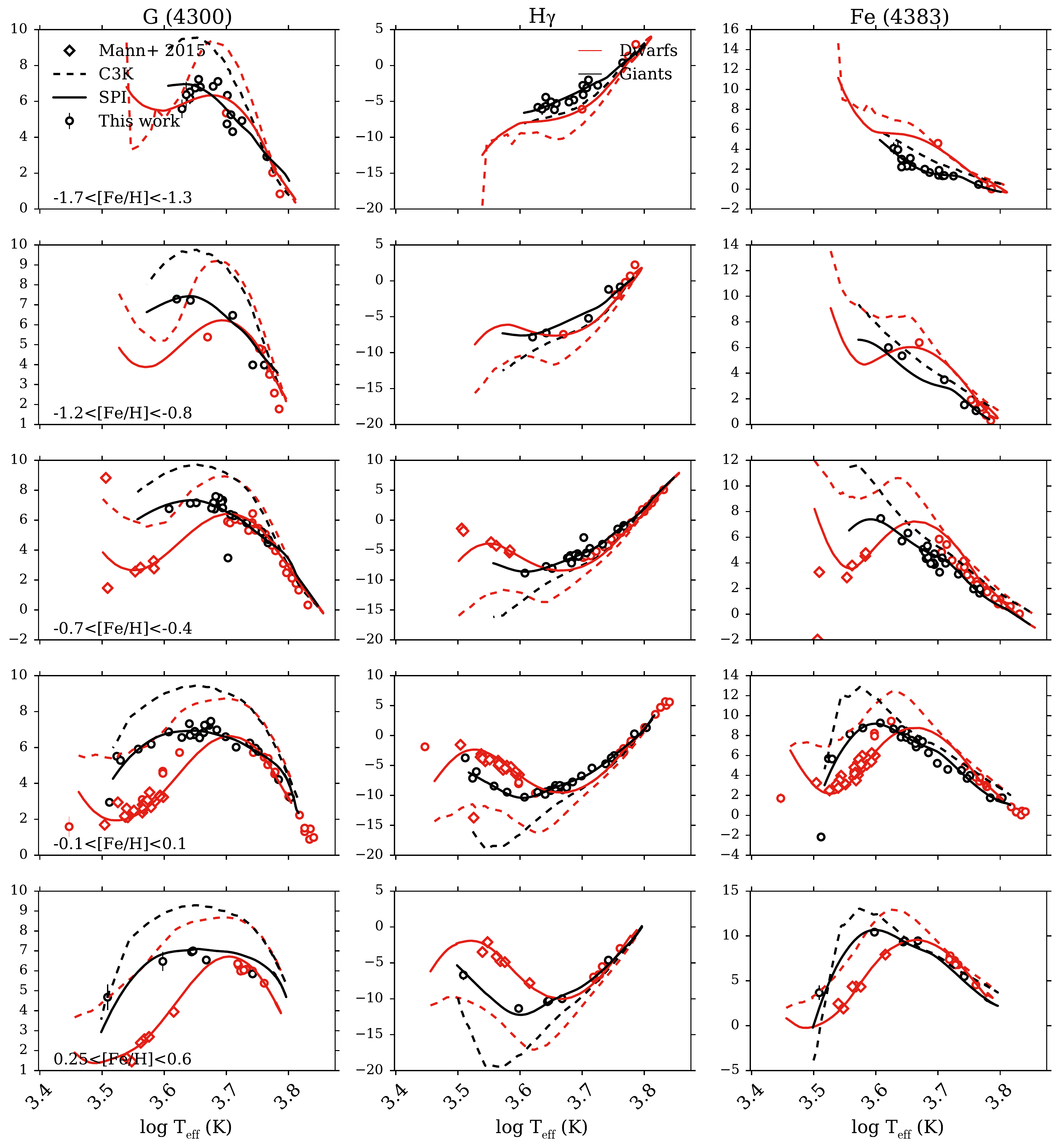}
	\caption{Continuation of Figure~\ref{fig:ew4}.}
	\label{fig:ew5}
\end{figure*}

\begin{figure*}[!th]
	\centering
	\includegraphics[width=1\textwidth]{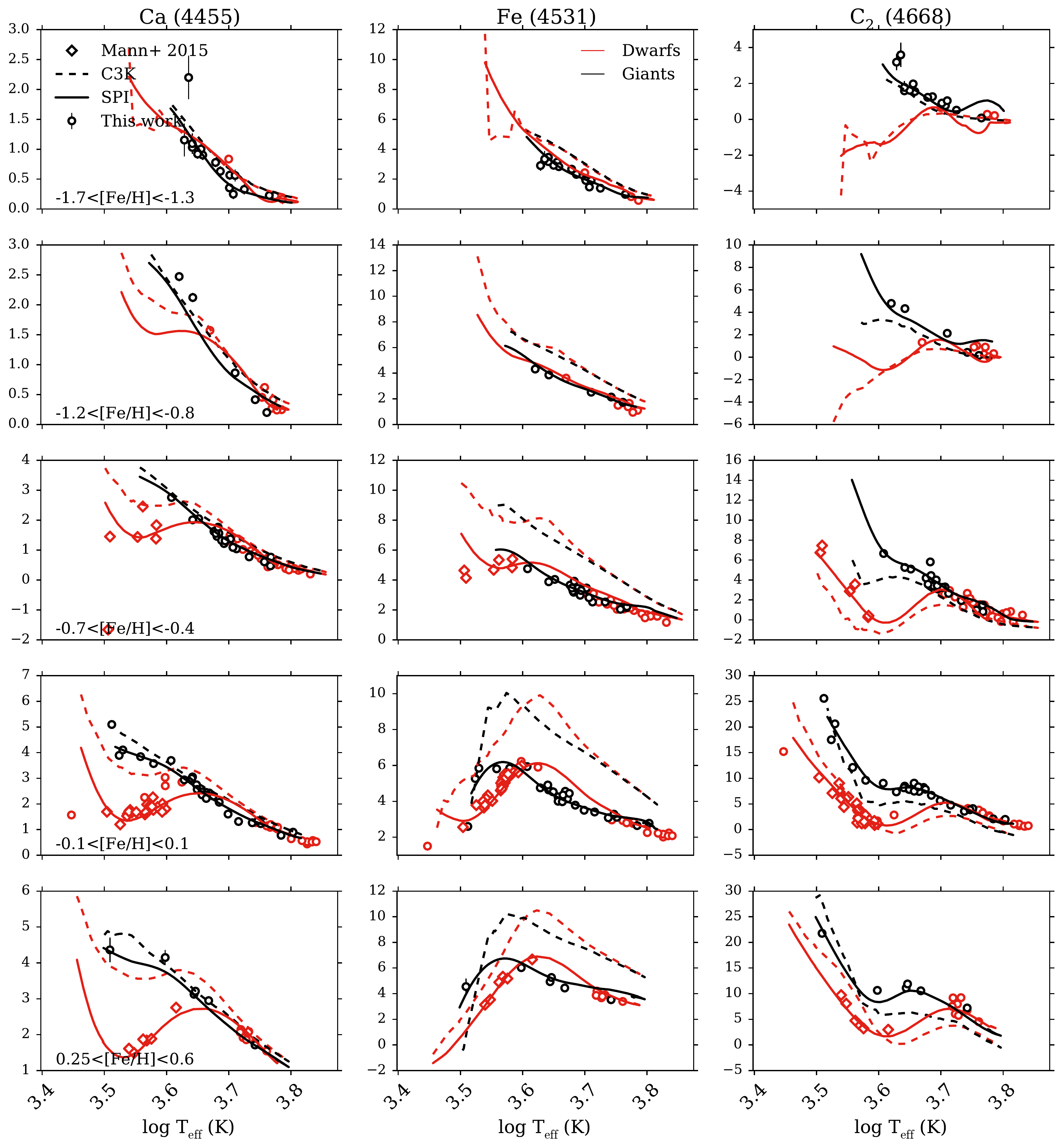}
	\caption{Continuation of Figure~\ref{fig:ew4}.}
	\label{fig:ew6}
\end{figure*}

\begin{figure*}[!th]
	\centering
	\includegraphics[width=1\textwidth]{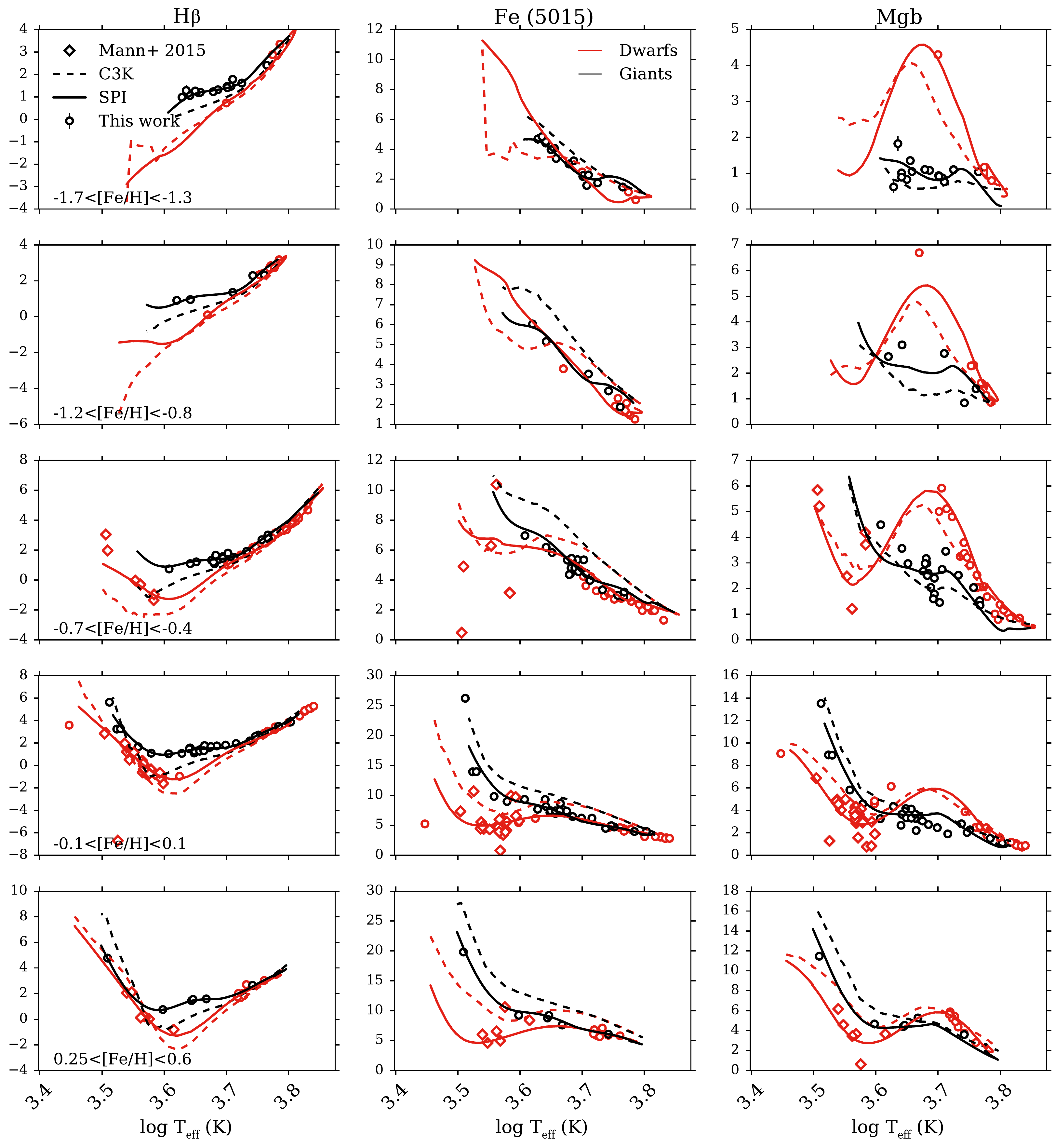}
	\caption{Continuation of Figure~\ref{fig:ew4}.}
	\label{fig:ew7}
\end{figure*}

\begin{figure*}[!th]
	\centering
	\includegraphics[width=1\textwidth]{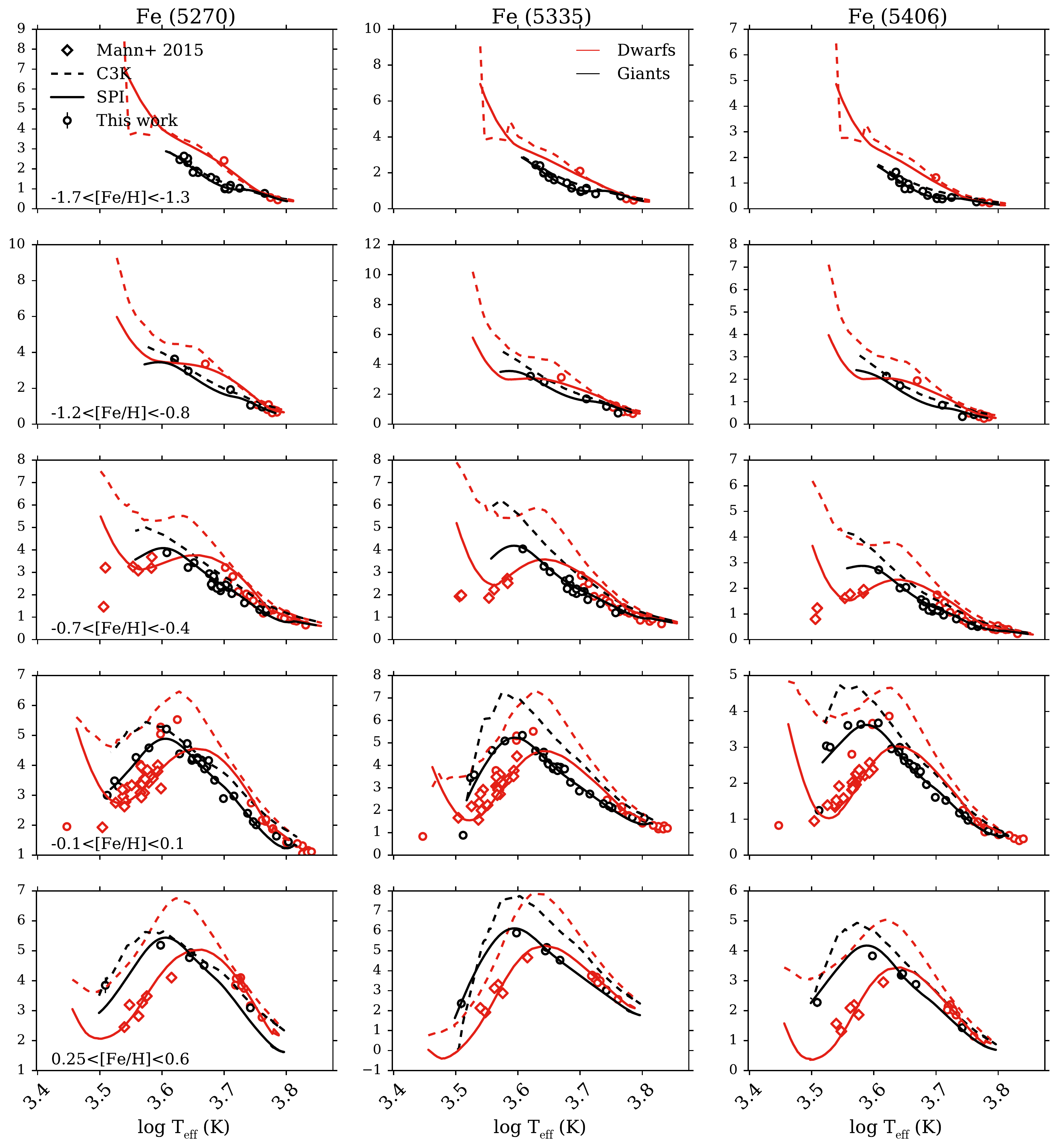}
	\caption{Continuation of Figure~\ref{fig:ew4}.}
	\label{fig:ew8}
\end{figure*}

\begin{figure*}[!th]
	\centering
	\includegraphics[width=1\textwidth]{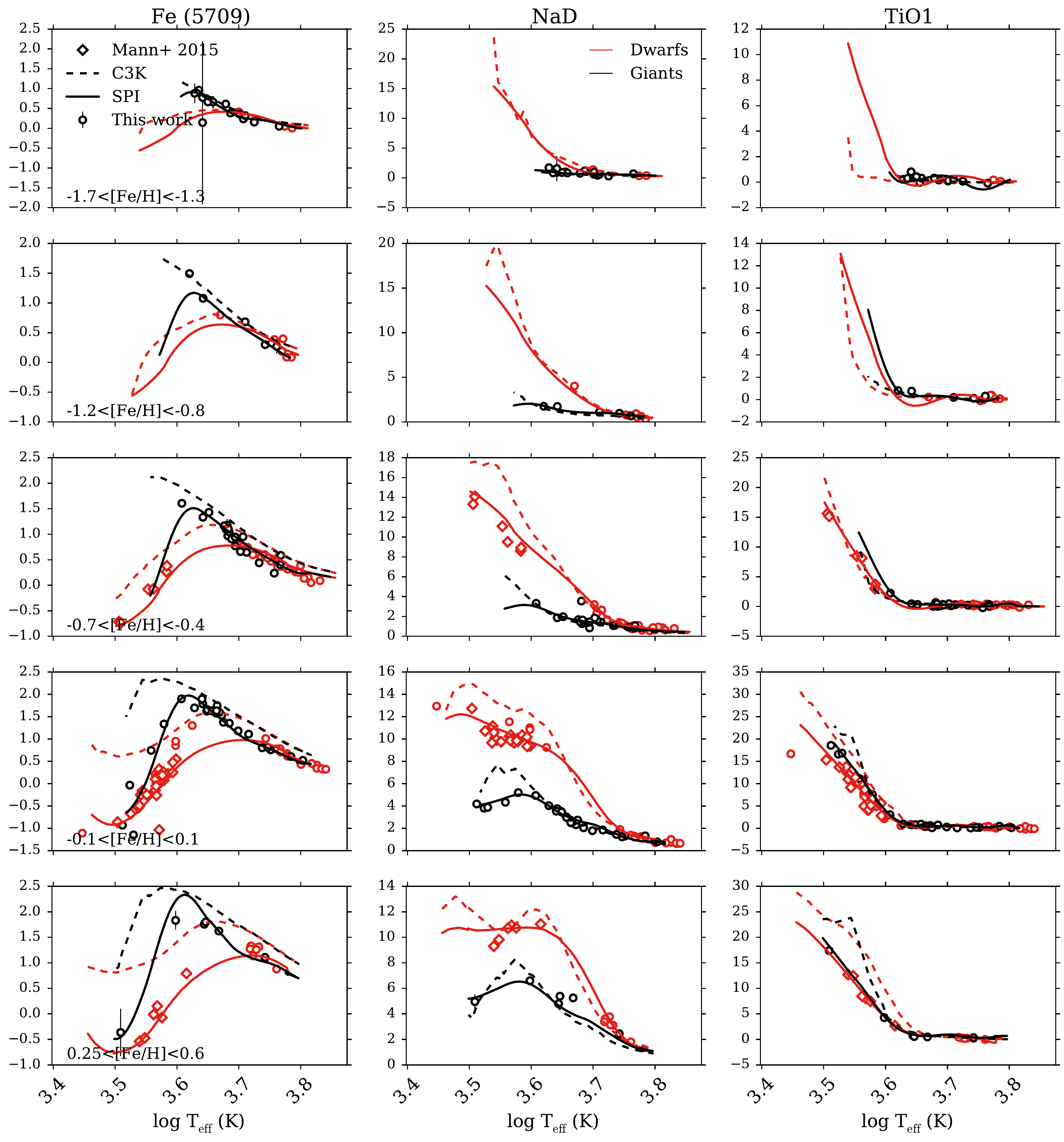}
	\caption{Continuation of Figure~\ref{fig:ew1}.}
	\label{fig:ew9}
\end{figure*}

We computed equivalent widths using the following equation, 

\begin{equation}
{\rm EW} = \lambda_2 - \lambda_1 -  \int_{\lambda_1}^{\lambda_2 } \frac{{ F_i}}{c_b + \left({ \lambda_i} - \lambda_b \right) \left(\frac{c_r - c_b }{\lambda_r - \lambda_b} \right)} d\lambda,
\label{eq:ew}
\end{equation}

 \noindent where $\lambda_1$ and $\lambda_2$ are the blue and red wavelength boundaries of the feature definition, $\lambda_b$ and $\lambda_r$ are the average wavelengths of the blue and red continuum definitions for each feature. The blue and red continuum values $c_b$ and $c_r$  are the integral of the flux over the wavelength range that defines the blue and red continuum.

For the observations, we made 100 realizations of each spectrum by Monte Carlo sampling the noise. For each realization we used Equation~\ref{eq:ew} to compute the equivalent width. The errors for the equivalent widths are given by the 1$\sigma$ confidence values from the distribution of equivalent widths.

In Figures \ref{fig:ew1}-\ref{fig:ew9} we show equivalent width strength versus effective temperature for different spectral features (columns) and different metallicity bins (rows). In every panel dwarf stars (log$g$ $> 4.0$, red) are distinguished from giant stars (log$g$ $\leq 4.0$, black).  The equivalent widths for the stars in the Extended IRTF Library are shown in the empty circles, the equivalent widths from the \citet{mann2015} data are shown in empty diamonds, the SPI equivalent widths are shown as the solid lines, and the C3K equivalent widths are shown in the dashed line. All data points have error bars though in most cases the error bars are smaller than the symbols. We do not expect a perfect match between the lines and the data points especially for the hot effective temperatures where there can be a range of log $g$ for fixed $T_{\rm eff}$ in the data.

\subsection{Data and empirical trends}

In this section we focus on the observations and the SPI predictions. The theoretical predictions will be discussed in the following section.

Several spectral features were highlighted in \citet{cvd2012} as means to discriminate between dwarf and giant stars: NaI0.82$\mu$m, CaT, FeH0.99$\mu$m KI1.17$\mu$m, CaI1.98$\mu$m, and CO2.30$\mu$m \citep[see also, e.g.,][]{spinrad1962, wing1969, cohen1978, frogel1978, kleinmann1986, diaz1989, ivanov2004, rayner2009}. This ability was only assessed for solar metallicity stars and it is of interest to know whether this behavior holds for an extended metallicity range. 

In Figures~\ref{fig:ew1}-\ref{fig:ew3} we show the temperature dependent equivalent width trends for several near-IR spectral features separated by dwarfs and giants. For the three highest-metallicity bins we can see the clear separation between dwarf and giant stars from the data alone.  In all of these features there is a metallicity dependence on the temperature range over which the separation between dwarf and giant stars occurs. All of the sodium lines, NaD, NaI0.82$\mu m$, NaI1.14$\mu m$, and NaI2.21$\mu m$ have equivalent widths that increase among the dwarf stars with decreasing temperature. For the dwarf stars  in the $-0.7 < {\rm [Fe/H]} < -0.4$ this is a precipitous increase for the coolest dwarf stars while at high metallicity there is a steadier increase over large range in temperature.

The CO2.30$\mu$m spectral feature remains a clear discriminator between giant and dwarf stars for the whole metallicity range and the temperature at which the separation occurs does not qualitatively change over the metallicity range. Likewise, the overall temperature dependent trend does not appear to change significantly for the KI1.17$\mu$m and CaT features. The strong FeH0.99$\mu$m feature remains {\it unique} to the cool dwarf stars for the entire metallicity range.

In addition to the near-IR features we show some of the Lick indices \citep{worthey1994} (Figures~\ref{fig:ew4}-\ref{fig:ew9}) that are the classic indicators of stellar population characteristics (e.g, age and metallicity). For some spectral features there is a noticeable difference in the temperature-dependent equivalent width trends from metallicity bin to metallicity bin (e.g., Fe4383, 4531, 5015, 5335, 5270, 5406, and 5782${\rm \AA}$,  Ca4455${\rm \AA}$, H$\beta$, C4668${\rm \AA}$). However, other spectral features ($H\gamma$, H$\delta$, Ca4277${\rm \AA}$, G4300${\rm \AA}$, and MgI0.88{\rm$\mu$m}) the temperature dependent trends remain similar from metallicity bin to metallicity bin.

As we would expect from the quality assessment given in Section~\ref{spi_quality}, SPI is consistent with the behavior of the data. Even where there is sparse data, e.g., the low-metallicity warm dwarf regime, SPI appears to make reasonable predictions of the behavior of all the spectral features displayed in Figures~\ref{fig:ew1} - \ref{fig:ew9}.

As shown in Figure~\ref{fig:psi_train} there are no dwarf stars cooler than 5000 K for $-1.7 < $ [Fe/H] $< -0.8$. This makes it impossible to know how accurate the SPI predicted spectra in this regime are. However, as described earlier the training and interpolating ranges are optimized so that SPI is able to use all three stellar parameter dimensions to make a prediction for this regime. That is, SPI is able to pull information from the predictions of both the hot and warm dwarf stars with $-1.7 < $ [Fe/H] $< -0.8$ and the higher metallicity cool dwarf stars. Indeed, the cool dwarfs in the lowest metallicity bins have temperature dependent trends that appear consistent with the trends in the higher metallicity regimes.

\subsection{Comparison with theoretical trends}

We now turn to a comparison between the theoretical C3K models and the empirical trends.  Comparisons between models and observations have been made previously. \citet{martins2007} compared the ATLAS9 \citep[][]{castelli2003}, MARCS \citep[][]{gustafsson2003}, and PHOENIX \citep[][]{brott2005} theoretical stellar libraries to various empirical stellar libraries. \citet{bertone2008} compared the ATLAS9 library to the ELODIE catalog \citep[][]{prugniel2001}, and \citet{cvd2012} compared equivalent width trends between theoretical and empirical stellar libraries. However, there are limitations to these studies that we can now overcome. \citet{martins2007} and \citet{cvd2012} were limited to solar metallicity stars and could not reach the coolest M dwarfs. The sample from \citet{bertone2008} spanned a wide range of metallicity but was limited to stars with  $T_{\rm eff} > 4000$K and \citet{bertone2008} did not compare specific feature strengths.

In the present case we have extended the library and the interpolator into the cool dwarf regime, and cover a much wider range in metallicity in the near-IR than was previously available.  This means that we can now make explicit comparisons between the theoretical and empirical behavior in this regime.

Starting our comparison with near-IR features of the solar metallicity stars (fourth row from the top in Figures~\ref{fig:ew1}-\ref{fig:ew3}) we see that the theoretical predictions are consistent with the empirical trends for the NaI0.82$\mu$m, MgI0.88$\mu$m, NaI1.14$\mu$m, KI1.17$\mu$m, and NaI2.21$\mu$m features. There is discrepancy between the theoretical predictions and empirical trends for the CaT, FeH, CaI1.98$\mu$m, and CO2.30$\mu$m features. Although, with the exception of the warm dwarf regime in the CaI1.98$\mu$m feature, the discrepancies are relatively small offsets in strength of the feature, with an overall similar trend with effective temperature. Furthermore, the discrepancies tend to be in the dwarf stars rather than the giant stars.  

We find some small differences when comparing the consistency between the empirical and theoretical trends shown in Figures~\ref{fig:ew1}-\ref{fig:ew3} with \citet{cvd2012}. First, the isochrones used in this work extend to cooler temperature than those used in \citet{cvd2012} which means we can now compare the equivalent widths of the coolest dwarfs to the theoretical libraries. For NaI0.82$\mu$m the prediction for the equivalent widths of the coolest dwarfs is inconsistent with the data. However, the prediction for NaI0.82$\mu$m feature strength in this work is completely consistent with the empirical trend. Similarly, in \citet{cvd2012} the theoretical prediction for the CaT was consistent with the data and here the CaT is slightly overpredicted. This most likely do the modest changes that have occurred in the theoretical libraries since \citet{cvd2012}.

Moving onto solar metallicity stars in the optical, Figures~\ref{fig:ew4}-\ref{fig:ew9}, we find that the theoretical predictions fare worse in the optical than the near-IR. We show that in the optical the overall trends predicted by the theoretical library are consistent with the empirical trend but are often offset in predicted strength. \citet{martins2007} compared the strengths of Lick indices of various empirical and theoretical libraries and we will compare the discrepancies of that work with those shown here. Here we focus on the comparison between the theoretical models from \citet{martins2005} and the MILES library.

The C3K models underpredict the index strength for all the Balmer lines: H$\delta$, H$\gamma$, and H$\beta$ with the H$\beta$ difference being less severe than the former two. This is consistent with the result from \citet{martins2007}. Furthermore, C3K overpredicts the G4300${\rm \AA}$ band which is again consistent with \citet{martins2007}. The theoretical predictions for the iron lines are also stronger than what is observed (with the exception of the feature at 5782 ${\rm \AA}$). However, \citet{martins2007} found that the inconsistency between the theoretical and empirical predictions for the iron lines was worse at cooler temperatures which is not consistent with what we find in our comparison. Additionally we find good agreement between the C3K predictions for the line strength of MgI0.88$\mu$m and Mgb at temperatures where \citet{martins2007} find disagreement in the magnesium lines at cool temperatures.  While both C3K and the theoretical stellar spectra in \citet{martins2005} use the ATLAS model atmospheres, C3K uses ATLAS12 and \citet{martins2005} uses ATLAS9 so it is not unexpected that the comparisons in \citet{martins2007} are different.

We now turn to a comparison of the metallicity-dependence of the models and observations. For the near-IR features the consistency seen at solar-metallicity holds at the other metallicities with the exception of some of the coolest stars, particularly dwarf stars, and especially at low-metallicity. For instance, the theoretical predictions for the low-metallicity cool dwarf stars is different from the empirical trend for the CaT, NaI (at $0.82 \mu$m and $1.14\mu$m), FeH0.99$\mu$m, CaI1.98$\mu$m, and KI1.17$\mu$m features. However, there is good consistency over this same regime for the CO2.30$\mu$m and MgI0.88$\mu$m features.

The theoretical predictions for the NaD feature are consistent with the empirical trends over all the metallicity bins, albeit slightly overpredicted in the solar metallicity bin. For some of the optical features the theoretical predictions seem to improve at lower metallicities. This includes many of the iron lines, CN1, G4300${\rm\AA}$, and the Balmer lines with the exception of the coolest dwarf stars. The theoretical predictions for the TiO band are in agreement with the empirical trends at solar metallicity but that agreement worsens at lower metallicities for the cool dwarf stars. 

We emphasize that SPI is not well constrained for the low-metallicity cool dwarf regime and  therefore apparent disagreement between the models and the data should not be over-interpreted. However, the pattern of discrepancy in feature strengths for the low metallicity cool dwarf stars does suggest that caution should be employed when supplementing the cool dwarf regime in empirical libraries with theoretical spectra. We will note that the stellar population models of \citet{cvd2012} and Lick index models \citep[e.g.,][]{trager2000, thomas2003, schiavon2007} use theoretical models only differentially, calculating relative changes with respect to a fiducial model (the response functions) and so they are less sensitive to the absolute limitations of the models.

\section{Summary}

In this paper we presented a new spectroscopic stellar library, the Extended IRTF Library, which in its entirety consists of 284 stars covering a wide range of stellar parameter space, including both low and high metallicities. The stars were observed on the SpeX instrument and the spectra cover a wavelength range of $0.7-2.5\mu$m. All the stars included in this library were selected from the MILES optical library, providing continuous coverage from $0.35-2.5\mu$m. 

In addition to the new library we have also created a Spectral Polynomial Interpolator (SPI). This is a tool that generates a data-driven model from a subset (194 out of 284) of the highest quality library stars and can be used to produce a stellar spectrum for arbitrary values for T$_{\rm eff}$, log$g$, and [Fe/H]. With the Extended IRTF Library and SPI:

\begin{itemize}

\item We find good agreement between observed and synthesized colors for all of the colors explored, including 2MASS H-K$_S$, J-K$_S$, Tycho B$_T$-V$_T$, and V$_T$-K$_S$.  This agreement means that we are recovering the overall spectral shapes to within 1-4\% percent, on average.

\item We find that the empirical uncertainty of the spectra is $ \sim0.5\%$  with the exception of regions heavily contaminated by telluric absorption or regions in the A0 V standard star heavily contaminated by hydrogen absorption lines. In these cases the uncertainty is on the order of a 1-2\% percent.

\item We measured the wavelength-dependent SpeX instrument resolution and found the median resolution of the stars in the library to be consistent with the nominal value of $R\approx2000$. We also independently measured the MILES resolution and found it to be consistent with the updated value from \citet{beifiori2011}.

\item We find that stellar features retain their characteristic properties at non-solar metallicities. This includes the surface gravity sensitive lines such as NaI (at 0.82$\mu$m and 1.14$\mu$m), CaT, FeH0.99$\mu$m, KI1.17$\mu$m, and CO2.30$\mu$m. 

\item We find the theoretical predictions for the spectral features qualitatively agree with the observed trends.  The C3K theoretical spectra in many cases reproduce the trends quantitatively as well, especially in the near-IR.  Nonetheless, there are many features that show significant quantitative discrepancies between models and observations, especially in the optical.

\end{itemize}

\acknowledgments

The authors wish to recognize and acknowledge the very significant cultural role and reverence that the summit of Maunakea has always had within the indigenous Hawaiian community. We are most fortunate to have the opportunity to conduct observations from this mountain. AV is supported in part by a NSF Graduate Research Fellowship. CC acknowledges support from NASA grant NNX15AK14G, NSF grant AST-1313280, and the Packard Foundation. The authors would like to thank the anonymous referee whose careful reading greatly improved the quality of this manuscript, M. Cushing for helpful discussion, and P. S{\'a}nchez-Bl{\'a}zquez for generously sharing data. This research has made use of the SIMBAD database, operated at CDS, Strasbourg, France, the VizieR catalogue access tool, CDS, Strasbourg, France, of Astropy, a community-developed core Python package for Astronomy \citep{2013A&A...558A..33A}, matplotlib, a Python library for publication quality graphics \citep{Hunter:2007}, and of SciPy \citep{jones_scipy_2001}. The atmospheric transmission spectra was provided by Gemini Observatory. 

\bibliographystyle{apj}
\bibliography{biblio}

\appendix
In Table~\ref{table:obs_log} we list all the stars in the Extended IRTF Library along with their corresponding MILES ID numbers. We also include the name and spectral type of the standard star used to telluric correct and flux calibrate each library star. We give basic properties of the stars including Right Ascension and Declination, the metallicities estimated by \citet{prugniel2011} and \citet{sharma2016} and 2MASS K$_S$ magnitude. The extinction values listed in the ``E(B-V)'' column are from the values compiled in \citet{sanchez2006}. We also list the velocities and bolometric luminosities we computed for this work. Finally, we indicate whether each star has been flagged for potential quality, stellar parameter, or shape issues.

\clearpage

\section{} \label{appendix}

\LongTables
\begin{landscape}

\begin{deluxetable*}{llllllclllrrrrrrrr}

\tablecolumns{13}
\tablecaption{The library stars. \label{table:obs_log}}
\tablehead{
\colhead{Name} & \colhead{MILES} & \colhead{Standard} & \colhead{} & \colhead{RA} &  \colhead{} &  \colhead{} & \colhead{$$} & \colhead{DEC}  &  \colhead{} & \colhead{[Fe/H]} & \colhead{K$_S$} & \colhead{E(B-V)} &  \colhead{Velocity} &\colhead{log L$_{\rm bol}$} & \colhead{Q$^1$} & \colhead{P$^2$} & \colhead{S$^3$} \\
\colhead{} & \colhead{ID} & \colhead{Star} & \colhead{hh}  & \colhead{mm}  & \colhead{ss}  & \colhead{$+/-$} &  \colhead{dd}  & \colhead{mm}  &  \colhead{ss} & \colhead{} & \colhead{(mag)} & \colhead{(mag)} & \colhead{(km ${\rm s}^{-1}$)} & \colhead{(L$_{\odot}$)} &  \colhead{Flag} & \colhead{Flag} & \colhead{Flag} } 
\startdata
\\

HD225239 & 0003F & HD007215-A0V & 00 & 04 & 53.7603 & + & 34 & 39 & 35.259 & -0.51 & 4.44 & 0.050 & -48.2 & 0.75 & 0 & 0 & 1 \\
HD000448 & 0009F & CCDM J00209+1059AB-A0V & 00 & 09 & 02.4241 & + & 18 & 12 & 43.066 & 0.02 & 3.32 & 0.025 & -15.9 & 1.55 & 0 & 1 & 0 \\
BD+130013 & 0010F & HD007215-A0V & 00 & 12 & 30.3109 & + & 14 & 33 & 48.678 & -0.75 & 6.24 & 0.011 & -24.0 & -0.15 & 0 & 0 & 0 \\
HD001326B & 0012F & HD001561-A0Vs & 00 & 18 & 25.498 & + & 44 & 01 & 37.64 & -0.57 & 5.95 & 0.001 & -7.3 & -2.45 & 1 & 0 & 0 \\
HD001461 & 0013F & HD013936-A0V & 00 & 18 & 41.8672 & - & 08 & 03 & 10.802 & 0.19 & 4.9 & 0.044 & -37.7 & 0.08 & 0 & 0 & 0 \\
HD001918 & 0014F & HD001561-A0Vs & 00 & 23 & 43.2412 & + & 45 & 05 & 20.892 & -0.4 & 5.23 & 0.037 & 59.5 & 2.05 & 0 & 0 & 0 \\
HD003567 & 0022F & HD001154-A0V & 00 & 38 & 31.9474 & - & 08 & 18 & 33.395 & -1.14 & 7.89 & 0.000 & -41.4 & 0.22 & 0 & 0 & 0 \\
HD003546 & 0023F & HD006457-A0Vn & 00 & 38 & 33.3461 & + & 29 & 18 & 42.313 & -0.66 & 2.07 & 0.000 & -58.0 & 1.68 & 0 & 0 & 0 \\
HD003574 & 0024F & HD001561-A0Vs & 00 & 39 & 09.8946 & + & 49 & 21 & 16.521 & 0.01 & 1.6 & 0.048 & -13.9 & 3.82 & 0 & 0 & 0 \\
HD004307 & 0028F & HD015130-A0V & 00 & 45 & 28.6878 & - & 12 & 52 & 50.914 & -0.24 & 4.62 & 0.000 & -25.2 & 0.44 & 0 & 0 & 0 \\
HD004744 & 0033F & HD006457-A0Vn & 00 & 49 & 52.7904 & + & 30 & 27 & 01.028 & -0.74 & 4.93 & 0.010 & -148.3 & 1.61 & 0 & 0 & 0 \\
HD004906 & 0034F & HD006229-G5III & 00 & 51 & 14.0528 & + & 18 & 47 & 25.153 & -0.66 & 6.77 & 0.028 & -127.4 & 0.65 & 0 & 0 & 0 \\
NGC288-77 & 0897C & HD004329-A0V & 00 & 52 & 43.00 & - & 26 & 27 & 28.0 & -1.32 & 10.01 & 0.030 & -31.8 & 2.85 & 0 & 0 & 0 \\
HD005384 & 0036F & HD013936-A0V & 00 & 55 & 42.3993 & - & 07 & 20 & 49.742 & 0.18 & 2.42 & 0.012 & -49.1 & 2.5 & 1 & 0 & 0 \\
HD005780 & 0038F & CCDM J00209+1059AB-A0V & 00 & 59 & 23.3126 & + & 00 & 46 & 43.563 & -0.71 & 4.22 & 0.000 & -90.0 & 2.41 & 0 & 0 & 1 \\
HD006229 & 0043F & HD006457-A0Vn & 01 & 03 & 36.4553 & + & 23 & 46 & 06.377 & -1.14 & 6.57 & 0.000 & -103.5 & 1.51 & 0 & 0 & 1 \\
HD006497 & 0045F & HD223386-A0V & 01 & 07 & 00.1738 & + & 56 & 56 & 05.904 & 0.04 & 3.88 & 0.014 & -75.5 & 1.42 & 0 & 0 & 0 \\
HD006834 & 0049F & HD007215-A0V & 01 & 09 & 35.2623 & + & 39 & 46 & 51.765 & -0.58 & 7.29 & 0.005 & 7.8 & 0.38 & 0 & 0 & 1 \\
HD006755 & 0050F & HD223386-A0V & 01 & 09 & 43.0646 & + & 61 & 32 & 50.190 & -1.58 & 5.66 & 0.060 & -286.6 & 1.19 & 0 & 0 & 0 \\
HD006833 & 0051F & HD001561-A0Vs & 01 & 09 & 52.2648 & + & 54 & 44 & 20.278 & -0.84 & 4.04 & 0.047 & -241.6 & 1.95 & 0 & 0 & 1 \\
HD007106 & 0052F & HD007215-A0V & 01 & 11 & 39.6364 & + & 30 & 05 & 22.690 & -0.02 & 2.15 & 0.011 & 19.1 & 1.59 & 0 & 0 & 0 \\
HD007351 & 0053F & HD007215-A0V & 01 & 14 & 04.9061 & + & 28 & 31 & 46.471 & -0.35 & 1.64 & 0.048 & -8.7 & 3.32 & 0 & 0 & 0 \\
HD007672 & 0056F & CCDM J00209+1059AB-A0V & 01 & 16 & 36.2886 & - & 02 & 30 & 01.321 & -0.42 & 3.38 & 0.076 & 23.7 & 1.55 & 0 & 1 & 0 \\
HD008724 & 0057F & HD006457-A0Vn & 01 & 26 & 17.5949 & + & 17 & 07 & 35.127 & -1.63 & 5.64 & 0.020 & -112.8 & 1.97 & 0 & 1 & 0 \\
HD009919 & 0064F & HD006457-A0Vn & 01 & 37 & 05.9152 & + & 12 & 08 & 29.518 & -0.35 & 4.58 & 0.006 & 38.4 & 0.72 & 0 & 0 & 0 \\
HD010700 & 0067F & HD013936-A0V & 01 & 44 & 04.0833 & - & 15 & 56 & 14.926 & -0.46 & 1.79 & 0.004 & -23.4 & -0.32 & 1 & 0 & 1 \\
HD011397 & 0072F & HD021875-A0V & 01 & 51 & 40.5274 & - & 16 & 19 & 03.526 & -0.58 & 7.27 & 0.070 & 2.7 & -0.16 & 0 & 0 & 0 \\
BD+290366 & 0077F & BD+320409-A0V & 02 & 10 & 24.5278 & + & 29 & 48 & 23.669 & -0.95 & 7.22 & 0.00 & 45.0 & -0.1 & 0 & 0 & 0 \\
HD013520 & 0080F & HD013869-A0V & 02 & 13 & 13.3238 & + & 44 & 13 & 53.954 & -0.27 & 1.33 & 0.020 & -46.7 & 2.77 & 0 & 0 & 0 \\
BD-010306 & 0081F & HD018571-A0V & 02 & 14 & 40.2986 & - & 01 & 12 & 05.119 & -0.89 & 7.52 & 0.000 & 18.0 & -0.11 & 0 & 0 & 0 \\
HD013783 & 0082F & HD012468-A0V & 02 & 16 & 49.2129 & + & 64 & 57 & 09.380 & -0.49 & 6.59 & 0.000 & -31.8 & -0.21 & 0 & 0 & 0 \\
HD014802 & 0084F & HD017224-A0V & 02 & 22 & 32.5464 & - & 23 & 48 & 58.779 & -0.07 & 3.74 & 0.000 & -14.2 & 0.47 & 0 & 1 & 0 \\
HD016673 & 0092F & HD015130-A0V & 02 & 40 & 12.4215 & - & 09 & 27 & 10.361 & 0.0 & 4.53 & 0.005 & -10.2 & 0.26 & 0 & 0 & 1 \\
HD017361 & 0098F & HD013869-A0V & 02 & 47 & 54.5414 & + & 29 & 14 & 49.613 & 0.02 & 2.1 & 0.020 & -18.9 & 1.6 & 0 & 0 & 0 \\
HD017491 & 0099F & HD013936-A0V & 02 & 47 & 55.9222 & - & 12 & 27 & 38.322 & -0.15 & 0.36 & 0.034 & -11.7 & 3.33 & 0 & 0 & 0 \\
HD017548 & 0101F & HD021686-A0Vn & 02 & 48 & 51.8520 & - & 01 & 30 & 34.747 & -0.53 & 6.76 & 0.008 & -35.5 & 0.13 & 0 & 0 & 0 \\
HD018191 & 0103F & HD016811-A0V & 02 & 55 & 48.4980 & + & 18 & 19 & 53.902 & -0.05 & -0.87 & 0.164 & 81.8 & 3.15 & 0 & 0 & 0 \\
HD019373 & 0108F & HD021038-A0V & 03 & 09 & 04.0198 & + & 49 & 36 & 47.799 & 0.11 & 2.72 & 0.034 & 56.3 & 0.29 & 0 & 0 & 0 \\
HD020893 & 0114F & HD023258-A0V & 03 & 22 & 45.2400 & + & 20 & 44 & 31.438 & 0.07 & 2.19 & 0.031 & -21.6 & 2.08 & 0 & 0 & 0 \\
BD+430699 & 0115F & HD021038-A0V & 03 & 23 & 33.4875 & + & 43 & 57 & 26.222 & -0.38 & 6.54 & 0.007 & 6.6 & -0.69 & 0 & 0 & 0 \\
HD021017 & 0116F & HD023258-A0V & 03 & 24 & 18.4745 & + & 24 & 43 & 26.617 & 0.07 & 2.89 & 0.004 & -24.5 & 1.45 & 1 & 0 & 0 \\
HD021197 & 0117F & HD025175-A0V & 03 & 24 & 59.7314 & - & 05 & 21 & 49.523 & 0.13 & 5.12 & 0.072 & -36.8 & -0.6 & 0 & 0 & 1 \\
HD021581 & 0118F & HD018571-A0V & 03 & 28 & 54.486 & - & 00 & 25 & 03.11 & -1.51 & 6.41 & 0.000 & 100.5 & 1.35 & 0 & 0 & 0 \\
HD022879 & 0123F & HD031295-A3Va & 03 & 40 & 22.0641 & - & 03 & 13 & 01.124 & -0.8 & 5.18 & 0.000 & 118.6 & 0.05 & 0 & 0 & 0 \\
HD023439A & 0127F & HD021038-A0V & 03 & 47 & 02.113 & + & 41 & 25 & 38.06 & -0.9 & 6.12 & 0.010 & 34.5 & -1.0 & 0 & 0 & 0 \\
HD023439B & 0128F & HD021038-A0V & 03 & 47 & 02.636 & + & 41 & 25 & 42.56 & -1.09 & 6.35 & 0.010 & 27.3 & -1.0 & 0 & 0 & 0 \\
HD023841 & 0130F & HD021686-A0Vn & 03 & 48 & 30.7665 & + & 09 & 38 & 45.410 & -0.66 & 3.8 & 0.053 & -94.1 & 2.42 & 0 & 0 & 1 \\
HD024341 & 0133F & HD029526-A0V & 03 & 54 & 51.1901 & + & 52 & 25 & 11.526 & -0.62 & 6.11 & 0.023 & 99.2 & 0.3 & 0 & 0 & 0 \\
HD024421 & 0134F & HD029526-A0V & 03 & 55 & 37.0801 & + & 52 & 13 & 36.550 & -0.29 & 5.47 & 0.003 & -56.4 & 0.32 & 0 & 0 & 0 \\
HD025673 & 0138F & HD021875-A0V & 04 & 04 & 20.3036 & - & 04 & 39 & 18.471 & -0.4 & 7.46 & 0.006 & 81.3 & -0.48 & 0 & 0 & 0 \\
HD026297 & 0139F & HD021875-A0V & 04 & 09 & 03.4176 & - & 15 & 53 & 27.063 & -1.79 & 4.63 & 0.017 & 25.1 & 2.73 & 0 & 0 & 0 \\
HD284248 & 0143F & HD023258-A0V & 04 & 14 & 35.5155 & + & 22 & 21 & 04.261 & -1.55 & 7.87 & 0.000 & 303.3 & 0.09 & 0 & 0 & 0 \\
BD-060855 & 0144F & HD025792-A0V & 04 & 14 & 58.1083 & - & 05 & 37 & 48.856 & -0.69 & 8.79 & 0.01 & 324.8 & -0.54 & 0 & 0 & 0 \\
HD027126 & 0147F & HD025152-A0V & 04 & 18 & 29.5492 & + & 35 & 59 & 30.080 & -0.38 & 6.64 & 0.000 & -56.5 & -0.09 & 0 & 0 & 1 \\
HD027371 & 0149F & HD032358-B6V & 04 & 19 & 47.6038 & + & 15 & 37 & 39.515 & 0.15 & 1.52 & 0.014 & 65.2 & 1.94 & 0 & 0 & 0 \\
HD027771 & 0150F & HD025175-A0V & 04 & 23 & 32.3310 & + & 14 & 40 & 13.717 & 0.27 & 7.14 & 0.000 & 20.1 & -0.33 & 0 & 0 & 0 \\
HD285773 & 0153F & HD025175-A0V & 04 & 29 & 31.6067 & + & 17 & 53 & 35.466 & 0.25 & 7.05 & 0.000 & 9.8 & -0.37 & 0 & 0 & 0 \\
BD+501021 & 0158F & HD029526-A0V & 04 & 38 & 07.8341 & + & 51 & 10 & 12.467 & -0.65 & 8.28 & 0.060 & -31.8 & -0.55 & 0 & 0 & 0 \\
HD030649 & 0162F & HD031069-A0V & 04 & 51 & 43.5572 & + & 45 & 50 & 02.971 & -0.48 & 5.44 & 0.000 & 20.0 & 0.08 & 0 & 0 & 0 \\
HD031128 & 0163F & HD034868-A0V & 04 & 52 & 09.9101 & - & 27 & 03 & 50.950 & -1.45 & 7.74 & 0.000 & 85.1 & -0.11 & 0 & 0 & 0 \\
HD031295 & 0166F & HD031411-A0V & 04 & 54 & 53.7287 & + & 10 & 09 & 02.995 & -0.73 & 4.42 & 0.100 & -17.0 & 1.15 & 0 & 1 & 0 \\
HD031767 & 0167F & HD034317-A0V & 04 & 58 & 32.9021 & + & 01 & 42 & 50.458 & -0.02 & 1.34 & 0.101 & 4.4 & 3.36 & 0 & 0 & 0 \\
HD033608 & 0173F & HD031411-A0V & 05 & 11 & 19.1768 & - & 02 & 29 & 26.817 & 0.21 & 4.82 & 0.000 & 62.2 & 0.62 & 0 & 0 & 0 \\
HD034411 & 0176F & HD031069-A0V & 05 & 19 & 08.4742 & + & 40 & 05 & 56.582 & 0.08 & 3.04 & 0.000 & 32.8 & 0.27 & 0 & 0 & 0 \\
HD035179 & 0178F & HD035505-A0V & 05 & 22 & 21.0066 & - & 14 & 23 & 50.061 & -0.6 & 7.1 & 0.000 & 13.7 & 1.96 & 0 & 0 & 0 \\
M79\_153 & 0915C & HD040972-A0V & 05 & 24 & 09.1300 & - & 24 & 34 & 16.1000 & -1.6 & 9.61 & 0.010 & -19.9 & -1.0 & 0 & 0 & 1 \\
M79\_160 & 0916C & HD040972-A0V & 05 & 24 & 09.4000 & - & 24 & 33 & 29.9000 & -1.6 & 10.26 & 0.010 & 216.4 & 3.06 & 0 & 0 & 0 \\
M79\_223 & 0917C & HD040972-A0V & 05 & 24 & 18.4860 & - & 24 & 31 & 43.3800 & -1.6 & 9.83 & 0.010 & 188.7 & 3.19 & 0 & 0 & 0 \\
HD035296 & 0180F & HD035036-A0V & 05 & 24 & 25.4638 & + & 17 & 23 & 00.716 & 0.01 & 4.04 & 0.000 & 78.5 & 0.1 & 0 & 0 & 1 \\
HD035620 & 0181F & HD035036-A0V & 05 & 27 & 38.8860 & + & 34 & 28 & 33.212 & 0.1 & 2.1 & 0.115 & 52.2 & 2.48 & 0 & 1 & 0 \\
HD036003 & 0182F & HD037887-B9.5V & 05 & 28 & 26.0960 & - & 03 & 29 & 58.399 & -0.15 & 4.88 & 0.011 & -19.9 & -0.71 & 0 & 0 & 0 \\
HD036395 & 0183F & HD037887-A0V & 05 & 31 & 27.3959 & - & 03 & 40 & 38.031 & -0.05 & 4.04 & 0.006 & -24.3 & -1.27 & 0 & 0 & 0 \\
HD037160 & 0184F & HD035656-B9V & 05 & 36 & 54.3881 & + & 09 & 17 & 26.409 & -0.64 & 1.81 & 0.000 & 124.2 & 1.47 & 0 & 0 & 0 \\
HD037828 & 0187F & HD035505-A0V & 05 & 40 & 54.6451 & - & 11 & 12 & 00.191 & -1.41 & 4.06 & 0.020 & 188.5 & 2.55 & 0 & 0 & 0 \\
HD037984 & 0189F & HD034317-A0V & 05 & 42 & 28.6327 & + & 01 & 28 & 28.665 & -0.52 & 2.21 & 0.010 & 80.4 & 2.02 & 0 & 0 & 0 \\
HD038393 & 0191F & * 10 lep-A0V & 05 & 44 & 27.7908 & - & 22 & 26 & 54.180 & -0.09 & 2.51 & 0.000 & -17.8 & 0.28 & 0 & 0 & 0 \\
HD039364 & 0196F & * 10 Lep-A0V & 05 & 51 & 19.2961 & - & 20 & 52 & 44.723 & -0.74 & 1.4 & 0.000 & 81.9 & 1.54 & 0 & 0 & 0 \\
HD039833 & 0198F & HD039953-A0V & 05 & 55 & 01.958 & - & 00 & 30 & 28.69 & 0.18 & 6.15 & 0.010 & 6.8 & 0.08 & 0 & 0 & 0 \\
HD041312 & 0203F & HD042729-A0V & 06 & 03 & 15.6045 & - & 26 & 17 & 04.355 & -0.75 & 1.69 & 0.030 & 149.4 & 2.43 & 1 & 0 & 0 \\
HD042543 & 0211F & HD046553-A0Vnn & 06 & 12 & 19.0986 & + & 22 & 54 & 30.642 & 0.18 & 0.81 & 0.606 & 51.6 & 5.19 & 0 & 0 & 0 \\
HD043380 & 0215F & HD045105-A0V & 06 & 18 & 16.8650 & + & 46 & 21 & 37.593 & -0.05 & 3.65 & 0.019 & -27.5 & 1.74 & 1 & 0 & 0 \\
HD044007 & 0216F & HD038206-A0V & 06 & 18 & 48.5271 & - & 14 & 50 & 43.425 & -1.53 & 5.7 & 0.000 & 162.1 & 1.3 & 0 & 0 & 1 \\
HD044030 & 0219F & HD046553-A0Vnn & 06 & 20 & 35.3327 & + & 25 & 36 & 29.306 & -0.51 & 4.14 & 0.107 & -132.7 & 2.65 & 0 & 0 & 0 \\
HD045282 & 0222F & HD045137-A0V & 06 & 26 & 40.7734 & + & 03 & 25 & 29.794 & -1.42 & 6.09 & 0.000 & 281.2 & 1.11 & 0 & 0 & 0 \\
HD045829 & 0223F & HD043583-A0V & 06 & 30 & 02.2945 & + & 07 & 55 & 16.009 & 0.11 & 3.35 & 0.228 & 58.3 & 4.06 & 0 & 0 & 0 \\
HD047914 & 0227F & HD045105-A0V & 06 & 43 & 04.9710 & + & 44 & 31 & 28.022 & 0.04 & 1.71 & 0.033 & -43.7 & 2.2 & 1 & 0 & 0 \\
HD048682 & 0232F & HD045105-A0V & 06 & 46 & 44.3381 & + & 43 & 34 & 38.726 & 0.11 & 4.13 & 0.000 & -1.0 & 0.13 & 0 & 1 & 0 \\
HD054719 & 0245F & HD058296-A0V & 07 & 11 & 08.3704 & + & 30 & 14 & 42.583 & 0.13 & 1.68 & 0.020 & -2.6 & 2.3 & 0 & 0 & 0 \\
HD055496 & 0246F & HD056341-A0V & 07 & 12 & 11.3771 & - & 22 & 59 & 00.619 & -1.48 & 5.93 & 0.020 & 305.7 & 2.15 & 0 & 0 & 0 \\
HD056274 & 0249F & HD067725-B8V & 07 & 15 & 50.7956 & - & 13 & 02 & 58.131 & -0.53 & 6.2 & 0.079 & 80.9 & -0.07 & 0 & 0 & 1 \\
HD058551 & 0255F & HD064648-A0Vs & 07 & 26 & 50.2521 & + & 21 & 32 & 08.320 & -0.42 & 5.24 & 0.000 & 14.3 & 0.25 & 0 & 0 & 0 \\
HD059374 & 0257F & CCDM J07274+1519AB-A0V & 07 & 30 & 29.0172 & + & 18 & 57 & 40.620 & -0.82 & 6.98 & 0.000 & 75.7 & -0.05 & 0 & 0 & 1 \\
HD059984 & 0259F & HD067725-B8V & 07 & 32 & 05.7628 & - & 08 & 52 & 52.776 & -0.68 & 4.48 & 0.023 & 14.9 & 0.37 & 0 & 1 & 0 \\
BD-011792 & 0266F & HD079107-A0V & 07 & 39 & 50.1104 & - & 01 & 31 & 20.372 & -0.81 & 7.04 & 0.000 & 56.8 & 0.89 & 0 & 0 & 0 \\
HD061935 & 0270F & HD056525-A0V & 07 & 41 & 14.8325 & - & 09 & 33 & 04.071 & -0.06 & 1.64 & 0.015 & 16.9 & 1.74 & 0 & 0 & 0 \\
BD+002058A & 0272F & HD079108-A0V & 07 & 43 & 43.9549 & - & 00 & 04 & 00.899 & -1.22 & 8.84 & 0.02 & -54.4 & -0.07 & 0 & 0 & 0 \\
BD-182065 & 0278F & HD067213-A0V & 07 & 52 & 38.0071 & - & 18 & 35 & 37.734 & -0.71 & 7.27 & 0.000 & 111.3 & -1.0 & 0 & 0 & 0 \\
HD063791 & 0281F & HD237611-A0V & 07 & 54 & 28.7238 & + & 62 & 08 & 10.761 & -1.62 & 5.43 & 0.03 & -128.8 & 2.89 & 0 & 0 & 1 \\
HD065583 & 0285F & HD064648-A0Vs & 08 & 00 & 32.1289 & + & 29 & 12 & 44.471 & -0.65 & 5.09 & 0.000 & -10.8 & -0.34 & 0 & 0 & 0 \\
HD073898 & 0307F & HD079290-A0V & 08 & 39 & 42.4741 & - & 29 & 33 & 39.898 & -0.56 & 2.9 & 0.043 & 1.2 & 1.75 & 0 & 0 & 1 \\
HD073394 & 0309F & HD063586-A0Vn & 08 & 40 & 22.5437 & + & 51 & 45 & 06.558 & -1.49 & 4.96 & 0.002 & -80.2 & 2.34 & 0 & 0 & 0 \\
HD074011 & 0312F & HD071906-A0V & 08 & 42 & 30.8192 & + & 34 & 11 & 15.699 & -0.56 & 5.79 & 0.006 & 83.0 & 0.31 & 0 & 0 & 0 \\
HD074377 & 0316F & HD071906-A0V & 08 & 45 & 10.3977 & + & 41 & 40 & 18.615 & -0.37 & 6.13 & 0.002 & -9.2 & -0.5 & 0 & 0 & 0 \\
HD074721 & 0317F & HD075137-A0V & 08 & 45 & 59.2615 & + & 13 & 15 & 48.607 & -1.32 & 8.52 & 0.032 & 75.7 & 2.84 & 0 & 0 & 0 \\
HD074462 & 0319F & * 39 UMa-A0III & 08 & 48 & 20.6521 & + & 67 & 26 & 59.893 & -1.4 & 6.05 & 0.020 & -206.7 & 2.05 & 0 & 0 & 0 \\
HD075318 & 0320F & HD079108-A0V & 08 & 49 & 21.2072 & + & 03 & 41 & 02.392 & -0.13 & 6.08 & 0.000 & 64.0 & -0.12 & 0 & 0 & 0 \\
HD075691 & 0321F & * eta pyx-A0V & 08 & 50 & 31.9228 & - & 27 & 42 & 35.442 & -0.16 & 1.14 & 0.033 & 15.0 & 2.11 & 0 & 0 & 0 \\
HD076151 & 0323F & HD073687-A0V & 08 & 54 & 17.9480 & - & 05 & 26 & 04.058 & 0.15 & 4.46 & 0.000 & 46.6 & 0.0 & 0 & 0 & 0 \\
HD076932 & 0325F & HD073687-A0V & 08 & 58 & 43.9331 & - & 16 & 07 & 57.813 & -0.82 & 4.36 & 0.000 & 91.2 & 0.21 & 0 & 0 & 0 \\
HD076813 & 0330F & HD071906-A0V & 08 & 59 & 32.6543 & + & 32 & 25 & 06.809 & -0.06 & 3.15 & 0.331 & 4.1 & 2.02 & 0 & 1 & 0 \\
HD077338 & 0331F & HD073495-A0V & 09 & 01 & 12.4941 & - & 25 & 31 & 37.426 & 0.36 & 6.76 & 0.029 & 8.3 & -0.31 & 0 & 1 & 0 \\
HD077236 & 0332F & HD079108-A0V & 09 & 01 & 14.7956 & - & 02 & 33 & 41.689 & -0.89 & 4.64 & 0.000 & 144.3 & 1.98 & 0 & 0 & 0 \\
HD078737 & 0337F & HD073495-A0V & 09 & 09 & 03.3696 & - & 27 & 01 & 49.314 & -0.46 & 7.74 & 0.005 & 1.5 & 0.65 & 0 & 1 & 0 \\
HD081029 & 0345F & HD079108-A0V & 09 & 23 & 15.7647 & + & 03 & 30 & 04.997 & -0.08 & 6.49 & 0.065 & 60.8 & 1.03 & 0 & 1 & 0 \\
HD081797 & 0347F & HD084886-A0V & 09 & 27 & 35.2427 & - & 08 & 39 & 30.958 & 0.01 & -1.13 & 0.012 & 0.3 & 2.89 & 1 & 0 & 0 \\
HD082074 & 0349F & HD079108-A0V & 09 & 29 & 32.4157 & - & 04 & 14 & 47.887 & -0.43 & 4.15 & 0.009 & -50.0 & 0.95 & 0 & 0 & 0 \\
HD082734 & 0351F & HD082724-A0V & 09 & 33 & 12.4596 & - & 21 & 06 & 56.601 & 0.2 & 2.83 & 0.031 & 40.7 & 2.09 & 0 & 0 & 0 \\
HD083212 & 0354F & HD083104-A0V & 09 & 36 & 19.9522 & - & 20 & 53 & 14.756 & -1.64 & 5.61 & 0.025 & 97.5 & 2.83 & 0 & 0 & 0 \\
HD083632 & 0358F & HD089239-A0V & 09 & 40 & 34.0571 & + & 26 & 00 & 13.967 & -0.85 & 4.72 & 0.069 & 98.3 & 3.4 & 0 & 0 & 0 \\
HD233666 & 0359F & HD092728-A0III & 09 & 42 & 19.4752 & + & 53 & 28 & 26.156 & -1.62 & 7.37 & 0.040 & -63.7 & 2.82 & 0 & 1 & 0 \\
HD086986 & 0368F & HD089239-A0V & 10 & 02 & 29.5658 & + & 14 & 33 & 25.189 & -1.7 & 7.5 & 0.030 & 42.5 & 1.24 & 0 & 0 & 0 \\
HD087141 & 0369F & * 39 ma-A0V & 10 & 04 & 36.3229 & + & 53 & 53 & 30.170 & 0.09 & 4.5 & 0.030 & -16.4 & 1.01 & 0 & 0 & 0 \\
HD088230 & 0373F & HD099966-A0V & 10 & 11 & 22.1405 & + & 49 & 27 & 15.256 & -0.01 & 2.96 & 0.000 & -64.0 & -0.96 & 0 & 0 & 0 \\
HD088446 & 0374F & HD023366-A0V & 10 & 12 & 19.0769 & + & 17 & 17 & 57.044 & -0.51 & 6.45 & 0.000 & 91.6 & 0.41 & 0 & 0 & 0 \\
HD088725 & 0375F & HD097585-A0V & 10 & 14 & 08.3341 & + & 03 & 09 & 04.677 & -0.64 & 6.15 & 0.000 & -60.1 & -0.1 & 0 & 0 & 0 \\
HD088986 & 0378F & HD089239-A0V & 10 & 16 & 28.0831 & + & 28 & 40 & 56.927 & 0.04 & 4.88 & 0.000 & 27.4 & 0.37 & 0 & 0 & 0 \\
HD089010 & 0379F & HD089239-A0V & 10 & 16 & 32.2887 & + & 23 & 30 & 11.144 & 0.0 & 4.34 & 0.025 & -48.0 & 0.53 & 0 & 0 & 0 \\
HD091347 & 0389F & HD099966-A0V & 10 & 33 & 50.5558 & + & 49 & 11 & 10.222 & -0.44 & 6.05 & 0.000 & -38.7 & 0.04 & 0 & 0 & 0 \\
HD094028 & 0394F & HD088960-A0Vn & 10 & 51 & 28.1244 & + & 20 & 16 & 38.960 & -1.3 & 6.83 & 0.000 & 85.1 & -0.03 & 0 & 0 & 0 \\
BD-103166 & 0395F & BD+172473-G5 & 10 & 58 & 28.7798 & - & 10 & 46 & 13.386 & 0.42 & 8.12 & 0.077 & 63.0 & -0.38 & 0 & 0 & 0 \\
HD095128 & 0396F & HD098989-A0V & 10 & 59 & 27.9728 & + & 40 & 25 & 48.920 & 0.02 & 3.75 & 0.000 & 18.9 & 0.14 & 0 & 0 & 0 \\
HD096360 & 0400F & * 86 UMa-A0V & 11 & 07 & 52.7864 & + & 68 & 21 & 58.864 & -0.02 & 2.77 & 0.139 & -124.5 & 2.61 & 0 & 0 & 0 \\
BD+362165 & 0401F & HD097585-A0V & 11 & 12 & 47.9995 & + & 35 & 43 & 43.899 & -1.45 & 8.5 & 0.00 + & -154.7 & 0.32 & 0 & 1 & 0 \\
HD097916 & 0405F & HD097585-A0V & 11 & 15 & 54.2297 & + & 02 & 05 & 12.083 & -0.73 & 8.02 & 0.000 & 61.5 & 0.3 & 0 & 0 & 1 \\
HD097855 & 0406F & HD092728-A0III & 11 & 16 & 04.0330 & + & 52 & 46 & 23.376 & -0.39 & 5.28 & 0.021 & -6.7 & 0.35 & 0 & 0 & 0 \\
HD098553 & 0408F & HD099459-A0V & 11 & 20 & 11.5950 & - & 19 & 34 & 40.457 & -0.44 & 6.08 & 0.000 & -38.0 & -0.1 & 0 & 0 & 1 \\
HD099109 & 0409F & HD097585-A0V & 11 & 24 & 17.3588 & - & 01 & 31 & 44.664 & 0.4 & 7.16 & 0.088 & 81.3 & -0.21 & 0 & 0 & 0 \\
HD100906 & 0414F & HD101369-A0V & 11 & 36 & 42.4113 & - & 18 & 58 & 10.585 & -0.46 & 7.49 & 0.000 & 77.7 & 2.06 & 0 & 0 & 0 \\
BD+511696 & 0419F & HD099966-A0V & 11 & 46 & 35.1545 & + & 50 & 52 & 54.682 & -1.3 & 8.31 & 0.005 & 102.7 & -0.26 & 0 & 0 & 0 \\
HD102328 & 0420F & HD092728-A0III & 11 & 46 & 55.6221 & + & 55 & 37 & 41.485 & 0.28 & 2.63 & 0.050 & 58.5 & 1.58 & 0 & 0 & 0 \\
HD102634 & 0421F & * 69 Leo-A0V & 11 & 49 & 01.281 & - & 00 & 19 & 07.22 & 0.22 & 4.92 & 0.000 & -7.4 & 0.53 & 0 & 0 & 1 \\
HD103095 & 0423F & HD107655-A0V & 11 & 52 & 58.7693 & + & 37 & 43 & 07.234 & -1.21 & 4.37 & 0.000 & -114.5 & -0.68 & 0 & 0 & 0 \\
HD104307 & 0427F & HD099922-A0V & 12 & 00 & 42.2223 & - & 21 & 50 & 15.006 & -0.07 & 3.15 & 0.042 & 96.0 & 2.21 & 0 & 0 & 0 \\
HD105452 & 0431F & HD110902-A0V & 12 & 08 & 24.8165 & - & 24 & 43 & 43.950 & -0.19 & 3.31 & 0.028 & -16.7 & 0.55 & 0 & 0 & 0 \\
HD105546 & 0432F & HD092728-A0III & 12 & 09 & 02.7208 & + & 59 & 01 & 05.134 & -1.46 & 6.67 & 0.000 & 84.0 & 1.57 & 0 & 0 & 0 \\
HD105740 & 0433F & HD107655-A0V & 12 & 10 & 16.9638 & + & 16 & 22 & 13.363 & -0.54 & 5.85 & 0.000 & -30.9 & 1.33 & 0 & 0 & 0 \\
HD106038 & 0434F & HD107655-A0V & 12 & 12 & 01.3700 & + & 13 & 15 & 40.622 & -1.25 & 8.76 & 0.000 & 124.5 & -0.18 & 0 & 0 & 0 \\
CD-2809374 & 0435F & HD112305-A0V & 12 & 14 & 29.7453 & - & 29 & 35 & 55.677 & -0.76 & 8.12 & 0.010 & 19.5 & 0.34 & 0 & 0 & 0 \\
HD107213 & 0438F & HD107655-A0V & 12 & 19 & 29.5259 & + & 28 & 09 & 24.902 & 0.24 & 5.13 & 0.000 & 5.3 & 0.81 & 0 & 0 & 0 \\
BD+172473 & 0439F & HD107655-A0V & 12 & 23 & 31.021 & + & 16 & 54 & 09.24 & -1.04 & 8.16 & 0.000 & 54.3 & -1.0 & 0 & 0 & 0 \\
BD+312360 & 0440F & HD107655-A0V & 12 & 24 & 17.400 & + & 30 & 46 & 48.73 & -0.79 & 8.83 & 0.00 & 51.8 & -1.0 & 0 & 0 & 0 \\
HD108177 & 0441F & HD109309-A0V & 12 & 25 & 34.9550 & + & 01 & 17 & 02.267 & -1.41 & 8.35 & 0.000 & 169.6 & 0.08 & 0 & 1 & 0 \\
HD108564 & 0442F & HD112304-A0V & 12 & 28 & 19.1200 & - & 16 & 54 & 39.771 & -1.0 & 6.9 & 0.016 & 78.0 & -0.8 & 0 & 0 & 0 \\
HD109443 & 0444F & HD101122-A0V & 12 & 34 & 46.732 & - & 23 & 28 & 32.20 & -0.6 & 8.16 & 0.035 & 76.6 & -1.0 & 0 & 0 & 0 \\
HD110885 & 0450F & HD116960-A0V & 12 & 45 & 19.2569 & + & 01 & 03 & 21.097 & -1.06 & 7.38 & 0.016 & -50.6 & 2.89 & 0 & 0 & 1 \\
HD111786 & 0453F & HD114345-A0V & 12 & 51 & 57.8954 & - & 26 & 44 & 17.783 & -1.5 & 5.4 & 0.023 & -8.6 & 1.09 & 0 & 1 & 0 \\
HD114038 & 0460F & HD109309-A0V & 13 & 07 & 53.8123 & - & 10 & 44 & 25.476 & -0.02 & 2.72 & 0.041 & 34.6 & 1.89 & 0 & 0 & 0 \\
HD114606 & 0462F & HD121996-A0Vs & 13 & 11 & 21.4043 & + & 09 & 37 & 33.519 & -0.53 & 7.09 & 0.000 & 13.2 & -0.06 & 0 & 0 & 0 \\
HD114710 & 0463F & HD107655-A0V & 13 & 11 & 52.3937 & + & 27 & 52 & 41.453 & 0.05 & 2.92 & 0.013 & 57.2 & 0.09 & 0 & 0 & 0 \\
HD115589 & 0467F & HD112304-A0V & 13 & 18 & 13.5353 & - & 14 & 34 & 55.555 & 0.28 & 7.62 & 0.082 & -45.2 & -0.28 & 0 & 0 & 0 \\
HD115617 & 0468F & HD119786-A0V & 13 & 18 & 24.3142 & - & 18 & 18 & 40.304 & 0.02 & 2.96 & 0.000 & 7.3 & -0.06 & 0 & 0 & 0 \\
HD116316 & 0471F & HD121996-A0Vs & 13 & 22 & 34.1508 & + & 26 & 06 & 56.513 & -0.51 & 6.42 & 0.002 & -59.3 & 0.31 & 0 & 0 & 0 \\
HD117200 & 0473F & HD143187-B8V & 13 & 27 & 04.6096 & + & 64 & 44 & 07.890 & 0.03 & 5.7 & 0.011 & -71.6 & 0.88 & 0 & 0 & 0 \\
HD117176 & 0474F & HD121996-A0Vs & 13 & 28 & 25.8086 & + & 13 & 46 & 43.638 & -0.1 & 3.5 & 0.000 & 23.6 & 0.38 & 0 & 0 & 0 \\
HD117635 & 0475F & HD116960-A0V & 13 & 31 & 39.9375 & - & 02 & 19 & 02.602 & -0.42 & 5.31 & 0.000 & -48.1 & 0.03 & 0 & 0 & 0 \\
HD117876 & 0476F & HD121996-A0Vs & 13 & 32 & 48.2137 & + & 24 & 20 & 48.280 & -0.51 & 3.87 & 0.003 & 20.5 & 1.77 & 0 & 0 & 0 \\
HD118244 & 0479F & HD121880-A0V & 13 & 35 & 11.4174 & + & 22 & 29 & 59.031 & -0.46 & 5.63 & 0.020 & 75.8 & 0.49 & 0 & 0 & 0 \\
HD119228 & 0481F & HD118214-A0V & 13 & 40 & 44.2733 & + & 54 & 40 & 53.889 & -0.1 & 0.34 & 0.008 & -11.4 & 3.09 & 1 & 0 & 0 \\
HD119288 & 0482F & HD121996-A0Vs & 13 & 42 & 12.7599 & + & 08 & 23 & 18.226 & -0.23 & 5.11 & 0.000 & -44.2 & 0.51 & 0 & 0 & 0 \\
HD120136 & 0485F & HD116960-A0V & 13 & 47 & 15.7434 & + & 17 & 27 & 24.855 & 0.24 & 3.51 & 0.000 & 30.0 & -1.0 & 0 & 0 & 0 \\
HD120933 & 0487F & HD122945-A0V & 13 & 51 & 47.4753 & + & 34 & 26 & 39.261 & -0.15 & -0.01 & 0.113 & -108.7 & 3.38 & 1 & 0 & 0 \\
HD121258 & 0490F & HD119752-A0V & 13 & 54 & 54.6217 & - & 26 & 00 & 56.758 & -0.2 & 8.8 & 0.283 & -29.9 & -1.0 & 0 & 1 & 0 \\
HD125451 & 0504F & HD121996-A0Vs & 14 & 19 & 16.2803 & + & 13 & 00 & 15.479 & 0.01 & 4.39 & 0.000 & 12.6 & 0.57 & 0 & 0 & 0 \\
BD+012916 & 0505F & HD126129-A0V & 14 & 21 & 45.2615 & + & 00 & 46 & 59.178 & -1.86 & 6.47 & 0.03 & -33.8 & 2.05 & 0 & 0 & 0 \\
HD126141 & 0506F & HD121996-A0Vs & 14 & 23 & 06.8213 & + & 25 & 20 & 17.433 & -0.05 & 5.23 & 0.000 & -6.0 & 0.53 & 0 & 0 & 0 \\
HD126053 & 0507F & HD126129-A0V & 14 & 23 & 15.2849 & + & 01 & 14 & 29.648 & -0.37 & 4.64 & 0.000 & 3.3 & -0.11 & 0 & 0 & 1 \\
HD126614 & 0511F & HD132072-A0V & 14 & 26 & 48.2807 & - & 05 & 10 & 40.007 & 0.52 & 7.06 & 0.055 & -53.9 & 0.18 & 0 & 0 & 0 \\
HD126778 & 0512F & HD127304-A0Vs & 14 & 26 & 54.1638 & + & 28 & 35 & 20.423 & -0.52 & 5.84 & 0.000 & -125.1 & 2.01 & 0 & 0 & 0 \\
HD127243 & 0514F & HD121409-A0V & 14 & 28 & 37.8131 & + & 49 & 50 & 41.458 & -0.78 & 3.16 & 0.060 & 41.2 & 1.9 & 0 & 1 & 1 \\
BD+182890 & 0516F & HD121996-A0Vs & 14 & 32 & 13.4854 & + & 17 & 25 & 24.290 & -1.54 & 7.74 & 0.020 & 84.5 & 1.34 & 0 & 0 & 0 \\
HD128959 & 0521F & HD131885-A0V & 14 & 40 & 43.3622 & - & 26 & 43 & 46.937 & -0.51 & 7.57 & 0.050 & 17.1 & 0.82 & 0 & 0 & 0 \\
HD130095 & 0523F & HD119752-A0V & 14 & 46 & 51.2120 & - & 27 & 14 & 53.958 & -1.7 & 7.8 & 0.068 & 112.6 & 1.31 & 0 & 0 & 0 \\
HD130322 & 0524F & HD131951-A0V & 14 & 47 & 32.724 & - & 00 & 16 & 53.32 & 0.12 & 6.23 & 0.025 & -43.5 & -0.26 & 0 & 0 & 0 \\
HD130817 & 0525F & HD127304-A0Vs & 14 & 49 & 06.7425 & + & 37 & 48 & 40.243 & -0.29 & 5.08 & 0.013 & -40.4 & 0.64 & 0 & 0 & 0 \\
HD130705 & 0526F & HD131951-A0V & 14 & 49 & 26.1554 & + & 10 & 02 & 38.965 & 0.34 & 3.95 & 0.000 & -53.0 & 1.79 & 0 & 0 & 0 \\
HD132142 & 0529F & HD121409-A0V & 14 & 55 & 11.0434 & + & 53 & 40 & 49.260 & -0.37 & 5.8 & 0.000 & 19.3 & -0.38 & 0 & 0 & 0 \\
HD132475 & 0534F & HD138813-A0V & 14 & 59 & 49.7651 & - & 22 & 00 & 45.809 & -1.37 & 6.91 & 0.000 & 143.4 & 0.49 & 0 & 0 & 0 \\
BD+302611 & 0538F & HD127304-A0Vs & 15 & 06 & 53.8278 & + & 30 & 00 & 36.941 & -1.45 & 6.09 & 0.01 & -287.5 & 2.47 & 0 & 0 & 0 \\
HD134063 & 0539F & HD131951-A0V & 15 & 07 & 15.4405 & + & 22 & 33 & 51.695 & -0.69 & 5.47 & 0.019 & -89.7 & 1.49 & 0 & 0 & 0 \\
HD134083 & 0540F & HD131951-A0V & 15 & 07 & 18.0660 & + & 24 & 52 & 09.101 & -0.01 & 3.86 & 0.000 & -9.4 & 0.47 & 0 & 0 & 0 \\
HD134169 & 0541F & HD140775-A0V & 15 & 08 & 18.0580 & + & 03 & 55 & 50.128 & -0.81 & 6.16 & 0.000 & 4.5 & 0.4 & 0 & 1 & 0 \\
HD134440 & 0542F & HD134013-A0V & 15 & 10 & 12.9689 & - & 16 & 27 & 46.516 & -1.34 & 7.19 & 0.000 & 322.2 & -0.84 & 0 & 0 & 0 \\
HD134439 & 0543F & HD134013-A0V & 15 & 10 & 13.0880 & - & 16 & 22 & 45.860 & -1.27 & 7.04 & 0.005 & 289.5 & -0.71 & 0 & 0 & 1 \\
HD134987 & 0544F & HD138295-A0V & 15 & 13 & 28.6670 & - & 25 & 18 & 33.646 & 0.26 & 4.88 & 0.026 & 2.9 & -1.0 & 0 & 0 & 0 \\
HD136064 & 0545F & HD143187-B8V & 15 & 14 & 38.3401 & + & 67 & 20 & 48.197 & 0.03 & 3.87 & 0.010 & -47.0 & 0.58 & 0 & 0 & 0 \\
HD136202 & 0550F & HD140755-A2 & 15 & 19 & 18.7971 & + & 01 & 45 & 55.468 & 0.05 & 4.01 & 0.030 & 33.5 & 0.61 & 0 & 1 & 0 \\
HD137391 & 0553F & HD127304-A0Vs & 15 & 24 & 29.4283 & + & 37 & 22 & 37.757 & 0.1 & 3.62 & 0.004 & -603.0 & -1.0 & 0 & 0 & 0 \\
HD137471 & 0555F & HD140729-A0V & 15 & 25 & 47.3975 & + & 15 & 25 & 40.930 & -0.04 & 1.04 & 0.034 & -28.7 & 3.07 & 0 & 0 & 0 \\
HD138290 & 0559F & HD140729-A0V & 15 & 30 & 55.4332 & + & 08 & 34 & 44.737 & -0.1 & 5.66 & 0.015 & -35.9 & 0.65 & 0 & 0 & 0 \\
HD138776 & 0562F & HD140775-A0V & 15 & 34 & 16.9188 & - & 02 & 43 & 26.643 & 0.35 & 7.11 & 0.069 & 7.0 & 0.17 & 0 & 0 & 0 \\
HD139641 & 0565F & HD127304-A0Vs & 15 & 37 & 49.5979 & + & 40 & 21 & 12.363 & -0.51 & 3.1 & 0.020 & 20.7 & 1.31 & 0 & 0 & 0 \\
BD+053080 & 0569F & HD140775-A2 & 15 & 45 & 52.4015 & + & 05 & 02 & 26.558 & -0.44 & 6.88 & 0.00 & -17.9 & -0.26 & 0 & 0 & 0 \\
HD142575 & 0574F & HD140775-A0V & 15 & 55 & 02.8379 & + & 05 & 04 & 12.148 & -0.7 & 7.51 & 0.019 & -55.0 & 0.9 & 0 & 0 & 0 \\
HD142703 & 0577F & HD138062-A0V & 15 & 56 & 33.3736 & - & 14 & 49 & 45.976 & -1.1 & 5.34 & 0.036 & 9.2 & 0.86 & 0 & 1 & 0 \\
HD144585 & 0584F & HD148968-A0V & 16 & 07 & 03.3694 & - & 14 & 04 & 16.665 & 0.29 & 4.8 & 0.014 & -45.1 & 0.28 & 0 & 0 & 0 \\
HD145148 & 0586F & HD140755-A2 & 16 & 09 & 11.2145 & + & 06 & 22 & 43.302 & 0.1 & 3.68 & 0.000 & -0.2 & 0.53 & 0 & 0 & 0 \\
HD145976 & 0589F & HD145122-A0Vnn & 16 & 12 & 45.4709 & + & 26 & 40 & 14.094 & -0.02 & 5.48 & 0.018 & -8.3 & 1.04 & 0 & 0 & 0 \\
BD+112998 & 0598F & HD140775-A0V & 16 & 30 & 16.7825 & + & 10 & 59 & 51.748 & -1.01 & 7.18 & 0.024 & 69.8 & 1.26 & 0 & 0 & 0 \\
HD149009 & 0603F & HD164899-A0Vs & 16 & 31 & 13.4310 & + & 22 & 11 & 43.649 & 0.09 & 2.04 & 0.075 & -28.0 & 2.84 & 1 & 0 & 0 \\
HD150177 & 0611F & HD159415-A0V & 16 & 39 & 39.1298 & - & 09 & 33 & 16.512 & -0.58 & 4.98 & 0.000 & -46.7 & 0.6 & 0 & 0 & 0 \\
HD152601 & 0616F & HD148968-A0V & 16 & 54 & 35.6932 & - & 06 & 09 & 14.333 & 0.1 & 2.8 & 0.023 & -45.0 & 1.65 & 0 & 0 & 0 \\
HD152781 & 0617F & HD149134-A0V & 16 & 56 & 01.8443 & - & 16 & 48 & 22.553 & 0.08 & 4.26 & 0.048 & -19.3 & 0.79 & 0 & 0 & 1 \\
HD154733 & 0620F & HD165029-A0V & 17 & 06 & 18.0476 & + & 22 & 05 & 02.955 & -0.09 & 2.48 & 0.018 & -127.4 & 1.98 & 0 & 0 & 0 \\
HD155763 & 0621F & HD143187-B8V & 17 & 08 & 47.1959 & + & 65 & 42 & 52.863 & -0.11 & 3.55 & 0.030 & 23.6 & 2.58 & 0 & 0 & 0 \\
HD155358 & 0622F & HD157778-A0Vn & 17 & 09 & 34.6174 & + & 33 & 21 & 21.085 & -0.63 & 5.81 & 0.000 & 15.7 & 0.3 & 0 & 0 & 1 \\
HD157089 & 0628F & HD161289-A0V & 17 & 21 & 07.0556 & + & 01 & 26 & 34.988 & -0.57 & 5.42 & 0.000 & -133.8 & 0.26 & 0 & 0 & 0 \\
HD157881 & 0630F & HD161289-A0V & 17 & 25 & 45.2326 & + & 02 & 06 & 41.120 & -0.01 & 4.37 & 0.000 & -54.4 & -0.95 & 0 & 0 & 0 \\
HD159482 & 0636F & HD161289-A0V & 17 & 34 & 43.0609 & + & 06 & 00 & 51.574 & -0.75 & 6.79 & 0.004 & -103.1 & -0.02 & 0 & 0 & 1 \\
HD160693 & 0639F & HD164899-A0Vs & 17 & 39 & 36.8724 & + & 37 & 11 & 01.506 & -0.56 & 6.9 & 0.000 & 23.7 & 0.06 & 0 & 0 & 1 \\
HD161149 & 0641F & HD165029-A0V & 17 & 43 & 22.0221 & + & 14 & 17 & 42.604 & 0.33 & 5.1 & 0.043 & -53.5 & -1.0 & 0 & 0 & 0 \\
HD162211 & 0648F & HD165029-A0V & 17 & 48 & 49.1464 & + & 25 & 37 & 22.329 & -0.06 & 2.4 & 0.001 & -42.5 & 1.65 & 0 & 0 & 0 \\
HD164349 & 0654F & HD165029-A0V & 18 & 00 & 03.4160 & + & 16 & 45 & 03.309 & -0.01 & 1.94 & 0.001 & -35.8 & 2.88 & 0 & 0 & 0 \\
HD166460 & 0667F & HD161289-A0V & 18 & 10 & 40.2995 & + & 03 & 19 & 27.325 & -0.05 & 2.58 & 0.023 & 10.1 & 2.15 & 0 & 0 & 0 \\
HD170737 & 0679F & HD174567-A0Vs & 18 & 29 & 54.1103 & + & 26 & 39 & 26.247 & -0.77 & 5.95 & 0.025 & -110.8 & 0.88 & 0 & 0 & 0 \\
HD171496 & 0683F & HD177213-A0V & 18 & 36 & 07.4940 & - & 24 & 26 & 11.320 & -0.73 & 5.59 & 0.151 & -18.1 & 1.55 & 0 & 0 & 0 \\
HD173740 & 0690F & HD172728-A0V & 18 & 42 & 46.9665 & + & 59 & 37 & 36.347 & -0.34 & 5.0 & 0.001 & 16.6 & -2.06 & 0 & 0 & 0 \\
HD173819 & 0697F & HD170654-A0V & 18 & 47 & 28.9498 & - & 05 & 42 & 18.541 & -0.67 & 2.16 & 0.260 & 19.3 & 4.27 & 1 & 0 & 0 \\
HD174912 & 0699F & HD174567-A0Vs & 18 & 51 & 25.1791 & + & 38 & 37 & 35.654 & -0.45 & 5.68 & 0.000 & 3.3 & 0.03 & 0 & 0 & 0 \\
HD175535 & 0703F & CCDM J19302+5639AB-A0V & 18 & 53 & 13.5416 & + & 50 & 42 & 29.585 & -0.07 & 2.82 & 0.007 & -18.6 & 2.25 & 0 & 0 & 0 \\
HD175588 & 0704F & HD174567-A0Vs & 18 & 54 & 30.2833 & + & 36 & 53 & 55.013 & -0.14 & -1.25 & 0.000 & -27.0 & 4.06 & 0 & 0 & 1 \\
HD175640 & 0706F & HD171149-A0Vn & 18 & 56 & 22.6604 & - & 01 & 47 & 59.505 & 0.17 & 6.25 & 0.044 & 24.4 & 1.84 & 0 & 0 & 0 \\
NGC6791-R4 & 0940C & HD174567-A0Vs & 19 & 20 & 49.7000 & + & 37 & 43 & 41.2 & 0.42 & 7.82 & 0.119 & -45.5 & 2.78 & 0 & 0 & 0 \\
NGC6791-R19 & 0943C & HD174567-A0Vs & 19 & 20 & 52.8000 & + & 37 & 44 & 26.900 & 0.42 & 10.19 & 0.119 & -41.2 & 2.03 & 0 & 0 & 0 \\
HD182572 & 0721F & HD182761-A0V & 19 & 24 & 58.2002 & + & 11 & 56 & 39.886 & 0.34 & 3.04 & 0.031 & 608.6 & -1.0 & 0 & 0 & 0 \\
CD-2415398 & 0723F & HD177120-A0V & 19 & 32 & 20.7821 & - & 23 & 51 & 12.755 & -1.17 & 9.06 & 0.097 & 17.7 & 3.17 & 0 & 0 & 1 \\
HD184499 & 0726F & HD174567-A0Vs & 19 & 33 & 27.0812 & + & 33 & 12 & 06.716 & -0.54 & 5.08 & 0.000 & -123.8 & 0.31 & 0 & 0 & 1 \\
HD185657 & 0730F & HD178207-A0Vn & 19 & 37 & 56.7133 & + & 49 & 17 & 03.902 & -0.19 & 4.09 & 0.000 & -55.6 & 1.74 & 0 & 0 & 0 \\
HD186408 & 0733F & HD194354-A0Vs & 19 & 41 & 48.9534 & + & 50 & 31 & 30.215 & 0.08 & 4.43 & 0.000 & -54.8 & 0.2 & 1 & 0 & 0 \\
M71\_1-77 & 0967C & * 5 Vul-A0V & 19 & 53 & 37.4000 & - & 1 & -1 & -1 & -0.22 & 12.11 & 0.270 & 26.4 & 1.58 & 1 & 0 & 0 \\
M71\_I & 0971C & * 5 Vul-A0V & 19 & 53 & 44.6000 & + & 18 & 46 & 35.0000 & -0.78 & 8.66 & 0.270 & -9.0 & 2.71 & 0 & 0 & 0 \\
HD188510 & 0742F & HD182919-A0V & 19 & 55 & 09.6783 & + & 10 & 44 & 27.372 & -1.43 & 7.13 & 0.000 & -134.7 & -0.39 & 0 & 1 & 0 \\
HD189558 & 0747F & HD182678-A0V & 20 & 01 & 00.2457 & - & 12 & 15 & 20.340 & -1.04 & 6.16 & 0.000 & -1.6 & 0.47 & 0 & 0 & 0 \\
HD190360 & 0749F & HD192538-A0V & 20 & 03 & 37.4058 & + & 29 & 53 & 48.494 & 0.21 & 4.08 & 0.000 & -55.4 & 0.01 & 0 & 0 & 0 \\
HD190178 & 0752F & HD195549-A0V & 20 & 04 & 49.4677 & - & 28 & 16 & 25.164 & -0.66 & 8.07 & 0.002 & 22.8 & 0.64 & 0 & 0 & 0 \\
HD191046 & 0755F & HD192538-A0V & 20 & 06 & 28.9887 & + & 36 & 13 & 35.918 & -0.75 & 4.19 & 0.040 & -84.5 & 2.5 & 0 & 0 & 1 \\
HD345957 & 0757F & HD182919-A0V & 20 & 10 & 48.1619 & + & 23 & 57 & 54.514 & -1.2 & 7.36 & 0.000 & -74.7 & 0.35 & 0 & 0 & 0 \\
HD192640 & 0759F & HD192538-A0V & 20 & 14 & 32.0333 & + & 36 & 48 & 22.692 & -0.8 & 4.42 & 0.026 & -0.6 & 1.18 & 0 & 0 & 0 \\
HD192909 & 0760F & HD199217-A0V & 20 & 15 & 28.3228 & + & 47 & 42 & 51.160 & -0.02 & 0.16 & 0.016 & -61.8 & 3.78 & 1 & 0 & 0 \\
HD196755 & 0768F & HD198070-A0Vn & 20 & 39 & 07.7843 & + & 10 & 05 & 10.338 & -0.02 & 3.61 & 0.014 & -85.0 & 0.76 & 0 & 0 & 0 \\
BD+044551 & 0777F & HD198070-A0Vn & 20 & 48 & 50.7212 & + & 05 & 11 & 58.801 & -1.26 & 8.25 & 0.00 & -82.7 & 1.63 & 1 & 0 & 0 \\
HD199191 & 0779F & HD205314-A0V & 20 & 53 & 18.3149 & + & 54 & 31 & 05.164 & -0.7 & 4.73 & 0.010 & -207.3 & 1.54 & 0 & 0 & 0 \\
HD200527 & 0782F & HD201076-A0V & 21 & 02 & 24.1998 & + & 44 & 47 & 27.523 & -0.07 & 1.09 & 0.000 & -9.7 & 3.21 & 0 & 0 & 0 \\
HD200779 & 0785F & HD210501-A0V & 21 & 05 & 19.7452 & + & 07 & 04 & 09.477 & 0.02 & 5.31 & 0.003 & -47.6 & -0.76 & 0 & 0 & 0 \\
HD201891 & 0790F & HD208108-A0Vs & 21 & 11 & 59.0315 & + & 17 & 43 & 39.890 & -1.05 & 5.93 & 0.000 & -86.8 & 0.03 & 0 & 0 & 0 \\
HD201889 & 0791F & HD210501-A0V & 21 & 11 & 59.5235 & + & 24 & 10 & 05.389 & -0.74 & 6.43 & 0.000 & -83.9 & 0.31 & 0 & 0 & 1 \\
HD203638 & 0795F & HD202025-A0V & 21 & 24 & 09.5924 & - & 20 & 51 & 06.740 & 0.12 & 2.69 & 0.010 & 6.4 & 1.7 & 0 & 0 & 0 \\
HD204041 & 0796F & HD198070-A0Vn & 21 & 25 & 51.5837 & + & 00 & 32 & 03.620 & -0.44 & 5.96 & 0.025 & 29.3 & 0.96 & 0 & 1 & 0 \\
HD204155 & 0798F & HD20722-A0 D & 21 & 26 & 42.9055 & + & 05 & 26 & 29.902 & -0.69 & 7.01 & 0.000 & -49.5 & 0.31 & 0 & 0 & 0 \\
HD204613 & 0799F & HD205314-A0V & 21 & 27 & 42.9678 & + & 57 & 19 & 18.861 & -0.38 & 6.79 & 0.008 & -102.9 & 0.33 & 0 & 0 & 0 \\
HD204587 & 0804F & HD203893-A0V & 21 & 30 & 02.7541 & - & 12 & 30 & 36.255 & -0.11 & 5.88 & 0.004 & -86.4 & -0.86 & 0 & 0 & 0 \\
HD205153 & 0806F & HD202941-A0V & 21 & 34 & 09.1873 & - & 27 & 54 & 04.333 & 0.07 & 6.83 & 0.025 & -27.5 & 0.97 & 0 & 0 & 0 \\
HD205512 & 0807F & HD209932-A0V & 21 & 34 & 46.5638 & + & 38 & 32 & 02.604 & 0.03 & 2.6 & 0.000 & -108.5 & 1.69 & 0 & 0 & 0 \\
HD206078 & 0808F & HD201184-A0V & 21 & 37 & 10.4225 & + & 62 & 18 & 15.283 & -0.59 & 4.72 & 0.002 & -120.4 & 1.61 & 0 & 1 & 0 \\
HD206453 & 0811F & HD201184-A0V & 21 & 42 & 39.5071 & - & 18 & 51 & 58.766 & -0.41 & 2.55 & 0.033 & 40.2 & -1.0 & 1 & 0 & 0 \\
HD207222 & 0818F & HD208108-A0Vs & 21 & 46 & 56.2591 & + & 21 & 17 & 50.768 & -0.36 & 8.37 & 0.057 & -7.0 & 1.11 & 0 & 0 & 0 \\
HD210295 & 0825F & HD215143 - B7.5V & 22 & 09 & 41.4440 & - & 13 & 36 & 19.478 & -1.26 & 7.18 & 0.000 & -23.9 & 0.98 & 0 & 0 & 0 \\
HD211075 & 0830F & HD208108-A0Vs & 22 & 14 & 20.0378 & + & 18 & 01 & 13.370 & -0.42 & 5.22 & 0.016 & 33.5 & 2.14 & 1 & 1 & 0 \\
HD213042 & 0835F & HD220455-A0V & 22 & 29 & 15.2354 & - & 30 & 01 & 06.271 & 0.12 & 5.11 & 0.076 & 13.5 & -0.63 & 0 & 0 & 0 \\
HD216174 & 0845F & HD219290-A0V & 22 & 49 & 46.3148 & + & 55 & 54 & 09.999 & -0.61 & 2.63 & 0.030 & -47.4 & 2.03 & 0 & 0 & 0 \\
HD216131 & 0846F & BD+3904890-A0V & 22 & 50 & 00.1931 & + & 24 & 36 & 05.698 & -0.07 & 1.18 & 0.000 & 87.9 & 1.69 & 0 & 0 & 0 \\
HD217107 & 0855F & HD215143-B7.5V & 22 & 58 & 15.5411 & - & 02 & 23 & 43.384 & 0.33 & 4.54 & 0.019 & 10.9 & 0.01 & 0 & 0 & 0 \\
HD218640 & 0861F & HD223352-A0Vnp & 23 & 09 & 54.896 & - & 22 & 27 & 27.44 & 0.36 & 2.95 & 0.021 & 25.0 & -1.0 & 0 & 0 & 0 \\
HD219449 & 0867F & HD219833-A0V & 23 & 15 & 53.4948 & - & 09 & 05 & 15.854 & 0.03 & 1.6 & 0.000 & -22.1 & 1.78 & 0 & 1 & 0 \\
HD219623 & 0868F & HD211301-A0V & 23 & 16 & 42.3033 & + & 53 & 12 & 48.510 & 0.07 & 4.31 & 0.000 & 17.0 & 0.1 & 0 & 0 & 1 \\
HD219615 & 0870F & HD215143-B7.5V & 23 & 17 & 09.9374 & + & 03 & 16 & 56.238 & -0.57 & 1.39 & 0.000 & 1.8 & 1.81 & 0 & 0 & 1 \\
HD219734 & 0871F & HD219290-A0V & 23 & 17 & 44.6471 & + & 49 & 00 & 55.081 & -0.04 & 0.51 & 0.019 & 12.1 & 3.08 & 0 & 0 & 0 \\
HD219978 & 0873F & HD223386-A0V & 23 & 19 & 23.7734 & + & 62 & 44 & 23.185 & 0.18 & 0.79 & 0.228 & -27.4 & 3.92 & 0 & 1 & 0 \\
HD220009 & 0874F & HD210501-A0V & 23 & 20 & 20.5831 & + & 05 & 22 & 52.701 & -0.8 & 1.99 & 0.000 & 85.4 & 2.43 & 0 & 1 & 0 \\
HD221148 & 0881F & HD215143-B7.5V & 23 & 29 & 32.0820 & - & 04 & 31 & 57.891 & 0.34 & 3.89 & 0.000 & 0.5 & 0.8 & 0 & 0 & 0 \\
HD221377 & 0883F & HD219290-A0V & 23 & 31 & 19.7259 & + & 52 & 24 & 38.504 & -0.6 & 6.37 & 0.005 & 37.3 & 0.64 & 0 & 0 & 0 \\
HD221830 & 0886F & HD210501-A0V & 23 & 35 & 28.8937 & + & 31 & 01 & 01.830 & -0.4 & 5.3 & 0.000 & -71.4 & 0.2 & 0 & 0 & 0 \\
G171-010 & 0890F & HD219290-A0V & 23 & 41 & 54.989 & + & 44 & 10 & 40.78 & 0.09 & 5.93 & 0.088 & -36.3 & -2.62 & 0 & 0 & 0 \\
HD223524 & 0893F & HD219833-A0V & 23 & 50 & 14.7298 & - & 09 & 58 & 26.893 & 0.02 & 3.22 & 0.013 & 2.5 & 1.67 & 0 & 0 & 0 \\
HD224458 & 0895F & HD006457-A0Vn & 23 & 58 & 06.2501 & + & 29 & 58 & 35.906 & -0.44 & 5.73 & 0.050 & -114.2 & 1.55 & 0 & 0 & 0 \\

\tablecomments{The Quality, Parameter, and Shape flags are indicators of good the spectra are. A star marked with `1' in any of those categories might have some issue with it: (1) The IRTF spectra are unacceptably noisy, (2) the stellar parameters seem inconsistent with the data, (3) something may be wrong with the flux calibration that is causing issues with the spectral shape.}

\end{deluxetable*}

\clearpage

\end{landscape}

\end{document}